\documentclass[12pt]{article}

\ifx\pdfoutput\undefined
\usepackage[dvips,bookmarks]{hyperref}
\else
\usepackage{hyperref}
\fi
\hypersetup{colorlinks=false,bookmarksopen,bookmarksnumbered,citecolor=blue,
   pdfstartview=FitH}

\usepackage{latexsym}
\usepackage{amssymb,amsfonts,amsmath}
\usepackage{graphicx} 
\usepackage{indentfirst}
\usepackage{bbm}
\usepackage{amssymb}
\usepackage{verbatim}
\usepackage{amsmath, amsthm,amssymb}
\usepackage{mathrsfs}
\usepackage{hyperref}
\usepackage{amsfonts}
\usepackage{dsfont}

\parskip=\medskipamount

\newcommand {\cD}{{\cal D}}
\newcommand {\cE}{{\cal E}}

\newcommand {\cG}{{\cal G}}
\newcommand {\cH}{{\cal H}}

\newcommand {\cJ}{{\cal J}}
\newcommand {\cK}{{\cal K}}
\newcommand {\cL}{{\cal L}}
\newcommand {\cM}{{\cal M}}
\newcommand {\cN}{{\cal N}}
\newcommand {\cO}{{\cal O}}

\newcommand {\cS}{{\cal S}}

\newcommand {\cV}{{\cal V}}

\def\a{\alpha}
\def\b{\beta}
\def\c{\chi}
\def\d{\delta}

\def\g{\gamma}

\def\k{\kappa}
\def\l{\lambda}
\def\m{\mu}

\def\o{\omega}
\def\p{\pi}
\def\q{\theta}

\def\s{\sigma}
\def\t{\tau}

\def\x{\xi}

\def\D{\Delta}

\def\L{\Lambda}

\def\S{\Sigma}

\def\X{\Xi}

\def\tr{{\rm tr}}

\def\ri{{\rm i}}

\newcommand{\ve}{\varepsilon}

\newcommand{\pa}{\partial}  
\newcommand{\hf}{\frac12}

\newcommand{\be}{\begin{equation}}
\newcommand{\ee}{\end{equation}}
\newcommand{\bea}{\begin{eqnarray}}
\newcommand{\eea}{\end{eqnarray}}
\newcommand{\non}{\nonumber}
\newcommand{\ba}{\begin{array}}
\newcommand{\ea}{\end{array}}

\newcommand{\bm}[1]{\mbox{\boldmath$#1$}}

\def\double #1{#1{\hbox{\kern-2pt $#1$}}}

\newcommand{\bsubeq}{\begin{subequations}}
\newcommand{\esubeq}{\end{subequations}}

\newcommand{\ul}{\underline}
\newcommand{\eps}{{\ve}}

\newcommand{\rd}{\mathrm d}
\numberwithin{equation}{section}

\renewcommand{\eps}{\ve}

\newcommand{\RM}{R(M)}
\newcommand{\RD}{R(\mathbb D)}
\newcommand{\RN}{R(N)}

\newcommand{\RS}{R(S)}
\newcommand{\RK}{R(K)}

\newcommand{\rL}{{\rm L}}
\newcommand{\rR}{{\rm R}}

\begin{document}
\begin{titlepage}
\begin{flushright}
January, 2014\\
\end{flushright}

\begin{center}
{\Large \bf 
Supergravity-matter actions 
in three dimensions
and Chern-Simons terms 
}
\end{center}

\begin{center}
{\bf
Sergei M. Kuzenko
and
Joseph Novak
} \\
\vspace{5mm}

\footnotesize{
{\it School of Physics M013, The University of Western Australia\\
35 Stirling Highway, Crawley W.A. 6009, Australia}}  
~\\
\texttt{joseph.novak@uwa.edu.au}\\
\vspace{2mm}

\end{center}

\begin{flushright}
{\it Dedicated to Professor Martin Ro\v{c}ek}\\
{\it on the occasion of his 60th birthday}
\end{flushright}

\begin{abstract}
\baselineskip=14pt
 We study off-shell $\cN$-extended Yang-Mills multiplets coupled to 
 conformal supergravity in three spacetime dimensions. 
 Superform formulations are presented for the  non-Abelian  Chern-Simons actions 
in the cases $\cN=1,\,2,\,3$, and the corresponding component actions are explicitly worked out.
Such a Chern-Simons action 
does not exist 
for $\cN=4$. In the latter case, a superform formulation 
is given for the $BF$ term that describes the coupling of two Abelian  vector multiplets 
with self-dual and anti-self-dual superfield strengths respectively.
The superform results obtained are used to construct linear multiplet action 
principles in the cases  $\cN=2,\,3,\,4$.  The $\cN=3$ and $\cN=4$ actions 
are demonstrated to be universal in the sense that all known off-shell 
supergravity-matter systems (with the exception of  pure conformal supergravity) 
may be described using such an action. Starting from  the $\cN=3$ and $\cN=4$
Abelian vector multiplets,  we also construct composite $\cO(2)$ multiplets which are 
analogues of the four-dimensional construction of an $\cN=2$ reduced chiral scalar engineered from the improved tensor multiplet. 
Using  these composites, we present the superfield equations of motion 
for $\cN=3$ and $\cN=4$ anti-de Sitter and topologically massive supergravity theories. We also sketch the construction of a 
large family of higher derivative couplings for $\cN=3$ and $\cN=4$ vector multiplets. 
\end{abstract}

\vfill
\end{titlepage}

\newpage
\renewcommand{\thefootnote}{\arabic{footnote}}
\setcounter{footnote}{0}

\tableofcontents{}
\vspace{1cm}
\bigskip\hrule


\section{Introduction}

The linear multiplet\footnote{In four-dimensional $\cN=2$ Poincar\'e supersymmetry, 
the linear multiplet was introduced by Sohnius  \cite{Sohnius1} as a superfield Lagrangian 
for the matter hypermultiplet \cite{Fayet} coupled to the Yang-Mills vector multiplet 
\cite{GSW}.  The linear multiplet action 
 was generalized  to $\cN=2$ supergravity by Breitenlohner and Sohnius \cite{BS}, and then 
reformulated by de Wit, van Holten and Van Proeyen
\cite{deWvHVP3} within the $\cN=2$ superconfomal tensor calculus
\cite{deWvHVP,BdeRdeW,deWvHVP2}, see \cite{FVP}
for a recent review. The linear multiplet actions, 
and their use, in five-dimensional $\cN=1$ and six-dimensional $\cN=(1,0)$
supergravity theories  were described in \cite{Zucker,Ohashi,Bergshoeff} 
and \cite{BSV} respectively. It should be mentioned that 
in five dimensions different authors use different notations, 
$\cN=1$ or $\cN=2$, for supersymmetric theories with eight supercharges.  
The notation $\cN=1$ is used, e.g., in Refs. \cite{KT-M08}. The rationale for its use
is that the case of eight supercharges corresponds to simple supersymmetry. 
The alternative notation $\cN=2$ is used in \cite{Zucker,Ohashi,Bergshoeff}. The rationale for 
this choice is that dimensional reduction of five-dimensional theories with eight 
supercharges leads to $\cN=2$ theories in four dimensions. 
Here we follow the conventions of \cite{KT-M08}.}
plays an important role in the context of matter-coupled
supergravity theories with eight supercharges in four, five and six dimensions. 
There are two 
reasons for the significance of this representation of supersymmetry  that 
can be attributed to its 
possible realizations as: 
(i) a dynamical multiplet; or 
(ii) a composite multiplet. In the first realization, the linear multiplet without 
central charge \cite{SSW2} (nowadays, often called the $\cO(2)$ multiplet 
\cite{LR2008,Kuzenko_lectures})
 provides a dual off-shell formulation for the massless hypermultiplet, 
in which one of the four physical scalars of the hypermultiplet is dualized into a 
gauge $(d-2)$-form in $d$ dimensions.  
In the $d=4$ case, the $\cO(2)$ multiplet 
describes the field strength of the massless $\cN=2$ tensor multiplet
\cite{Wess,deWvHVP2}.
In the second realization, which is most relevant for this paper, 
the linear multiplet takes on the role of a Lagrangian
for a locally supersymmetric action \cite{BS,deWvHVP3}. 
This action principle turns out to be  
universal in the sense that 
it supports general off-shell supergravity-matter theories.\footnote{Its universality 
may be readily justified in the case of 4D $ \cN=2$ supergravity. 
Within  the off-shell formulation for supergravity-matter systems
given in \cite{KLRT-M}, any dynamical system can be described 
using the curved projective superspace action. 
This action can be recast as a chiral action with specially chosen
Lagrangian \cite{KT-M09}.  
The latter may equivalently be rewritten, using a simple transformation, 
as a linear multiplet action.}
Different theories correspond to different  composite linear multiplets. 
In this paper we present three-dimensional (3D) analogues of the linear multiplet action. 

The linear multiplet action actually involves   two building blocks: 
an Abelian vector multiplet and a linear multiplet, the latter  with or without central charge (no central charge is possible in six dimensions). 
The vector multiplet is dynamical and model-independent. 
The linear multiplet is composite and contains all the information about 
the dynamical system under consideration. 
Within the superconformal 
tensor calculus, the action is formulated in terms of the component fields
\cite{deWvHVP3}, which is useful for many applications. However, 
this component approach obscures a geometric origin of the action. 
On the other hand, the action acquires a simple geometric 
interpretation as a supersymmetric $BF$ term 
when formulated in  curved 4D $\cN=2$ harmonic superspace \cite{KT} 
(as an extension of the rigid supersymmetric construction given in \cite{DIKST})
or, in the case of the linear multiplet without central charge,  
in curved 4D $\cN=2$, 5D $\cN=1$ and 6D $\cN=(1,0)$  projective superspaces 
\cite{KLRT-M,KT-M08,LT-M}.\footnote{The harmonic \cite{GIKOS,GIOS} 
and projective \cite{KLR,LR-proj} superspaces are powerful 
approaches to engineer off-shell supersymmetric theories
with eight supercharges.} 
From the viewpoint of $x$-space practitioners, a 
disadvantage of these superspace approaches 
is that some work is required in order to reduce the action
to components. Recently, there has appeared 
a new formulation for the 4D $\cN=2$ linear multiplet action \cite{BKN} 
that combines the advantages of both the superconformal tensor calculus
and the powerful superspace techniques. It made use of 
4D $\cN=2$ conformal superspace \cite{Butter-N=2} in conjunction with 
 the superform approach to the  construction of  supersymmetric invariants 
\cite{Castellani,Hasler,Ectoplasm,GGKS}. 

The superform formulation given in  \cite{BKN},  
and its extension to describe 3D $\cN=1$ conformal supergravity \cite{KT-M12}, 
has recently been applied to derive off-shell $\cN$-extended conformal 
supergravity actions in three dimensions for the cases 
$\cN \leq 6$ \cite{BKNT-M2,KNT-M}.\footnote{The off-shell action for 3D $\cN=6$ conformal supergravity  was independently constructed in 
\cite{NT}.} In the  past, the off-shell actions were known only 
for $\cN=1$ \cite{vN} (see also \cite{KT-M12}) and $\cN=2$ \cite{RvN}  
 conformal supergravities.
Refs. \cite{BKNT-M2,KNT-M} made use of the novel off-shell formulation for 
3D $\cN$-extended conformal supergravity \cite{BKNT-M1} called
conformal superspace.\footnote{The conventional off-shell formulation 
for 3D $\cN$-extended conformal supergravity \cite{HIPT,KLT-M11}, 
also known as SO$(\cN)$ superspace, is obtained from conformal superspace 
by gauge fixing some of the local symmetries, see \cite{BKNT-M1} for more details.
Within the SO$(\cN)$ superspace setting, 
the most general off-shell supergravity-matter couplings were constructed 
in \cite{KLT-M11} for the cases $1\leq \cN\leq 4$.}
Within the superspace setting of \cite{BKNT-M1}, 
conformal supergravity is simply a  gauge theory 
of the $\cN$-extended superconformal group. 
Conceptually, this supergravity formulation 
is very similar to that for $\cN$-extended Yang-Mills multiplets in superspace. 
Here we use this analogy to develop a superform realization,
in conformal superspace, for $\cN$-extended supersymmetric 
Chern-Simons actions,  with $1 \leq \cN \leq 4$. 
 Using different techniques, the supersymmetric Chern-Simons actions were 
originally constructed in \cite{Siegel,GGRS} for the case $\cN=1$, 
in \cite{ZP, Ivanov91} for $\cN=2$,\footnote{The Abelian $\cN=2$ Chern-Simons
action was first constructed by Siegel \cite{Siegel}.} 
in \cite{ZH} for $\cN=3$. 
The $\cN=4$ supersymmetric $BF$ term was first constructed 
in components \cite{BrooksG}, 
then in $\cN=2$ superspace \cite{KS},   $\cN=4$ harmonic 
superspace \cite{Zupnik99,Zupnik2009} and also in $\cN=3$ harmonic superspace
\cite{Zupnik2010}.  
The $\cN=4$ case is actually very special, 
since a non-Abelian $\cN=4$ Chern-Simons action does not exist. 
This will be discussed in more detail in the main body of our paper.  

Using the superform realization of the Chern-Simons actions given, it becomes 
trivial to construct linear multiplet actions for  the cases $\cN=2,\,3,\,4$; 
the relevant constructions are given in the main body of our paper. 
We demonstrate that 
these actions are actually universal for $\cN=3$ and $\cN=4$ in the sense that 
the most general off-shell $\cN=3$ and $\cN=4$ supergravity-matter systems presented  in \cite{KLT-M11} may be described using the 
appropriate linear multiplet action. 
This simplifies the problem of constructing component actions for 
$\cN = 3$ and $\cN = 4$ off-shell supergravity-matter systems.
We should emphasize that our statement of universality concerns 
the off-shell locally supersymmetric theories. 
The on-shell locally supersymmetric nonlinear sigma models in three dimensions
have been described, e.g.,  in 
\cite{dWNT,dWHS,dWNS,BCSS,NR02}.

This paper is organized as follows. 
In section \ref{VM} 
we describe 
the $\cN$-extended non-Abelian vector multiplet in  
conformal superspace. 
In section \ref{C3FCSA} our method to construct supersymmetric Chern-Simons 
actions is briefly described. 
In section \ref{CSCI3F} we derive the curvature induced 
three-forms for $\cN \leq 4$. The component expressions for the supersymmetric 
Chern-Simons actions 
with 
$\cN \leq 4$ are given in section \ref{ComponentActions}. 
Section \ref{Section6} is devoted to the $\cN=2$ linear multiplet action. 
In section \ref{N=3composite} we work out the $\cN=3$ linear multiplet action 
and apply this construction to the cases of (2,1) anti-de Sitter supergravity and 
$\cN=3$ topologically massive supergravity. 
In section \ref{N=4composite} we work out two $\cN=4$ linear multiplet actions 
and make use of these actions to study  (2,2) anti-de Sitter supergravity and 
$\cN=4$ topologically massive supergravity. 
Some implications of our results and open problems 
are briefly discussed in section \ref{conclusion}.

We have included a couple of technical appendices. Appendix \ref{geometry} includes some salient facts about the conformal superspace 
of \cite{BKNT-M1}. In Appendix \ref{SUSYTrans} we give the supersymmetry transformations for vector multiplets with $\cN \leq 4$.
In Appendix \ref{AppendixC} we briefly review
covariant projective  $\cN=3$ supermultiplets and demonstrate universality 
of the $\cN=3$ linear multiplet action. In Appendix \ref{AppendixD} 
we sketch the structure of left and right covariant projective $\cN=4$ 
supermultiplets and  demonstrate universality 
of the two $\cN=4$ linear multiplet actions. 


\section{Vector multiplets in conformal superspace}\label{VM}

In this section we show how to describe Yang-Mills multiplets within the superspace formulation 
of \cite{BKNT-M1}, known as conformal superspace. Conformal superspace is based on 
gauging the entire superconformal algebra. 
Its essential aspects are summarized 
in Appendix \ref{geometry}.

To describe a Yang-Mills multiplet in the 3D 
$\cN$-extended conformal superspace $\cM^{3|2\cN}$ of \cite{BKNT-M1}, 
 parametrized by coordinates $z^M = (x^m, \ \q^\mu_I)$, 
we introduce gauge covariant derivatives
\bea
\bm \nabla = E^A \bm \nabla_A \ , \quad {\bm\nabla}_A := \nabla_A - \ri V_A~,
\eea
with $E_A =E_A{}^M \pa_M$ the inverse vielbein, 
$\nabla_A$ the superspace covariant derivatives obeying the (anti-)commutation relations \eqref{nablanabla} and 
$V = E^A V_A$ the gauge connection taking its values in the Lie algebra 
of the Yang-Mills gauge group $G_{\rm YM}$. The generators of $G_{\rm YM}$
commutes with all the generators of the superconformal algebra \eqref{SCA}. 
The Yang-Mills gauge transformation acts on the gauge covariant 
derivatives as
\be 
\bm \nabla_A \rightarrow e^{\ri \t} \bm \nabla_A e^{- \ri \t } , \quad \t^\dag = \t \ ,
\label{2.2}
\ee
where the gauge parameter $\t (z)$ takes its values in the Lie algebra 
of $G_{\rm YM}$. 

The gauge covariant derivative algebra is
\begin{align} [{\bm \nabla}_A, {\bm \nabla}_B\} &= -T_{AB}{}^C{\bm \nabla}_C
-\hf \RM_{AB}{}^{cd} M_{cd}
-\hf \RN_{AB}{}^{PQ} N_{PQ}
- \RD_{AB} \mathbb D \non\\
&\quad - \RS_{AB}{}^\g_I S_\g^I
	- \RK_{AB}{}^c K_c
	- \ri F_{AB} \ ,
\end{align}
where the torsion and curvatures are those of conformal superspace but with $F_{AB}$ corresponding 
to the gauge covariant field strength $F = \hf E^B \wedge E^A F_{AB}$.
The field strength $F_{AB}$ 
satisfies the Bianchi identity
\be \bm \nabla F = 0 \ , \quad \bm \nabla_{[A} F_{BC\}} + T_{[AB}{}^D F_{|D| C\}} = 0
\ee
and must be subject to covariant constraints to describe an irreducible vector multiplet. The structure 
of the constraints and their consequence is different for $\cN = 1$ and for $\cN > 1$. Below we describe 
the various cases.

\subsection{The $\cN = 1$ case}

In the $\cN=1$ case, 
one imposes the covariant constraint \cite{Siegel,GGRS}
\bea
F_{\a\b}=0~. \label{N=1FConst}
\eea
Then one derives from the Bianchi identities the remaining components
\bsubeq
\bea
F_{a\b}&=&\hf (\g_a)_\b{}^\g G_\g
~,  \label{2.6a}\\
F_{ab}&=&-\frac{\ri}{4}\ve_{abc}(\g^c)^{\g\d} \bm \nabla_\g G_\d
~,
\eea
\esubeq
together with the dimension-2 differential constraint on the spinor field strength 
\bea
\bm \nabla^\a G_\a=0~.\label{N=1Const}
\eea
Furthermore, the Jacobi identities require $G_\a$ to be primary and of dimension-3/2:
\be S_\b G_\a = 0 \ , \quad K_b G_\a =0\ , \quad
\mathbb D G_\a = \frac{3}{2} G_\a \ .
\ee

\subsection{The $\cN > 1$ case}

For  $\cN>1$ one imposes the following dimension-1 covariant  constraint 
\cite{HitchinKLR,ZP,ZH}
\be
F_{\a}^I{}_\b^J =  -2\ri\ve_{\a\b}G^{IJ} \ , \label{Fconst}
\ee
where $G^{IJ}$ is antisymmetric, primary and of dimension-$1$
\be
G^{IJ}=-G^{JI}~, \quad S_\a^I G^{JK} = 0 \ , \quad K_a G^{IJ}=0\ , 
\quad \mathbb D G^{IJ} = G^{IJ} \ .
\ee
These constraints are a natural generalization of the 
$\cN>1$ constraints in four dimensions \cite{GSW,Sohnius}.
The Bianchi identities then give the remaining field strength components:
\bsubeq \label{FSComps}
\bea
F_{a}{}_\a^I&=&
\frac{1}{ (\cN-1)}(\g_a)_\a{}^{\b} \bm \nabla_{\b J} G^{I J}
~,
\\
F_{ab}&=&
-\frac{\ri}{ 4\cN(\cN-1)}\ve_{abc}(\g^c)^{\a\b}[\bm \nabla_{\a}^{ K}, \bm \nabla_{\b}^{ L}] G_{ K L}
~.
\eea
\esubeq

The $\cN=2$ case is special because $G^{IJ}$ becomes proportional to the 
antisymmetric tensor $\ve^{IJ}$
\bea
G^{IJ}= \ve^{IJ} G~.
\eea
The components of $F_{AB}$ then become
\bsubeq
\bea
F_\a^I{}_\b^J&=& -2\ri\ve_{\a\b}\ve^{IJ}G
~,
\\
F_a{}_\b^J&=&\ve^{JK}(\g_a)_\b{}^\g \bm \nabla_{\g K}G
~,
\\
F_{ab}&=&
-\frac{\ri}{4} \ve_{abc}
(\g^c)^{\g\d}\ve^{KL} \bm \nabla_{\g K} \bm \nabla_{\d L} G~.
\eea
\esubeq
The Bianchi identities imply a constraint on $G$ at dimension-2
\bea
\bm \nabla^{\g I} \bm \nabla_\g^J G = \hf \d^{IJ} \bm \nabla^{\g}_K \bm \nabla_\g^K G \ .  \label{N=2Const}
\label{2.32}
\eea

Unlike for $\cN = 2$, in the case  $\cN>2$ the field strength $G^{I J}$ 
is constrained by the dimension-3/2 Bianchi identity 
\bea
\bm \nabla_{\g}^{I} G^{ J K}&=&
\bm \nabla_{\g}^{[I} G^{ J K]}
- \frac{2}{ \cN-1} \d^{I [J} \bm \nabla_{\g L} G^{ K] L}
~.
\label{VMBI}
\eea
This constraint may be shown to define an off-shell 
supermultiplet, see e.g. \cite{GGHN, BKNT-M1}.


\section{Chern-Simons and curvature induced three-forms}\label{C3FCSA}

In this section our method to construct supersymmetric Chern-Simons actions 
is outlined. 
This method heavily builds on 
the superform formalism for the construction of supersymmetric invariants 
\cite{Castellani,Hasler,Ectoplasm,GGKS}.
 First of all, we sketch its salient points in the framework 
of  3D $\cN$-extended conformal superspace. 
The formalism 
makes use of a closed three-form
\be \frak{J} = \frac{1}{3!}E^{C} \wedge E^{B} \wedge E^{A} \frak{J}_{ABC} \ , \quad 
\rd \frak{J} = 0 \ .
\ee
Under an infinitesimal coordinate transformation generated by a vector field 
$\xi = \xi^M \partial_M =\x^A E_A$, the three-form varies as
\be \d_{\xi} \frak{J} = \cL_{\xi} \frak{J} \equiv i_\xi \rd \frak{J} + \rd i_\xi \frak{J} = \rd i_\xi \frak{J} \ .
\ee
We note that $\d_\x  \frak{J} =\delta_{\rm gct} \frak{J}$,
where  $\delta_{\rm gct}$ stands for the general coordinate transformation
associated with $\x$. 
As discussed in Appendix A, 
the gauge group  of conformal supergravity, $\cG$, 
is generated by two types of transformations:
(i) covariant general coordinate transformations, 
$\delta_{\rm cgct}$, associated with a parameter $\xi^A$;
 and (ii)  standard superconformal transformations, 
$\delta_{\cH}$, associated with a parameter $\L^{\ul a}$. 
The covariant diffeomorphism $\delta_{\rm cgct} (\x) $
is related to the ordinary one $\delta_{\rm gct} (\x) $ by the rule 
\cite{BKNT-M1}
\begin{align}
\delta_{\rm cgct}(\xi^A) = \delta_{\textrm{gct}} (\xi^A E_A{}^M) - \delta_{\cH}(\xi^A \omega_A{}^{\ul a})~.
\end{align}
The closed three-form $\frak J$ is required to  transform
by an exact three-form under the standard superconformal  transformations, 
\bea
 \d_{\cH} \frak{J} =  \rd \Theta (\L^{\underline{a}} ) \ . 
\eea
If we assume the components $\xi^M$ and $ \L^{\underline{a}}$ 
vanish at spacetime infinity,
 then we have the supersymmetric invariant
\bea 
S = \int_{\cM^3} i{}^*\frak{J} 
 \ .
\label{ectoS}
\eea
Here $\cM^3$ denotes the bosonic body of the curved superspace $\cM^{3|2\cN}$ and $i : \cM^3 \to \cM^{3|2\cN}$ is the 
inclusion map.

Suitable  actions must also be gauge invariant 
for any additional gauge symmetries of the theory under consideration. 
If the closed three-form $\frak{J}$ 
transforms by an exact three-form under the gauge transformations, 
\bea
 \d \frak{J} =  \rd \Theta \ ,
\eea
then the functional \eqref{ectoS} is a suitable candidate for an action.

Our method to construct Chern-Simons actions is 
analogous to the one used in \cite{BKNT-M2,KNT-M}
to derive the conformal supergravity actions for $\cN \leq 6$. 
In the super Yang-Mills case, following \cite{BKNT-M2}, 
we will construct a 
closed three-form $\frak J$ by  finding two solutions to the superform equation
\be
\rd \S = \langle F^2 \rangle := \tr \big\{ F \wedge F \big\} \ . \label{superformMaster}
\ee
The first of these solutions is the Chern-Simons three-form $\S_{\rm CS}$,
\be \S_{\rm CS} = \tr \big\{ V \wedge F - \frac{\ri}{3} V \wedge V \wedge V \big\} \ . \label{CSFORM}
\ee
It changes by an exact three-form under the Yang-Mills gauge 
transformation \eqref{2.2},
\be 
\d_\t \S_{\rm CS}  = \rd \,
\tr \big\{\ri \, \rd \t \wedge V \big\} 
~.
\ee
It is invariant under the standard superconformal transformations, 
\bea
\d_\cH \S_{\rm CS} = 0~.
\eea
The other solution,
 the so-called curvature induced form $ \S_R$, 
 is  defined to be such that its components are constructed 
 in terms of the field strength $F_{AB}$ and its covariant derivatives.
 This three-form is required to be invariant under the Yang-Mills gauge 
transformations \eqref{2.2} and under the the standard superconformal ones, 
\begin{subequations}
\bea
\d_\t \S_R &=&0~, \\
\d_\cH \S_R &=&0~. \label{3.10b}
\eea
\end{subequations}
The existence of $\S_R$ is not guaranteed for arbitrary $\cN$ and 
crucially depends on the explicit structure of the constraints obeyed by the field strength. 
If  $\S_R$ exists, the properties of $\S_{\rm CS} $ and $\S_R$  imply that 
their difference 
\be \frak{J} = \S_{\rm CS} - \S_{R} 
= \tr \big\{ V \wedge F - \frac{\ri}{3} V \wedge V \wedge V \big\} - \S_R
\ee
is an appropriate closed three-form that constitutes a supersymmetric action.

We would like to emphasize that the three-form $\S_R$ is required 
to be conformally invariant, eq. \eqref{3.10b}. 
Actually, it turns out that 
the only non-trivial invariance condition on $\S_R$ is with respect to the special conformal generators $K_A$. 
It is equivalent to the condition \cite{BKNT-M2}
\be S_\b^J \S_{a_1 \cdots a_{n}}{}_{\a_1}^{I_1}{\cdots}{}_{\a_{p - n}}^{I_{p-n}} 
= \ri n (\g_{[a_{1}})_\b{}^\g \S_\g^J{}_{a_2 \cdots a_{n}]}{}_{\a_1}^{I_1}{\cdots}{}_{\a_{p - n}}^{I_{p-n}}{}  \ . \label{recurs}
\ee

The above scheme is  an example of a known construction 
where an invariant derived from a closed super
$d$-form can be generated from a closed, gauge-invariant super $(d+1)$-form
provided that the latter is Weil trivial, i.e. exact in invariant cohomology
(a concept introduced by Bonora, Pasti and Tonin \cite{BPT} in the
context of anomalies in supersymmetric theories). Examples of this include
higher-order invariants in other supersymmetric theories which were studied, e.g.,
 in \cite{BHLSW,BHS13}.


\section{Non-Abelian curvature induced three-form} \label{CSCI3F}

We introduce the curvature induced form 
$\S_R = \frac{1}{3!}E^{C} \wedge E^{B} \wedge E^{A} \S_{ABC}$ 
as the {\it covariant} solution 
to the superform equation\footnote{When referring to 
the components of the curvature induced form we will use $\S$ instead 
of $\S_R$ to avoid awkward notation.}
\be \rd \S_R = \tr \{F \wedge F\} \ , \quad 4 \nabla_{[A} \S_{BCD \}} + 6 T_{[AB}{}^E \S_{|E|CD \}} = (\tr \{F \wedge F\})_{ABCD} \ . \label{superformEq}
\ee
By covariant we mean that the components  $\S_{ABC}$ are 
directly expressible in terms of $F_{AB}$ and their covariant derivatives.
It should be emphasized that the curvature induced form can only exist if the field strength $F$ is constrained in a such a way that eq. \eqref{superformEq} 
can be satisfied.

To see this, consider the $\cN > 1$ case where one finds at the lowest dimension the condition
\be E^\d_L \wedge E^\g_K \wedge E^\b_J \wedge E^\a_I \Big( -24 \eps_{\a\b} \eps_{\g\d} \tr \{ G^{IJ} G^{KL} \} - 4 \nabla_\d^L \S_\a^I{}_\b^J{}_\g^K + 12 \ri (\g^a)_{\a\b} \d^{IJ} \S_a{}_\g^K{}_\d^L \Big) = 0 \ .
\ee
On dimensional grounds, the most general ansatz to take for $\S_R$ is\footnote{This is analogous to the ansatz taken for conformal supergravity \cite{BKNT-M2}.}
\be \S_\a^I{}_\b^J{}_\g^K = 0 \ , \quad \S_a{}_\b^J{}_\g^K = \ri (\g_a)_{\b\g} \tr \{ A \d^{JK} G^{PQ} G_{PQ} + B G^{JL} G^{K}{}_L \} \ . \label{ansatzN>1}
\ee
It will turn out that the curvature induced three-form, 
based on the ansatz \eqref{ansatzN>1}, can only be found for $\cN < 4$. 
It is in these cases that we have
\be \tr \{ G^{IJ} G^{KL} \} = A \d^{K [I} \d^{J] L} \tr \{ G^{PQ} G_{PQ} \} + \frac{B}{2} \d^{K [I} \tr \{ G^{J] P} G^L{}_P \} - \frac{B}{2} \d^{L[I} \tr \{ G^{J] P} G^K{}_P \} \label{traceCond}
\ee
for some $A$ and $B$.

Below we give the solution to eq. \eqref{superformEq} on a case by case basis.


\subsection{The $\cN = 1$ case}

Since $F$ is constrained by eq. \eqref{N=1FConst}, solving \eqref{superformEq} is straightforward. One finds\footnote{Keep in mind that eq. \eqref{superformEq} 
is identically satisfied once it is solved up to and including the level of the highest dimension component, see \cite{Novak1}.}
\bsubeq \label{CIFN=1}
\begin{align}
\S_{\a\b\g} &= \S_{a \b \g} = \S_{ab \g} = 0 \ , \\
\S_{abc} &= - \frac{\ri}{4} \eps_{abc} \tr \{ G^\g G_\g \} \ .
\end{align}
\esubeq
Since the only non-zero component of this three-form is primary, 
 $\S_R$ is indeed conformally invariant 
 by virtue of equation \eqref{3.10b}.

\subsection{The $\cN = 2$ case}

In the $\cN = 2$ case, we can replace $G^{IJ}$ with its Hodge-dual:
\be 
G := \hf \eps^{IJ} G_{IJ}  \ , \quad 
G^{IJ} = \eps^{IJ} G\ .
\ee
Then we have
\be \tr \{ G^{IJ} G^{KL} \} = 2 \d^{I [K} \d^{L] J} \tr\{G^2\} = \d^{K [I} \d^{J] L} \tr\{G^{PQ} G_{PQ} \} \ .
\ee
Using the constraint \eqref{N=2Const} one finds the solution
\bsubeq \label{CIFN=2}
\begin{align}
\S_\a^I{}_\b^J{}_\g^K &= 0 \ , \\
\S_a{}_\b^J{}_\g^K &= 2 \ri (\g_a)_{\b\g} \d^{JK} \tr \{G^2\} \ , \\
\S_{ab}{}_\g^K &= - \eps_{abc} (\g^c)_{\g\d} \tr \{ \bm \nabla^{\d K} G^2\} \ , \\
\S_{abc} &= -  \frac{\ri}{2} \eps_{abc} \tr \{ 2 \bm \nabla^\g_K G \bm \nabla_\g^K G + G \bm \nabla^\g_K \bm \nabla_\g^K G \} \ .
\end{align}
\esubeq
The curvature induced three-form can be shown to 
obey equation 
\eqref{3.10b}.

It is often 
advantageous
to make use of the complex basis for the $\cN=2$ covariant derivatives, see \cite{KLT-M11, BKNT-M1} for 
details. In this basis, the field strength is given by
\bea \label{4.9}
 F = \bar{E}^\b \wedge E^\a  F_{\a \b} + E^\b \wedge E^a  F_{a \b} + \bar{E}^\b \wedge E^a \bar{ F}_{a \b} + \hf E^b \wedge E^a  F_{ab}~, 
\eea 
where its components are
\bsubeq \label{4.10}
\begin{align}
 F_{\a \b} &= - 2 \eps_{\a\b}  G \ , \\
 F_{a \b} &= \ri (\g_a)_\b{}^\g \bm \nabla_\g  G \ , \\
\bar{ F}_{a \b} &= - \ri (\g_a)_\b{}^\g \bar{\bm \nabla}_\g  G \ , \\
 F_{ab} &= - \frac{\ri}{8} \eps_{abc} (\g^c)^{\g\d} [\bm \nabla_\g , \bar{\bm \nabla}_\d]  G~.
\end{align}
\esubeq
The corresponding curvature induced form \eqref{CIFN=2} may be expressed as
\begin{align} \label{4.11}
\S &= \bar{E}^\g \wedge E^\b \wedge E^a \S_{a \b \g} + \hf E^\g \wedge E^b \wedge E^a \S_{ab \g}
+ \hf \bar{E}^\g \wedge E^b \wedge E^a \bar{\S}_{ab \g} \non\\
&\quad+ \frac{1}{6} E^c \wedge E^b \wedge E^a \S_{abc} \ ,
\end{align}
where the components have the explicit form
\bsubeq \label{4.12}
\begin{align}
\S_{a \b \g} &= - 2 \ri (\g_a)_{\b\g} \tr\{ G^2 \} \ , \\
\S_{ab \g} &= - \eps_{abc} (\g^c)_{\g\d} \tr \{\bm \nabla^\d G^2\} \ , \\
\bar{\S}_{ab \g} &= - \eps_{abc} (\g^c)_{\g\d} \tr \{\bar{\bm \nabla}^\d G^2\} \ , \\
\S_{abc} &= \ri \eps_{abc} \tr \{2 \bm \nabla^\g G \bar{\bm \nabla}_\g G + G \bar{\bm \nabla}^\g \bm \nabla_\g G \}~.
\end{align}
\esubeq


\subsection{The $\cN = 3$ case}

In the $\cN = 3$ case we define the Hodge-dual of $G^{IJ}$ as
\be 
 G_I := \hf \eps_{IJK} G^{JK}  \ , \quad
G^{IJ} = \eps^{IJK} G_K\ ,
\ee
which implies
\be \tr \{ G^{IJ} G^{KL} \} = 2 \d^{K [I} \d^{J]L} \tr \{G^P G_P\} - 2 \d^{K [I} \tr\{G^{J]} G^L \}+ 2 \d^{L[I} \tr\{G^{J]} G^K \}\ .
\ee
Furthermore, the Bianchi identity \eqref{VMBI} gives
\bsubeq
\begin{align} \bm \nabla_\g^I G^J &= \bm \nabla_\g^{[I} G^{J]} + \frac{1}{3} \d^{IJ} \bm \nabla_\g^K G_K \ , \label{N=3BI} \\
\bm \nabla^{\g J} \bm \nabla_{\g [J} G_{I]} &= 2 \bm \nabla^{\g J} \bm \nabla_{\g J} G_I - 8 \eps_{IJK} [G^J , G^K ]\ , \\
\bm \nabla_\a^I \bm \nabla_\b^J G_J &= - \frac{3}{2} \eps_{\a\b} \bm \nabla^\g_P \bm \nabla^P_\g G^I + 3 \ri \bm \nabla_{\a\b} G^I + 9 \eps_{\a\b} \eps^{IJK} [G_J , G_K] \ .
\end{align}
\esubeq
Using the above identities one finds the solution
\bsubeq \label{CIFN=3}
\begin{align}
\S_\a^I{}_\b^J{}_\g^K &= 0 \ , \\
\S_a{}_\b^J{}_\g^K &= 2 \ri (\g_a)_{\b\g} \tr \{ \d^{JK} G^I G_I - 2 G^{J} G^{K} \} \ , \\
\S_{ab}{}_\g^K &= 2 \eps_{abc} (\g^c)_{\g}{}^\d \tr \{ \bm \nabla_\d^{[K} G^{I]} G_I - \frac{1}{3} \bm \nabla_\d^I G_I G^K \} \ , \\
\S_{abc} &= \frac{\ri}{2} \eps_{abc} \tr \{ \frac{2}{9} (\bm \nabla^{\g K} G_K) (\bm \nabla^L_\g G_L) - (\bm \nabla^\g_{[K} G_{L]}) (\bm \nabla^{[K}_\g G^{L]}) \non\\
&\quad - 2 (\bm \nabla^\g_K \bm \nabla^K_\g G^L) G_L + 8 \eps^{IJK} G_I G_J G_K \} \ .
\end{align}
\esubeq
Conformal invariance follows since $\S_R$ obeys equation \eqref{3.10b}.


\subsection{The special case of $\cN = 4$} \label{SCN=4}

In the previous subsections, we found that our approach struck an obstacle 
at the $\cN = 4$ case. In particular, eq. \eqref{traceCond} no longer holds. 
Actually, the $\cN=4$ case requires some additional consideration. 
It is well known that the constraint \eqref{VMBI} does not define an irreducible 
off-shell supermultiplet for $\cN=4$. In this case,  the Hodge-dual of $G^{IJ}$, 
\be
\tilde{G}^{IJ} := \hf \ve^{IJKL}G_{KL} \ , 
\label{4.17}
\ee
obeys the same constraint as $G^{IJ}$ does, 
\begin{subequations}
\bea
\bm \nabla_{\g}^{I} G^{ J K}&=&
\bm \nabla_{\g}^{[I} G^{ J K]}
- \frac{2}{3} \d^{I [J} \bm \nabla_{\g L} G^{ K] L}~,  \label{4.18a}\\
\bm \nabla_{\g}^{I} \tilde{G}^{ J K}&=&
\bm \nabla_{\g}^{[I} \tilde{G}^{ J K]}
- \frac{2}{3} \d^{I [J} \bm \nabla_{\g L} \tilde{G}^{ K] L} \ .
\eea
\end{subequations} 
As a result, one may constrain 
the field strength $G^{IJ}$ to be self-dual, 
\begin{subequations}
\bea 
\tilde{G}^{IJ} = {G}^{IJ}~,
\label{14sd}
\eea
or anti-self-dual, 
 \bea 
\tilde{G}^{IJ} = -{G}^{IJ}~.
\label{14asd}
\eea
\end{subequations}
These choices correspond to two different off-shell $\cN=4$ vector multiplets, 
the left and right ones, see \cite{KLT-M11} for more details.

Now, if we consider an irreducible $\cN=4$ vector multiplet  obeying either
\eqref{14sd} or  \eqref{14asd}, it may be seen that eq. \eqref{traceCond} 
still does not hold. 
A possible way out is to consider two vector multiplets
and a generalization of  eq. \eqref{superformMaster} 
of the form 
$\rd \S= \tr \big\{ F_1 \wedge F_2 \big\} $.
However, this poses a problem for non-Abelian vector multiplets, since the 
two-form field strengths $F_1$ and $F_2$ are not gauge invariant; 
instead, they transform covariantly under 
the two different gauge groups.\footnote{It should be mentioned that there is 
an alternative approach to the problem of 
constructing the $\cN = 4$ Chern-Simons action \cite{NG93}. 
It is based on dualizing two scalars in the vector multiplet 
into vector fields and constructing a theory involving three  different vectors! 
However, as mentioned in \cite{BrooksG}, 
such an approach is on-shell and cannot be used to construct 
matter couplings.}
In this section, we therefore restrict ourselves to Abelian 
vector multiplets.

We will consider the general case of two Abelian vector multiplets 
$G_+^{IJ}$ and $G_-^{IJ}$ with 
the two-form field strengths $F_+$ and $F_-$ respectively.  
In this case the superform equation \eqref{superformMaster} is replaced by
\be \rd \S = F_+ \wedge F_- \ , \quad 4 \nabla_{[A} \S_{BCD \}} + 6 T_{[AB}{}^E \S_{|E|CD \}} = (F_+ \wedge F_-)_{ABCD} \ . \label{superformEqLR}
\ee
The Chern-Simons solution $\S_{\rm CS}$ to the above is
\be \S_{\rm CS} = F_+ \wedge V_- = V_+ \wedge F_- + {\rm closed \ form} \ , \label{CSFORMLR}
\ee
where $V_{\pm}$ are gauge one-forms associated with 
the two-form field strengths, $F_{\pm} = \rd V_{\pm} \ $. 

The curvature induced three-form $\S_R$ is the {\it covariant} solution to 
the superform equation \eqref{superformEqLR} (when it exists). 
For $\cN = 4$ one finds at the lowest dimension of \eqref{superformEqLR} 
the condition
\be E^\d_L \wedge E^\g_K \wedge E^\b_J \wedge E^\a_I (-24 \eps_{\a\b} \eps_{\g\d} G_+^{IJ} G_-^{KL} 
- 4 \nabla_\d^L \S_\a^I{}_\b^J{}_\g^K + 12 \ri (\g^a)_{\a\b} \d^{IJ} \S_a{}_\g^K{}_\d^L) 
= 0 \ .
\ee
The most general ansatz to take for $\S_R$ is
\be \S_\a^I{}_\b^J{}_\g^K = 0 \ , \quad \S_a{}_\b^J{}_\g^K = \ri (\g_a)_{\b\g} (A \d^{JK} G_+^{PQ} G_{- PQ} - B G_+^{L (J} G_-^{K)}{}_L) \ ,
\ee
which will lead to a solution if
\be G_+^{IJ} G_-^{KL} = A \d^{K[I} \d^{J]L} G_+^{PQ} G_{- PQ} + \frac{B}{2} \d^{K [I} G_+^{J] P} G_-^L{}_P - \frac{B}{2} \d^{L[I} G_+^{J] P} G_-^K{}_P \ . \label{traceCondLR}
\ee

It is easy to see that if we let both $G_{\pm}^{IJ}$ be (anti-)self-dual then we cannot satisfy eq. \eqref{traceCondLR} for any $A$ and $B$. However, 
imposing opposite duality conditions gives us a way out. Taking $G_{+}^{IJ}$ to be self-dual and $G_-^{IJ}$ to be anti-self-dual, 
\begin{align} \hf \eps^{IJKL} G_{\pm KL} &= \pm G_{\pm}^{IJ}  \,
\label{N=4duality}
\end{align}
gives\footnote{It is clear that $G_+^{IJ} G_{- IJ} = 0$.}
\be G_+^{IJ} G_-^{KL}= \d^{K [I} G_+^{J] P} G_-^L{}_P - \d^{L [I} G_+^{J] P} G_-^K{}_P \ .
\ee

Using the Bianchi identity \eqref{VMBI} 
and the (anti-)self-duality conditions \eqref{N=4duality}, one finds the curvature induced form to be
\bsubeq \label{CIFN=4Abelian}
\begin{align}
\S_\a^I{}_\b^J{}_\g^K &= 0 \ , \\
\S_a{}_\b^J{}_\g^K &= - 2 \ri (\g_a)_{\b\g} 
G_+^{L(J} G_-^{K)}{}_L
 \ , \\
\S_{ab}{}_\g^K &= - \frac{1}{3}\eps_{abc} (\g^c)_\g{}^\d (\nabla_{\d I} G_+^{IJ} G_{-J}{}^K + \nabla_{\d I} G_-^{IJ} G_{+J}{}^K) \ , \\
\S_{abc} &= \ri \eps_{abc} \big(\frac{1}{24} \nabla^{\g}_J \nabla_{\g}^I G_{+ IK} G_-^{KJ} + \frac{1}{24} \nabla^{\g}_J \nabla_{\g}^I G_{-IK} G_+^{KJ} + \frac{1}{9} \nabla^{\g}_I G_+^{IJ} \nabla_\g^K G_{-JK}\big) \ .
\end{align}
\esubeq
One can check that eq. \eqref{recurs} holds.

As is known, the group isomorphism 
${\rm SO}(4) \cong  \big( {\rm SU}(2)_{\rL}\times {\rm SU}(2)_{\rR} \big)/{\mathbb Z}_2$
allows us to convert 
each SO(4) vector index into a pair of SU(2) spinor ones, 
for instance $\nabla_\a^I \to \nabla_\a^{i \bar i}$, see \cite{KLT-M11} for more details.
It is instructive to look at  some of the above results in the isospinor notation.
The SO(4) bivector $G^{IJ} =-G^{JI}$ 
is equivalently described by two symmetric second-rank isospinors, $G^{ij}$ and $G^{\bar i\bar j}$,
which are defined as 
\bea
G^{IJ}~\to~G^{i\bar i , j\bar j }=-\ve^{\bar i\bar j}G^{ij}-\ve^{ij}G^{\bar i\bar j}
~,\qquad
G^{ij}=G^{ji}~,\quad
G^{\bar i\bar j}=G^{\bar j\bar i}
\eea
and transform under the local groups ${\rm SU}(2)_{\rL} $ and ${\rm SU}(2)_{\rR}$, respectively. For the Hodge-dual SO(4) bivector
$\tilde{G}^{IJ}$ defined by  \eqref{4.17}, we get 
\be \tilde{G}^{IJ}~\to~\tilde{G}^{i\bar i , j\bar j }
= \ve^{\bar i\bar j} G^{ij}-\ve^{ij}G^{\bar i\bar j}\ .
\ee
The Bianchi identity \eqref{4.18a} is equivalent to 
the two  analyticity constraints \cite{KLT-M11}
\bsubeq
\bea
\nabla_\a^{(i\bar i}G^{kl)}&=&0
~,
\label{a0-L}
\\
\nabla_\a^{i(\bar i}G^{\bar k\bar l)}&=&0
~.
\label{a0-R}
\eea
\esubeq
Thus the field strengths  
$G^{ij}$ and $G^{\bar i\bar j}$ 
are independent of each other. 
The (anti-)self-duality conditions \eqref{N=4duality} are equivalent to 
\begin{subequations}
\bea
G_-^{IJ}~& \to &~G_-^{i\bar i , j\bar j }=-\ve^{\bar i\bar j}G^{ij}
~,\qquad
G^{ij}=G^{ji}~, \\
G_+^{IJ}~ & \to &~G_+^{i\bar i , j\bar j }=
-\ve^{ij}G^{\bar i\bar j}
~,\qquad
G^{\bar i\bar j}=G^{\bar j\bar i}\ .
\eea
\end{subequations}
In accordance with \cite{KLT-M11}, 
a symmetric isospinor superfield $G^{ij}$ under the constraint 
(\ref{a0-L}) is called a left linear multiplet or, equivalently, a  left $\cO(2)$ multiplet. 
Similarly, eq. (\ref{a0-R}) defines a  right linear multiplet or, equivalently, a right 
$\cO(2)$ multiplet.


\section{Component actions} \label{ComponentActions}

In the previous sections we have given a complete superspace description of 
the  Chern-Simons actions for non-Abelian vector multiplets with $\cN < 4$ and of the $BF$ action 
for Abelian vector multiplets in the  $\cN = 4$ case. 
In this section we will derive the corresponding component action. 
To do so we will need to elaborate on the component 
structure of the theory. For a complete description of the component fields 
of the Weyl multiplet including their supersymmetry transformations we refer the reader to \cite{BKNT-M2}. Here we outline some of the salient details.

The Weyl multiplet contains
a set of gauge one-forms 
which appear 
explicitly in the actions. These  include the vielbein $e_m{}^a$, the gravitino $\psi_m{}^\a_I$, the $\rm SO(\cN)$ gauge field $V_m{}^{IJ}$ and the dilatation 
gauge field $b_m$  defined as
\begin{align}\label{eq:GaugeFields}
e_m{}^a &:= E_m{}^a| \ , \qquad \psi_m{}^\a_I := 2 E_m{}^\a_I| ~, \qquad
V_m{}^{IJ} := \Phi_m{}^{IJ}| \ ,\qquad b_m := B_m| ~,
\end{align}
where the bar-projection \cite{WZ2, WB, GGRS} of a superfield $V(z) =V(x,\q)$ is defined by 
the standard rule $V| := V(x,\q)|_{\theta = 0}$. 
The remaining gauge fields are the spin connection  $\omega_m{}^{ab}$,
the special conformal and $S$-supersymmetry connections $\frak{f}_m{}^a$ and $\phi_m{}_\a^I$ defined as
\begin{align}\label{eq:compGaugeFields}
\omega_m{}^{ab} := \Omega_m{}^{ab}| \ , \qquad  
\frak{f}_m{}^a := \frak{F}_m{}^a| \ , \qquad
\phi_m{}^I_\a := 2 \frak F_m{}^I_\a| \ .
\end{align}
These connections turn out to be composite and their expressions are 
given in \cite{BKNT-M2}. 

The Weyl multiplet also contains some auxiliary fields for $\cN>2$. 
In the $\cN = 3$ case, 
there is a single fermionic auxiliary field defined by 
\be w_\a = W_\a| \ .
\ee
In the $\cN = 4$ case, the Weyl multiplet contains 
 both bosonic and fermionic auxiliary fields,
\begin{align} 
w &= W| \ , \qquad y = - \frac{\ri}{4} \nabla^\a_I \nabla_\a^I W| \ , 
\qquad
w_{\a }^I = - \frac{\ri}{2} \nabla_{\a}^I W | \ ,
\end{align}
where $W$ denotes the Hodge-dual of $W^{IJKL}$, 
\be 
W^{IJKL} = \ve^{IJKL} W \ .
\label{5.7}
\ee


\subsection{Vector multiplets in components}

The component fields of vector multiplets may be extracted from the 
field strength $G^{IJ}$. 
For $\cN > 1$, we define the matter fields as follows\footnote{The 
coefficients are chosen so that the $\cN = 1$ case may be derived via the higher $\cN$ cases.}
\allowdisplaybreaks{
\bsubeq \label{components}
\begin{align}
g^{IJ} &:= G^{IJ}| \ , \\
\l_\a^I &:= \frac{2}{\cN - 1} \bm \nabla_{\a J} G^{IJ}| \ , \\
h^{IJ} &:= \frac{\ri}{\cN - 1} \bm \nabla^{\g [I} \bm \nabla_{\g K} G^{J] K}| \ , \\
\c_{\a_1 \cdots \a_n}{}^{I_1 \cdots I_{n+2}} &:= I(n) \bm \nabla_{(\a_1}^{[I_1} \cdots \bm \nabla_{\a_n)}^{I_n} G^{I_{n+1} I_{n+2}]}| \ ,
\end{align}
\esubeq}
where
\be \label{Ifunct}
I(n)=
\begin{cases}
\ri \ , & n = 1,2 \ ({\rm mod} \ 4) \\
1 \ , & n = 3,4 \ ({\rm mod} \ 4) ~.
\end{cases}
\ee
A final component field $v_m$ is given by the bar-projection of the corresponding superspace connection, 
\be v_a = e_a{}^m v_m \ , \quad v_m := V_m| \ .
\ee

The covariant field strength may be constructed from the bar-projection of the two-form $F = \hf E^B \wedge E^A F_{AB}$. Making use of the identity
\be F_{mn} = E_m{}^A E_n{}^B F_{AB} (-1)^{\eps_A \eps_B}
\ee
and performing a component projection, we find
\be \hat{F}_{ab} := F_{ab}| = f_{ab} + \hf (\psi_{[a}{}^K \g_{b]} \l_K) - \frac{\ri}{2} \psi_a{}^{\g K} \psi_b{}_\g^L g_{KL} \ ,
\ee
where
\be f_{ab} = e_a{}^m e_b{}^n f_{mn} \ , \quad f_{mn} := F_{mn}| = 2 (\partial_{[m} V_{n]} - \ri V_{[m} V_{n]})| = 2 (\partial_{[m} v_{n]} - \ri v_{[m} v_{n]}) \ .
\ee

The component fields of the vector multiplet form the following tower \cite{GGHN}:\footnote{The tower is analogous to the one for the 
$\cN$-extended super Cotton tensor \cite{GGHN, BKNT-M2}.}\\
\begin{minipage}[t]{\textwidth}
\begin{picture}(430,195)
\put(197,170){$g^{I J}$}
\put(205,165){\vector(-1,-1){20}}
\put(210,165){\vector(1,-1){20}}
\put(165,130){$\c_\a{}^{I J K}$}
\put(230,130){$\l_\a{}^{I}$}
\put(165,125){\vector(-1,-1){20}}
\put(175,125){\vector(1,-1){20}}
\put(240,125){\vector(-1,-1){20}}
\put(250,125){\vector(1,-1){20}}
\put(125,90){$\c_{\a_1\a_2}{}^{I_1 \cdots I_4}$}
\put(197,90){$y^{I J}$}
\put(270,90){$\hat{F}_{\a_1 \a_2}$}
\put(122,80){\vector(-1,-1){20}}
\put(83,45){$\cdots$}
\put(80,38){\vector(-1,-1){20}}
\put(40,0){$\c_{\a_1\cdots\a_{\cN-2}}{}^{I_1\cdots I_{\cN}}$
}
\end{picture}
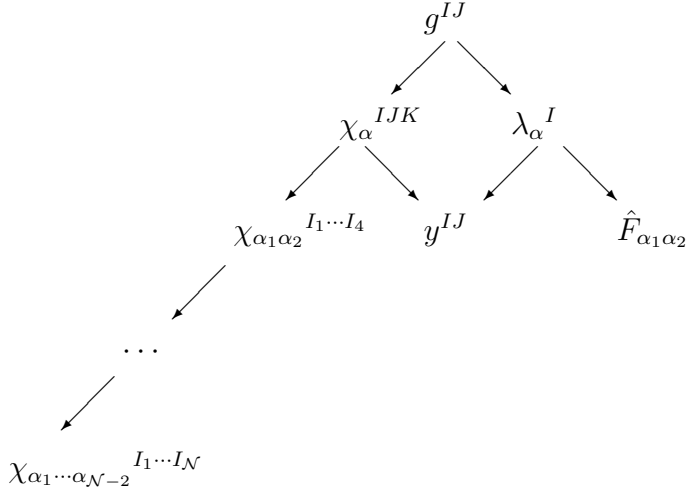
\begin{center}{\bf Figure 1.} Component fields of the $\cN$-extended vector multiplet\end{center}
\end{minipage} \\

The coefficients chosen in eq. \eqref{components} allow 
for a straightforward truncation of higher $\cN$ 
cases to the lower $\cN$ ones via a procedure analogous to the one 
described in \cite{BKNT-M2}. For the $\cN = 1$ case 
we have to switch off all matter fields except
\be \l_\a^I = \l_\a = G_\a| \ ,
\ee
with the field strength $G_\a$ defined in \eqref{2.6a}.

The $\cN=4$ case is special, since it allows for two inequivalent off-shell vector multiplets 
with field strengths $G_+^{IJ}$ and $G_{-}^{IJ}$ obeying the self-duality 
condition \eqref{14sd}
and the anti-self-duality condition \eqref{14asd} respectively. In this case we define the 
the component fields of the vector multiplets as
\bsubeq \label{N=4comps}
\begin{align}
g_\pm^{IJ} &:= G_\pm^{IJ}| \ , \\
\l_{(\pm)}{}_\a^I &:= \frac{2}{3} \bm \nabla_{\a J} G_\pm^{IJ}| \ , \\
h_{(\pm)}{}^{IJ} &:= \frac{\ri}{3} \bm \nabla^{\g [I} \bm \nabla_{\g K} G_\pm^{J] K}| \ , \\
\c_{(\pm) \a_1 \cdots \a_n}{}^{I_1 \cdots I_{n+2}} &:= I(n) \bm \nabla_{(\a_1}^{[I_1} \cdots \bm \nabla_{\a_n)}^{I_n} G_{\pm}^{I_{n+1} I_{n+2}]}| \ , \quad n = 1, 2 \ ,
\end{align}
\esubeq
where $g_{\pm}{}^{IJ}$ is (anti-)self-dual
\be \hf \eps^{IJKL} g_{\pm KL} = \pm g_{\pm}{}^{IJ} \ . \label{5.14}
\ee
The component one-forms are given by
\be v_{(\pm) a} = e_a{}^m v_{(\pm) m} \ , \quad v_{(\pm) m} := V_{\pm m}| \ ,
\ee
where $V_{\pm}$ is the gauge one-form associated with the field strength $G_{\pm}^{IJ}$.

The (anti-)self-duality property of $G_{\pm}^{IJ}$, eq. \eqref{N=4duality}, reduces the degrees of freedom for 
each vector multiplet by half. To see this, it is useful to replace $h_{(\pm)}{}^{IJ}$ by the fields
\begin{align} \hat{h}_{\pm}{}^{IJ} &= \hf (h_{(\mp)}{}^{IJ} + \tilde{h}_{(\mp)}{}^{IJ}) \non\\
&= h_{(\mp)}{}^{IJ} \mp 2 w g_{\mp}{}^{IJ} \mp \frac{\ri}{4} [g_+{}^{P [I} , g_-{}^{J]}{}_P] \ ,
\end{align}
which prove to be (anti-)self-dual
\bsubeq
\be \hf \eps^{IJKL} \hat{h}_{\pm KL} = \pm \hat{h}_{\pm}^{IJ} \ .
\ee
The other implications of (anti-)self-duality are
\begin{align} \label{compDuality}
\c_{\pm \a}{}^{IJK} &= \mp \frac{\ri}{2} \eps^{IJKL} \l_{\pm \a L} \ , \\
\chi_{\pm \a\b}{}^{IJKL} &= \mp \eps^{IJKL} \hat{F}_{\pm \a\b} \ .
\end{align}
\esubeq
Diagrammatically, this means that the components on the left hand side of Figure 1 are related to those on the 
right hand side via (anti-)self-duality. One can see that each vector multiplet constitutes 8+8 degrees of freedom.


\subsection{Off-shell component actions}

Now we have all the ingredients to construct the component actions corresponding to the closed forms
\be \frak{J} = \S_{\rm CS} - \S_R
\ee
found in the previous sections. To do so we just need to apply the action principle \eqref{ectoS},
\bea
S= \int \rd^3 x \,e \,{}^* \frak{J} |_{\q=0}\ , 
\qquad 
{}^*\frak{J} = \frac{1}{3!} \eps^{mnp} \frak{J}_{mnp} \ , \quad e= \det (e_m{}^a)\ ,
\eea
 and make use of the formula
\begin{align}\label{eq:SigmaProjection}
\frac{1}{3 !} \eps^{mnp} {{\S}}_{mnp}| &=
	\frac{1}{3 !} \eps^{mnp} E_p{}^C E_n{}^B E_m{}^A {\S}_{ABC}| \non \\
	&= \frac{1}{3 !} \eps^{abc} \big( {{\S}}_{abc}| + \frac{3}{2} \psi_a{}^\a_I {{\S}}_\a^I{}_{bc}| + \frac{3}{4} \psi_b{}^\b_J \psi_a{}^\a_I {{\S}}_\a^I{}_\b^J{}_{c}|
	\non\\&\quad
	+ \frac{1}{8} \psi_c{}^\g_K \psi_b{}^\b_J \psi_a{}^\a_I {{\S}}_\a^I{}_\b^J{}_\g^K| \big) \ .
\end{align}
Here we present the resulting actions on a case by case basis.

Although all our actions are automatically supersymmetric, we give the supersymmetry transformations of the component 
fields in Appendix \ref{SUSYTrans}.


\subsubsection{The non-Abelian $\cN = 1$ case}

The action is constructed using eqs. \eqref{CSFORM} and \eqref{CIFN=1}. From eq. \eqref{eq:SigmaProjection} we find
\be \frac{1}{3!} \eps^{mnp} \S_{mnp}| =  \frac{\ri}{4} \tr \{ \l^\g \l_\g \} \ .
\ee
Combining this with the contribution coming from the Chern-Simons form \eqref{CSFORM} gives
\begin{align}
S =& \hf \int \rd^3 x \,e \, \tr \Big\{
	\eps^{abc} (v_a f_{bc} + \frac{2 \ri}{3} v_a v_b v_ c)
	- \frac{\ri}{2} \,\l^\g \l_\g \Big\} \ .
\end{align}


\subsubsection{The non-Abelian $\cN = 2$ case}

Using eqs. \eqref{eq:SigmaProjection} and \eqref{CIFN=2} we find
\be \frac{1}{3!} \eps^{mnp} \S_{mnp}| =  \frac{1}{2} \tr \{ \frac{\ri}{2} \tilde{\l}^\g_K \tilde{\l}_\g^K - 2 g h \} + \hf (\g^a)_{\g\d} \psi_a{}^\g_I \tr \{ g \tilde{\l}^{\d I} \} 
- \frac{\ri}{4} \eps^{abc} (\g_a)_{\g\d} \psi_b{}^\g_K \psi_c{}^{\d K} \tr \{ g^2 \} \ ,
\ee
where we have defined
\begin{align}
\tilde{\l}_{\a I} := \eps_{JI} \l_\a^J = -2 \bm \nabla_{\a I} G| \ , \quad h :=  \hf \eps_{IJ} h^{IJ} = - \frac{\ri}{2} \bm \nabla^{\g K} \bm \nabla_{\g K} G | \ .
\end{align}
Using the above result and incorporating the Chern-Simons form \eqref{CSFORM} gives the action
\begin{align}
S =& \hf \int \rd^3 x \,e \, \tr \Big\{
	\eps^{abc} (v_a f_{bc} + \frac{2 \ri}{3} v_a v_b v_ c)
	-   \frac{\ri}{2} \tilde{\l}^\g_K \tilde{\l}_\g^K  + 2 g h - (\g^a)_{\g\d} \psi_a{}^\g_I g \tilde{\l}^{\d I} \non\\
	&+ \frac{\ri}{2} \eps^{abc} (\g_a)_{\g\d} \psi_b{}^\g_K \psi_c{}^{\d K} g^2  \Big\} \ .
\end{align}

Expressing the above action in terms of the complex basis gives
\begin{align}
S =& \hf \int \rd^3 x \,e \, \tr \Big\{
	\eps^{abc} v_a f_{bc} 
	+ \ri \l^\g \bar{\l}_\g
	+ 2 g h \non\\
	&- (\g^a)_{\g\d} \psi_a{}^\g g \l^\d + (\g^a)_{\g\d} \bar{\psi}_a{}^\g g \bar{\l}^\d 
	- \ri \eps^{abc} (\g_a)_{\g\d} \psi_b{}^\g \bar{\psi}_c{}^{\d} g^2  \Big\} \ ,
\end{align}
where we have made use of the component fields in the complex basis:
\bsubeq
\begin{align}
\l_\a &=  -2 \nabla_\a G| \ , \quad \bar{\l}_{\a} =  -2 \bar{\nabla}_\a G| \ , \quad h = \ri \bar{\nabla}^\g \nabla_\g G| \ .
\end{align}
\esubeq


\subsubsection{The non-Abelian $\cN = 3$ case}

Using eqs. \eqref{eq:SigmaProjection} and \eqref{CIFN=3} we find
\begin{align} \frac{1}{3!} \eps^{mnp} \S_{mnp}| &= - \frac{\ri}{2} \tr \{ - 2 \c^\g \c_\g - \frac{1}{4} \l^{\g IJ} \l_{\g IJ}  - 2 \ri h^I g_I - 8 \eps^{IJK} g_I g_J g_K \} \non\\
&+ \hf (\g^a)_{\g\d} \psi_a{}^\g_I \tr \{ \l^{\d IJ} g_J + 2 \ri \c^\d g^I \}\non\\
& - \frac{\ri}{4} \eps^{abc} (\g_a)_{\g\d} \psi_b{}^\g_K \psi_c{}^\d_L \tr \{ \d^{KL} g^P g_P - 2 g^K g^L \} \ ,
\end{align}
where we have defined
\bsubeq \label{N=3Comps}
\begin{align}
g_I &:=  \frac{1}{2} \eps_{IJK} G^{JK}| = G_I| \ , \\
\l_{\a}{}^{IJ} &:= \eps^{IJK} \l_{\a K} = 2 \bm \nabla_{\a}^{[I} G^{J]}| \ , \\
\c_\a :&= \frac{1}{3!} \eps_{IJK} \c_\a{}^{IJK} = \frac{\ri}{3} \bm \nabla_\a^I G_I | \ , \\
h_I :&= \frac{1}{2!} \eps_{IJK} h^{JK} = \frac{\ri}{2} \bm \nabla^{\g J} \bm \nabla_{\g [I} G_{J]}| = - \ri \bm \nabla^{\g J} \bm \nabla_{\g J} G_I | + 8 \ri \eps_{IJK} g^J g^K \ .
\end{align}
\esubeq
Combining this with the contribution coming from the Chern-Simons form \eqref{CSFORM} gives
\begin{align}
S =& \hf \int \rd^3 x \,e \, \tr \Big\{
	\eps^{abc} (v_a f_{bc} + \frac{2 \ri}{3} v_a v_b v_ c)
	- 2 \ri \c^\g \c_\g  - \frac{\ri}{4} \l^{\g IJ} \l_{\g IJ} + 2 g^I h_I - 8 \ri \eps^{IJK} g_I g_J g_K \non\\
	&- (\g^a)_{\g\d} \psi_a{}^\g_I (\l^{\d IJ} g_J + 2 \ri \c^\d g^I ) \non\\
	&+ \frac{\ri}{2} \eps^{abc} (\g_a)_{\g\d} \psi_b{}^\g_K \psi_c{}^{\d}_L (\d^{KL} g^P g_P - 2 g^K g^L)  \Big\} \ .
\end{align}

As in \cite{BKNT-M2}, our choice of normalization for the component fields allows for a simple truncation
to the actions for lower values of $\cN$. For example, from the above action one can truncate the
auxiliary fields to $\cN=2$ by taking (with $ I,J=1,2 $)
\begin{align}\label{eq:5to4Trunc}
\left.\begin{gathered}
g_{I} \longrightarrow 0~, \qquad 
\l_\alpha{}^{IJ} \longrightarrow 0~, \qquad
\c_\alpha \longrightarrow 0~, \qquad
h^{I} \longrightarrow 0~, \\
g^{3} \longrightarrow g~, \qquad 
\l_\alpha{}^{I3} \longrightarrow \tilde{\l}_\alpha{}^{I} ~, \qquad
h^{3} \longrightarrow h ~
\end{gathered} \quad \right\} 
 \ .
\end{align}
For the fields of the Weyl multiplet one performs a similar truncation, which is given in \cite{BKNT-M2}.


\subsubsection{The Abelian $\cN = 4$ case}\label{subsub5.2.4}

Using eqs. \eqref{eq:SigmaProjection} and \eqref{CIFN=4Abelian} we find
\begin{align} \frac{1}{3!} \eps^{mnp} \S_{mnp}| &= - \frac{1}{8} \hat{h}_+^{IJ} g_{- IJ} - \frac{1}{8} \hat{h}_-^{IJ} g_{+IJ} + \frac{\ri}{4} \l_{(+)}^{\a I} \l_{(-)\a I} \non\\
& + \frac{1}{4} (\g^a)_{\g\d} \psi_a{}^\g_I (\l_{(+)}^{\d J} g_{-J}{}^I + \l_{(-)}^{\d J} g_{+ J}{}^I ) \non\\
& - \frac{\ri}{4} \eps^{abc} (\g_a)_{\g\d} \psi_b{}^\g_K \psi_c{}^\d_L g_+{}^{KP} g_-{}^L{}_P \ .
\end{align}
Combining this with the contribution coming from the Chern-Simons form \eqref{CSFORMLR} gives
\begin{align}
S =& \hf \int \rd^3 x \,e \Big( \eps^{abc} v_{(+)a} f_{(-) bc} + \frac{1}{4} \hat{h}_+{}^{IJ} g_{+ IJ} + \frac{1}{4} \hat{h}_{-}{}^{IJ} g_{-IJ} - \frac{\ri}{2} \l_{(+)}^{\a I} \l_{(-) \a I} \non\\
&\quad- \frac{1}{2} (\g^a)_{\g\d} \psi_a{}^\g_I (\l_{(+)}^{\d J} g_{-J}{}^I + \l_{(-)}^{\d J} g_{+ J}{}^I ) \non\\
&\quad + \frac{\ri}{2} \eps^{abc} (\g_a)_{\g\d} \psi_b{}^\g_K \psi_c{}^\d_L g_+{}^{KP} g_-{}^L{}_P \Big) \ .
\end{align}


\section{Matter-coupled $\cN=2$ supergravity}\label{Section6}

The results of sections \ref{CSCI3F} and \ref{ComponentActions} may be used 
to generate locally supersymmetric actions. This idea 
can be illustrated, in a simple and transparent way, by considering the $\cN=2$ 
case which we discuss  below. Unlike in section  \ref{ComponentActions}, here we use the complex basis for the $\cN=2$ covariant derivatives, 
see \cite{KLT-M11} for details.

Let us consider a locally supersymmetric $BF$ term described by the action 
\bea
S_{BF}&=&\int\rd^3x\rd^2\q\rd^2\bar \q \,E \,\cV\, \bm G
~,~~~
\qquad E^{-1}= {\rm Ber}(E_A{}^M)~. 
\label{N=2BF}
\eea
Here $\cV=\bar \cV$ is the gauge prepotential of an Abelian vector multiplet, 
and $\bm G = \bar {\bm G}$ a real linear superfield, 
$\nabla^2 {\bm G} = \bar \nabla^2 {\bm G} =0$.\footnote{The constraints on $\bm G$ may be solved as $\bm G = \ri \nabla^\a \bar \nabla_\a \bm \cV$, for 
some $\bm \cV$. In 
certain cases,  $\bm \cV$ is not a well-defined local operator.}  
The action \eqref{N=2BF} is invariant under gauge transformations 
\bea
\d \cV = \l +\bar \l~, \qquad \bar \nabla_\a \l =0~, 
\eea
with the gauge parameter $\l$ being an arbitrary covariantly chiral 
dimensionless scalar.  
Eq.  \eqref{N=2BF} defines  the $\cN=2$ 
linear multiplet action. 

It turns out that the action \eqref{N=2BF}
may be recast in terms of a closed three-form
\be
J = V \wedge \bm F - \S \ ,\qquad \rd J =0 
\label{6.2}
\ee
that involves three building blocks. 
First of all, $ \bm F = \hf E^B\wedge E^A \bm F_{AB}$ is a closed two-form, 
$\rd \bm F =0$, associated with $\bm G$.
Its components are defined as in eqs. \eqref{4.9} and \eqref{4.10} by 
\bea 
\bm F = \bar{E}^\b \wedge E^\a \bm F_{\a \b} + E^\b \wedge E^a \bm F_{a \b} + \bar{E}^\b \wedge E^a \bar{\bm F}_{a \b} + \hf E^b \wedge E^a \bm F_{ab}~,
\label{6.4}
\eea 
and are explicitly given as follows:
\bsubeq \label{6.5}
\begin{align}
\bm F_{\a \b} &= - 2 \eps_{\a\b} \bm G \ , \\
\bm F_{a \b} &= \ri (\g_a)_\b{}^\g \nabla_\g \bm G \ , \\
\bar{\bm F}_{a \b} &= - \ri (\g_a)_\b{}^\g \bar{\nabla}_\g \bm G \ , \\
\bm F_{ab} &= - \frac{\ri}{8} \eps_{abc} (\g^c)^{\g\d} [\nabla_\g , \bar{\nabla}_\d] \bm G~.
\end{align}
\esubeq
The second building block,  
$V = E^A V_A$, is the gauge one-form describing the vector multiplet 
associated with $\cV$. Modulo an exact one-form, we can choose 
the components of $V$ as follows:
\be V_\a = \ri \nabla_\a \cV \ , \quad \bar{V}_\a = - \ri \bar{\nabla}_\a \cV \ , \quad V_a = - \frac{1}{4} (\g_a)^{\a \b} [\nabla_\a , \bar{\nabla}_\b] \cV \ .
\ee
The corresponding gauge-invariant field strength $F = \rd V$  
has the explicit  structure given by eqs. \eqref{6.4} and \eqref{6.5} 
with $\bm F$ and $\bm G$ replaced with $F$ and $G$ respectively, 
where $G$ denotes 
the gauge-invariant field strength
\be 
G = \ri \nabla^\a \bar \nabla_\a \cV 
\ee
associated  with the prepotential $\cV$. 
Finally, the three-form $\S$ is chosen to obey the equation 
\bea
\rd \S = F \wedge \bm F~.
\eea
Its components are defined by 
\begin{align} \S &= \bar{E}^\g \wedge E^\b \wedge E^a \S_{a \b \g} + \hf E^\g \wedge E^b \wedge E^a \S_{ab \g}
+ \hf \bar{E}^\g \wedge E^b \wedge E^a \bar{\S}_{ab \g} \non\\
&\quad+ \frac{1}{6} E^c \wedge E^b \wedge E^a \S_{abc}
\end{align}
and have the following explicit form:
\bsubeq
\begin{align}
\S_{a \b \g} &= - 2 \ri (\g_a)_{\b\g} G \bm G \ , \\
\S_{ab \g} &= - \eps_{abc} (\g^c)_{\g\d} (\bm G \nabla^\d G + G \nabla^\d \bm G)\ , \\
\bar{\S}_{ab \g} &= - \eps_{abc} (\g^c)_{\g\d} (\bm G \bar{\nabla}^\d G + G \bar{\nabla}^\d \bm G ) \ , \\
\S_{abc} &= \frac{\ri}{2} \eps_{abc} (4 \nabla^\g \bm G \bar{\nabla}_\g G + \bm G \bar{\nabla}^\g \nabla_\g G + G \bar{\nabla}^\g \nabla_\g \bm G )~.
\end{align}
\esubeq
The components of $\S$ are symmetric under the interchange 
$G \leftrightarrow \bm G$. When $G = \bm G$ we have agreement with eqs. \eqref{4.11} and \eqref{4.12} in the Abelian case.

Associated with the closed three-form $J$, eq. \eqref{6.2}, 
is the component action 
\begin{align}
S =& \hf \int \rd^3 x \,e \, \Big(
	\eps^{abc} v_a \bm f_{bc} 
	+ \frac{\ri}{2} \l^\g \bar{\bm \l}_\g + \frac{\ri}{2} \bm \l^\g \bar{\l}_\g
	+ g \bm h + \bm g h \non\\
	&- \hf (\g^a)_{\g\d} \psi_a{}^\g (g \bm \l^\d +  \bm g \l^\d) + \hf (\g^a)_{\g\d} \bar{\psi}_a{}^\g (g \bar{\bm \l}^\d + \bm g \bar{\l}^\d) \non\\
	&- \ri \eps^{abc} (\g_a)_{\g\d} \psi_b{}^\g \bar{\psi}_c{}^{\d} g \bm g  \Big) \ ,
	\label{6.9}
\end{align}
where the component fields are defined as in section \ref{ComponentActions} (in the complex basis):
\bsubeq
\begin{align}
g &= G| \ , \quad \l_\a =  -2 \nabla_\a G| \ , \quad \bar{\l}_{\a} =  -2 \bar{\nabla}_\a G| \ , \quad h = \ri \bar{\nabla}^\g \nabla_\g G| \ , \\
\bm g &= \bm G| \ , \quad  {\bm \l}_\a =  -2 \nabla_\a \bm G| \ , \quad \bar{\bm \l}_{\a} =  -2 \bar{\nabla}_\a \bm G| \ , \quad \bm h = \ri \bar{\nabla}^\g \nabla_\g \bm G| \ ,  \\
v_a &= e_a{}^m V_m|  \  , \\
\bm f_{ab} &= \bm F_{ab}| - \psi_{[a}{}^\b \bm F_{b] \b}| - \bar{\psi}_{[a}{}^\b \bar{\bm F}_{b] \b}| - \hf \psi_{[a}{}^\a \bar{\psi}_{b]}{}^\b \bm F_{\a\b} \non\\
&= - \frac{\ri}{8} \eps_{abc} (\g^c)^{\g\d} [\nabla_\g , \bar{\nabla}_\d] \bm G| + \frac{\ri}{2} \psi_{[a}{}^\b (\g_{b]})_\b{}^\g {\bm \l}_\g
- \frac{\ri}{2} \bar{\psi}_{[a}{}^\b (\g_{b]})_\b{}^\g \bar{\bm \l}_\g + \psi_{[a}{}^\a \bar{\psi}_{b] \a} \bm g \ .
\end{align}
\esubeq
Eq. \eqref{6.9} is exactly the component form of the action \eqref{N=2BF}.

Let us recall that the most general $\cN=2$ supergravity-matter system 
(see \cite{KLT-M11,KT-M11} for more details)
is described by an action of the form 
\bea
S&=&\int\rd^3x\rd^2\q\rd^2\bar \q \,E \,\cL
+
\int  \rd^3x \rd^2\q\, \cE \,\cL_{\rm c} 
+ \int  \rd^3x \rd^2 \bar \q\, \bar \cE \, \bar \cL_{\rm c} 
~,
\label{N=2action}
\eea
for some real scalar $\cL$ and covariantly chiral scalar  $\cL_{\rm c}$ Lagrangians, 
$\bar \nabla_\a  \cL_{\rm c} =0$. Here $\cE$ denotes the chiral density.\footnote{The 
explicit expression for $\cE$ in terms of the supergravity prepotentials is given in 
\cite{Kuzenko12}.}  
We assume that  the dynamical supermultiplets include 
an Abelian vector multiplet described by prepotential $\cV$ with  
nowhere vanishing field strength $G =  \ri \nabla^\a \bar \nabla_\a \cV$.
This is the case for Type II minimal supergravity \cite{KLT-M11,KT-M11}. 
Then, the first term in \eqref{N=2action} may be represented in the $BF$-form
\eqref{N=2BF}, specifically:
\bea
\int\rd^3x\rd^2\q\rd^2\bar \q \,E \,\cL 
= \int\rd^3x\rd^2\q\rd^2\bar \q \,E \,\cV \,\bm G ~, \qquad
\bm G := \ri \nabla^\a \bar \nabla_\a \frac{\cL}{G}~.
\label{6.12}
\eea

We see that the linear multiplet action \eqref{N=2BF} allows us to describe 
a broad class of locally supersymmetric models. 
However, this action principle is not universal for,  in general, 
it cannot be used
to describe the chiral term in 
\eqref{N=2action} and its conjugated antichiral one.  
On the other hand, the (anti)chiral action is truly universal in $\cN=2$ supersymmetry,
due to the identity \cite{KLT-M11}
\bea
\int\rd^3x\rd^2\q\rd^2\bar \q\, E \,\cL
= \int\rd^3x\rd^2 \bar \q\, \bar \cE \, \bar \cL_{\rm c} ~, \qquad \bar \cL_{\rm c} :=
-\frac{1}{4}
\nabla^\a \nabla_\a
\cL~.
\label{N=2Ac-2}
\eea
As demonstrated in \cite{KLRST-M}, 
this action can equivalently be described
in terms of a closed three-form $\X$, $\rd \X=0$,  
such that its components
\bea
\X 
=\hf E^\g \wedge E^\b \wedge E^a\, \X_{a\b\g} 
+\hf E^\g \wedge E^b \wedge E^a\, \X_{ab\g} 
+\frac{1}{6} E^c \wedge E^b \wedge E^a \, \X_{abc} ~~~~
\eea
are expressed via $ \bar \cL_{\rm c} $ as follows:
\begin{subequations}
\bea
 \X_{a\b\g} &=& 4 (\g_a)_{\b\g} \bar \cL_{\rm c}~, \\
 \X_{ab\g} &=& -\ri \ve_{abd} (\g^d)_{\g\d} \bar \nabla^\d \bar \cL_{\rm c} ~,\\
\X_{a b c} &=&\frac{1}{4} \ve_{abc}\bar \nabla_\d \bar \nabla^\d 
\bar  \cL_{\rm c}~.
\eea
\end{subequations}
In summary,  the $\cN=2$ linear multiplet action \eqref{N=2BF} is useful but not universal. 
As will be shown in the next section, the situation in $\cN=3$ supersymmetry is 
conceptually different.


\section{Matter-coupled $\cN=3$ supergravity}\label{N=3composite}

General off-shell matter couplings in $\cN=3$ supergravity were constructed in 
\cite{KLT-M11}.  Given a supergravity-matter system, its 
dynamics can be described by a Lagrangian $\cL^{(2)}(v)$ 
which is a real weight-two projective supermultiplet,\footnote{In what follows, 
we do not indicate explicitly the $z$-dependence of $\cN=3$ and $\cN=4$ superfields.} 
with $v^i$ the homogeneous  coordinates for ${\mathbb C}P^1$.
The corresponding action is given by eq. \eqref{InvarAc}.
We assume that the dynamical supermultiplets include an Abelian vector multiplet such that 
its gauge invariant field strength $G^{ij}$ is nowhere vanishing, 
that is  $G:=\sqrt{G^{ij}G_{ij}}\neq 0$. 
As shown in Appendix \ref{AppendixC}, the action functional \eqref{InvarAc}
 can be rewritten as a $BF$ term
\begin{subequations} \label{7.1+2}
\bea
S_{\rm LM}&=&
\frac{1}{2\pi\ri} \oint_\g (v, \rd v)
\int \rd^3 x \,{\rm d}^6\q\,E\, 
C^{(-4)}
 \cV \,\bm G^{(2)}~, 
\label{7.1}
\eea
where $\cV(v)$ is the tropical prepotential for the vector multiplet,  
$\nabla^{(2)}_\a \cV =0$, and 
\bea
{\bm G}^{(2)}(v) := {\bm G}_{ij} v^iv^j
~, \qquad
\nabla^{(2)}_\a {\bm G}^{(2)} =0 \quad \Longleftrightarrow \quad
\nabla^{(ij}_\a {\bm G}^{kl)} =0
\label{7.2}
\eea
\end{subequations}
is a composite real $\cO (2) $ or linear multiplet. 
The explicit expression for ${\bm G}^{(2)}(v) $ in terms of the superfield 
Lagrangian $\cL^{(2)}$ is given by eq. \eqref{C15.b}.
Different theories correspond to different choices of the composite linear multiplet 
$ {\bm G}_{ij}$.
The action \eqref{7.1} is invariant under gauge transformations 
\bea
\d \cV = \l + \breve{\l} ~,\qquad \nabla^{(2)}_\a \l =0~,
\label{7.3}
\eea
where the gauge parameter $\l$ is an arbitrary weight-0 arctic multiplet, 
and $\breve \l$ its smile-conjugate, see \cite{KLT-M11} for more details.  
Eq. \eqref{7.1} defines the $\cN=3$ linear multiplet action. 

Instead of dealing with the symmetric spinors $G^{ij}$ and $\bm G^{ij}$, 
we can equivalently work with the isovectors
\bea
G^I := (\S^I)_{ij} G^{ij}~, \qquad \bm G^I := (\S^I)_{ij} \bm G^{ij}~,
\eea
where the sigma-matrices are defined by 
\bea
(\S_I)_{i j}=({\mathbbm 1},\ri\s_1, \ri\s_3) =(\S_I)_{ji}~.
\eea

\subsection{Linear multiplet action} 

It turns out that the action \eqref{7.1} may be recast in terms of a closed three-form:
\be
J = V \wedge \bm F - \S \ , \qquad \rd J =0~, 
\ee
where $\bm F$ is the two-form field strength associated with $\bm G^I$ and $V = E^A V_A$ is the gauge one-form associated with the prepotential $\cV$. 
The three-form
$\S = \frac{1}{3!} E^C \wedge E^B \wedge E^A \S_{ABC}$ is given by
\bsubeq
\begin{align}
\S_\a^I{}_\b^J{}_\g^K &= 0 \ , \\
\S_a{}_\b^J{}_\g^K &= 2 \ri (\g_a)_{\b\g} \tr \{ \d^{JK} G^I \bm G_I - G^{J} \bm G^{K} - \bm G^J G^K \} \ , \\
\S_{ab}{}_\g^K &= \eps_{abc} (\g^c)_{\g}{}^\d \tr \{  \nabla_\d^{[K} G^{I]} \bm G_I + \nabla_\d^{[K} \bm G^{I]} G_I - \frac{1}{3}  \nabla_\d^I G_I \bm G^K -  \frac{1}{3}  \nabla_\d^I \bm G_I G^K \} \ , \\
\S_{abc} &= - \frac{\ri}{2} \eps_{abc} \tr \{ ( \nabla^\g_K  \nabla^K_\g G^L) \bm G_L + ( \nabla^\g_K  \nabla^K_\g \bm G^L) G_L  + ( \nabla^\g_{[K} G_{L]}) ( \nabla^{[K}_\g \bm G^{L]}) \non\\
&\qquad\qquad - \frac{2}{9} ( \nabla^{\g K} G_K) ( \nabla^L_\g \bm G_L) \} \ .
\end{align}
\esubeq

The component action generated by $J$ is
\begin{align}
S_{\rm LM} =& \hf \int \rd^3 x \,e \, \Big(
	\eps^{abc} v_a \bm f_{bc}
	- 2 \ri \c^\g \c_\g  - \frac{\ri}{4} \l^{\g IJ} \bm \l_{\g IJ} + g^I \bm h_I + \bm g^I h_I \non\\
	&- \hf (\g^a)_{\g\d} \psi_a{}^\g_I (\l^{\d IJ} \bm g_J + \bm \l^{\d IJ} g_J  + \ri \c^\d \bm g^I + \ri \bm \c^\d g^I) \non\\
	&+ \frac{\ri}{2} \eps^{abc} (\g_a)_{\g\d} \psi_b{}^\g_K \psi_c{}^{\d}_L (\d^{KL} g^P \bm g_P - 2 g^K \bm g^L)  \Big) \ ,
\label{7.8}
\end{align}
where the component fields are defined as in section \ref{ComponentActions}. They are explicitly given by
\bsubeq
\begin{align}
g_I &= G_I| \ , \quad \l_{\a}{}^{IJ} = 2 \nabla_{\a}^{[I} G^{J]}| \ , \quad \c_\a = \frac{\ri}{3}  \nabla_\a^I G_I | \ , \quad h_I = - \ri  \nabla^{\g J}  \nabla_{\g J} G_I | \ , \\
\bm g_I &= \bm G_I| \ , \quad \bm \l_{\a}{}^{IJ} = 2 \nabla_{\a}^{[I} \bm G^{J]}| \ , \quad \bm \c_\a = \frac{\ri}{3}  \nabla_\a^I \bm G_I | \ , \quad \bm h_I = - \ri  \nabla^{\g J}  \nabla_{\g J} \bm G_I | \ , \\
v_a &= e_a{}^m V_m| = V_a| + \hf \psi_a{}^\a_I V_\a^I | \ , \\
\bm f_{ab} &= \bm F_{ab}| - \hf (\psi_{[a}{}^K \g_{b]} \bm \l_K) + \frac{\ri}{2} \psi_a{}^{\g K} \psi_b{}_\g^L \bm g_{KL} \non\\
&= - \frac{\ri}{12} \eps_{abc} (\g^c)^{\a\b} \eps^{IJK} 
\nabla_{\a I} \nabla_{\b J} \bm G_K |
- \frac{1}{4} \eps^{IJK} (\psi_{[a I} \g_{b]} \bm \l_{JK}) + \frac{\ri}{2} \eps^{IJK} \psi_a{}^{\g}_I \psi_b{}_{\g J}  \bm g_{K} \ .
\end{align}
\esubeq

To prove that the $\cN=3$  linear multiplet action \eqref{7.1} 
has the component form \eqref{7.8}, it suffices to redo, in a 3D setting,  
the 4D $\cN=2$ analysis given in \cite{BN}.


\subsection{Composite  $\cO(2)$ multiplet} \label{subsection7.2}

We now present a special example of the composite $\cO(2)$ multiplet 
defined by \eqref{C15.b}. 
We consider a vector multiplet Lagrangian of the form
\bea
\cL^{(2)} \propto
G^{(2)} \ln \frac{G^{(2)} }{\ri \Upsilon^{(1)} \breve\Upsilon^{(1)}} ~,
\label{7.10}
\eea
where $\Upsilon^{(1)} (v)$ is a weight-1 arctic multiplet and  
$\breve\Upsilon^{(1)} (v)$ its smile-conjugated antarctic multiplet. 
The superfields  $\Upsilon^{(1)} $ and  $\breve\Upsilon^{(1)} $ are pure gauge degrees 
of freedom \cite{KLT-M11}. In the rigid supersymmetric limit, the Lagrangian
\eqref{7.10} describes a superconformal vector multiplet, which is the 3D $\cN=3$ 
analogue of the 4D $\cN=2$  
improved tensor multiplet \cite{deWPV,LR83}.\footnote{The 
4D $\cN=1$  improved tensor multiplet was introduced in \cite{deWitRocek}.
The $\cN=2$ construction of  \cite{deWPV,LR83} is a natural extension 
of the one given in \cite{deWitRocek}.}

With the Lagrangian \eqref{7.10}, 
the contour integral in \eqref{C15.b} can be evaluated 
using the techniques of \cite{BK11}. 
Alternatively, one may look for a dimension-1 primary superfield 
that obeys the Bianchi identity \eqref{N=3BI}. The resulting composite 
$\cO(2)$ multiplet is
\be 
\bm G^I = \ri \frac{G_J}{G^2} \nabla^{\a (I} \nabla_\a^{J)} G
- \frac{\ri}{4 G^3} G^I \nabla^{\a [J} G^{K]} \nabla_{\a[J} G_{K]} - \frac{\ri}{18 G^3} G^I \nabla^\a_J G^J \nabla_\a^K G_K \ ,
\label{7.11}
\ee
where
\be G^2 := G^I G_I = \hf G^{IJ} G_{IJ} = G^{ij} G_{ij}
\ee
is required to be nowhere vanishing. The $\cO(2)$ multiplet may be expressed in terms of $\rm SU(2)$ indices as follows
\be \bm G^{ij} = \ri \frac{G_{kl}}{G^2} \nabla^{\a ij} \nabla_\a^{kl} G - \frac{\ri}{4 G^3} G^{ij} \nabla^{\a kp} G_k{}^q \nabla_\a{}^l{}_{(p} G_{q) l}
- \frac{\ri}{18 G^3} G^{ij} \nabla^\a_{kl} G^{kl} \nabla_\a^{pq} G_{pq} \ .
\ee


\subsection{Supercurrent}\label{subsection7.3}

Before turning to a consideration of specific supergravity models, 
it is worth giving a few  remarks concerning matter couplings
to $\cN=3$ conformal supergravity (see also \cite{BKNT-M1,KNT-M}). 
In general, 
matter-coupled conformal supergravity 
is described by an action of the form
\bea 
S = \frac{1}{\widetilde \m} S_{\rm CSG} + S_{\rm matter} ~.
\label{SUGRA-matteraction}
\eea
Here 
 $S_{\rm CSG}$ denotes the $\cN=3$ conformal supergravity action \cite{BKNT-M2} 
 and $S_{\rm matter}$ the matter action. The equation of motion for 
the Weyl multiplet is\footnote{The coupling constants 
$\widetilde{\mu}$ and $\m$ differ from each other by some numerical coefficient.}
\be \frac{1}{ \m} W_\a + T_\a = 0 \ ,
\ee
where $T_\a$ is the matter supercurrent. As a result the supercurrent $T_\a$ must 
have the same properties 
as the super Cotton tensor $W_\a$. Specifically, $T_\a$ must be a primary superfield 
of dimension 3/2, 
\bea
S^J_\b T_\a =0~, \qquad {\mathbb D} T_\a = \frac{3}{2} T_\a~,
\eea
and obey the conservation equation
\be 
\nabla^\a_I T_\a = 0 \ .
\label{7.16}
\ee
The latter holds provided the matter equations of motion  are satisfied. 

Matter-coupled Poincar\'e or anti-de Sitter supergravities can also be described 
by actions of the type \eqref{SUGRA-matteraction} 
with $1/\widetilde{\m} = 0$. The matter supermultiplets
have to include a conformal compensator. 
In what follows, the latter is assumed to be the vector multiplet 
described by the field strength $G^I$. The supergravity equation of motion is 
\bea
T^I_\a=0~.
\eea

As an example, consider $\cN=3$ AdS supergravity.
It can be described by the Lagrangian  \cite{KLT-M11} 
\bea
 \cL^{(2)}_{\rm SG} = \frac{1}{\k} \Big\{ 
G^{(2)} \ln \frac{G^{(2)} }{\ri \Upsilon^{(1)} \breve\Upsilon^{(1)}} 
+\hf \x \,
\cV G^{(2)} \Big\}~,
\label{7.18}
\eea
with $\k$ and  $\x$ the gravitational and  cosmological constants respectively. 
The cosmological term is  a U(1) Chern-Simons term.
The choice $\x=0$ corresponds to Poincar\'e supergravity.  
The corresponding  supercurrent is
\be 
\k \,T_\a 
= \frac{\ri}{G} \eps^{IJK} G_I \nabla_{\a J} G_K \ .
\ee
One can show that if $G^I$ satisfies the equation of motion for $\cV$,
$\bm G^I + \xi G^I = 0$,
the supercurrent does obey eq. \eqref{7.16}.


\subsection{(2,1) anti-de Sitter supergravity}

It was discovered by Ach\'ucarro and Townsend  \cite{AT} that 
three-dimensional $\cN$-extended anti-de Sitter (AdS) supergravity 
exists in $[\cN/2] +1$ different versions, 
with $[\cN/2]$ the integer part of $\cN/2$.
These were called the  $(p,q)$  supergravity theories  
where the  non-negative integers $p \geq q$ are such that 
$\cN=p+q$. 

We wish to demonstrate that the Lagrangian \eqref{7.18}
describes (2,1) AdS supergravity. 
 To see this we will degauge, following the procedure described in \cite{BKNT-M1}, 
 the corresponding equations of motion, 
\begin{subequations} 
\bea
\bm G^I + \xi G^I &=& 0 \ , 
\label{N=3EOM} \\
 \k\,T_\a 
= \frac{\ri}{G} \eps^{IJK} G_I \nabla_{\a J} G_K &=&0\ ,
\label{7.20b}
\eea
\end{subequations}
to  $\rm SO(3)$ superspace \cite{HIPT, KLT-M11}.
As in \cite{BKNT-M1}, the covariant derivatives of SO(3) superspace are denoted 
$\cD_A =(\cD_a,\cD^I_\a)$. 

Using the results of \cite{BKNT-M1} we can degauge our expression for $\bm G^I$,  
eq. \eqref{7.11}, to
\begin{align} 
\bm G^I &= \ri \frac{G_J}{G^2} (\cD^{\a (I} \cD_\a^{J)} - 4 \ri S^{IJ}) G 
- \frac{\ri}{4 G^3} G^I \cD^{\a [J} G^{K]} \cD_{\a[J} G_{K]} \non\\
&\quad- \frac{\ri}{18 G^3} G^I \cD^\a_J G^J \cD_\a^K G_K \ .
\end{align}
The covariant derivatives $\cD_A$ no longer contain the dilatation and special conformal 
generators, 
\bea
\cD_A 
= E_A{}^M \pa_M - \hf \Omega_A{}^{ab} M_{ab} - \hf \Phi_A{}^{PQ} N_{PQ}\ .
\eea
The original local dilatation symmetry 
is now realized in terms of the  super-Weyl transformations. 
In SO(3) superspace, there are two dimension-1 real torsion tensors, 
$S^{IJ} = S^{JI}$ and $C_a{}^{IJ} = - C_a{}^{JI}$. 
The super Cotton tensor $W_\a$ becomes a descendant of $C_a{}^{IJ} $,
\be
W_\a 
= \frac{\ri}{12} \eps_{IJK} \cD^{\b I} C_{\a\b}{}^{JK} \ .
\ee
We refer the reader to \cite{BKNT-M1} for more details about the degauging procedure. 

Using the super-Weyl transformation of $G$ \cite{KLT-M11},
\be 
G' = e^\s G~,
\ee
we can impose
the gauge condition
\be G = 1 \ .
\ee
Taking a spinor derivative of $G$ then gives
\be G_K \cD_\a^{[I} G^{K]} + \frac{1}{3} G^I \cD_\a^K  G_{K} = 0 \ , \label{5.10}
\ee
which requires
\be \cD_\a^K G_K = 0 \ , \quad G_J \cD_\a^{[I} G^{J]} = 0 \ . \label{5.11}
\ee
Note that the Bianchi identity now simplifies to
\be \cD_\a^{I} G^{J} = \cD_\a^{[I} G^{J]} \ . \label{dim3/2antisym}
\ee

Since the supercurrent vanishes, eq. \eqref{7.20b}, we must also have
\be G_{[I} \cD_{\a J} G_{K]} = 0 \ .
\ee
On the other hand, using eq. \eqref{5.10} we find
\begin{align}
G_I \cD_{\a J} G_K &= G_{[I} \cD_{\a J} G_{K]} + \frac{2}{3} G_I \cD_{\a J} G_K - \frac{2}{3} G_{[J} \cD_{\a K]} G_I \non\\
&= \frac{2}{3} G_I \cD_{\a J} G_K - \frac{2}{3} G_{[J} \cD_{\a K]} G_I \ .
\end{align}
Contracting the above with $G^I$ and implementing eq. \eqref{5.11}  
tells us that $G^I$ is covariantly constant,
\be \cD_{\a J} G_K = 0 \ .
\ee

The fact that $G^I$ is covariantly constant strongly constrains the superspace geometry. In particular, we have
\begin{align}
0 &= \{ \cD_\a^I , \cD_\b^J \} G^K \non\\
&= 2 \ri \d^{IJ} \cD_{\a\b} G^K - 4 \ri  \eps_{\a\b} S^{K[I} G^{J]} - 4 \ri \eps_{\a\b} \d^{K[I} S^{J] L} G_L \non\\
&\quad - 4 \ri C_{\a\b}{}^{K(I} G^{J)} + 4 \ri C_{\a\b}{}^{L(I} \d^{J) K} G_L \ ,
\end{align}
which 
fixes the form of the curvature components as
\begin{subequations}
\bea 
S^{IJ} &=& S (\d^{IJ} - 2 G^I G^J) \ , \qquad S := S^K{}_K\ , \\
C_{\a\b}{}^{IJ} &=& 0 \ .
\eea
\end{subequations}

The composite vector multiplet now reduces to
\bea 
\bm G^I = - 4 S G^I  \quad \Longrightarrow \quad 4S =\x~.
\eea
Due to the equation of motion \eqref{N=3EOM}, $S$ is seen to be constant,
\be \cD_\a^I S = 0 \ .
\ee
As a result, 
the covariant derivatives corresponds 
to $(2 , 1)$ $\rm AdS$ superspace \cite{AdSpq}. Therefore the 
theory \eqref{7.18} indeed describes (2,1) AdS supergravity. 

Without  a cosmological constant, $\xi = 0$, we find
\be 
S^{IJ} = 0~,
\ee
and the resulting geometry corresponds to Minkowski superspace. 


\subsection{Topologically massive supergravity}

Topologically massive $\cN=3$ supergravity\footnote{Topologically massive 
$\cN=1$ supergravity was introduced in \cite{DK,Deser}. The off-shell versions
of topologically massive $\cN=2$ supergravity were presented in \cite{KLRST-M}.}
can be described by the action 
\bea 
S_{\rm TMSG} = \frac{1}{\widetilde \m} S_{\rm CSG} + S_{\rm SG} ~,
\eea
where $S_{\rm SG} $ corresponds to the supergravity Lagrangian \eqref{7.18}.

The 
equation of motion for the Weyl multiplet now becomes 
\be 
T_\a + \frac{1}{ \m} W_\a = 0 ~,
\label{7.37}
\ee
compare with \eqref{7.20b}. The equation of motion for $\cV$ coincides with 
\eqref{N=3EOM}.

We would like to degauge the equations of motion to SO(3) superspace. 
Eq. \eqref{7.37} tells us that
\be 
\frac{\ri}{\k G} \eps^{IJK} G_I \cD_{\a J} G_{K} = - \frac{1}{ \m} W_\a \ ,
\ee
where upon degauging the Cotton tensor is
\be W_\a = \frac{\ri}{12} \eps_{IJK} \cD^{\b I} C_{\a\b}{}^{JK} = \frac{\ri}{6} \cD^{\b I} C_{\a\b I} \ .
\ee
Using the gauge condition $G = 1$ we find
\be G_{[I} \cD_{\a J} G_{K]} = \frac{\ri}{6 \hat{\mu}} \eps_{IJK} W_\a \ ,
\ee
where we have defined the constant
\be \hat{\m} = \frac{\m}{\kappa} \ .
\ee
Contracting with $G_K$ gives
\be \cD_{\a [I} G_{J]} = \frac{\ri}{2 \hat{\m}} \eps_{IJK} G^K W_\a \ .
\ee
Therefore the Bianchi identity becomes
\be \cD_\a^I G^J = \frac{\ri}{2 \hat{\m}} \eps^{IJK} G_K W_\a = - \frac{1}{12 \hat{\m}} \eps^{IJK} G_K \cD^{\b L} C_{\a\b L} \ . \label{BI7.44}
\ee

Next, we notice that on the one hand, due to eq. \eqref{BI7.44}, we have
\be \{ \cD_\a^I , \cD_\b^J \} G^K = - \frac{1}{4 \hat{\m}^2} \eps_{\a\b} \d^{K[I} G^{J]} W^\g W_\g - \frac{\ri}{\hat{\m}} G^{K(I} \cD_{(\a}^{J)} W_{\b )} \ ,
\ee
while on the other we have
\begin{align} \{ \cD_\a^I , \cD_\b^J \} G^K &= 2 \ri \d^{IJ} \cD_{\a\b} G^K - 4 \ri  \eps_{\a\b} S^{K[I} G^{J]} - 4 \ri \eps_{\a\b} \d^{K[I} S^{J] L} G_L \non\\
&\quad - 4 \ri C_{\a\b}{}^{K(I} G^{J)} + 4 \ri C_{\a\b}{}^{L(I} \d^{J) K} G_L \ .
\end{align}
Combining the two results gives
\bsubeq \label{cond7.46}
\begin{align}
S^{IJ} &= S (\d^{IJ} - 2 G^I G^J) - \frac{\ri}{8 \hat{\mu}^2} W^\g W_\g G^I G^J \ , \quad S = S_K{}^K \ , \\
\cD_{\a\b} G^I &= - \frac{1}{2 \hat{\m}} G^{IJ} \cD_{(\a J} W_{\b)} \ , \\
C_{\a\b}{}^{IJ} &= \frac{1}{4 \hat{\m}} \eps^{IJK} \cD_{(\a K} W_{\b )} \ .
\end{align}
\esubeq
These imply the equation of motion on $C_{\a\b}{}^I$
\be \cD_{(\a}^I \cD^{\g J} C_{\b ) \g J} + 24 \ri \hat{\m} C_{\a\b}{}^I = 0
\ee
and the corresponding equation of motion on $W_\a$
\begin{subequations} \label{7.48}
\bea 
\cD^\b_J \cD_\b^J W_\a + 24 \ri \hat{\mu} W_\a = 0 \ . 
\label{EOMWspinor}
\eea
In addition to \eqref{EOMWspinor}, the Cotton superfield must obey the 
Bianchi identity
\bea
\cD^{\a I} W_\a =0~.
\eea
\end{subequations}

Due to the conditions \eqref{cond7.46}, the composite vector multiplet may be expressed as follows
\be \bm G^I = 4 S^{IJ} G_J + \frac{\ri}{8 \hat{\m}^2} W^\a W_\a G^I = - 4 S G^I - \frac{3 \ri}{8 \hat{\m}^2} W^\g W_\g G^I \ .
\ee
Furthermore, from the equation of motion \eqref{N=3EOM} we see that $S$ can be expressed in terms of the Cotton tensor as
\be S = \frac{\x}{4} - \frac{3 \ri}{32 \hat{\m}^2} W^\a W_\a \ .
\ee

For $\x =0$, a solution of the equations of motion for topologically massive supergravity 
is obtained by setting $W_\a =0$ in the above relations. 
This solution describes a flat superspace. 
Linearizing the equations \eqref{7.48} around Minkowski superspace, 
it may be shown that $W_\a$ obeys the Klein-Gordon equation
\be 
(\Box -m^2) W_\a 
= 0 \ , \qquad m=4\hat \m = 4 \frac{\m}{\k}
\ ,
\ee
with $\Box := \partial^a \partial_a$. 
For $\x\neq 0$,  the equations of motion for topologically massive supergravity 
are solved by setting $W_\a =0$. Locally it describes  
(2,1) AdS superspace \cite{AdSpq}. 


\section{Matter-coupled $\cN=4$ supergravity}\label{N=4composite}

The off-shell matter couplings in $\cN=4$ supergravity were constructed in 
\cite{KLT-M11}.  In general, the   action for a supergravity-matter system
may be represented as a sum of two terms, $S= S_{\rm L} + S_{\rm R}$, 
the left  $S_{\rm L}$ and right  $S_{\rm R}$ actions, which are naturally 
formulated in  curved $\cN=4$ projective superspace 
$\cM^{3|8} \times {\mathbb C}P^1_\rL \times {\mathbb C}P^1_\rR$.
The left action is given by eq.  \eqref{Action-left},  
where the Lagrangian 
$\cL^{(2)}_\rL (v_\rL )$ is a real left projective multiplet of weight two, 
with $v_\rL = v^i$ the homogeneous coordinates for ${\mathbb C}P^1_\rL$.
The structure of $S_\rR$ is analogous. 

We assume that the dynamical supermultiplets include two 
Abelian vector multiplets such that their field strengths $G^{IJ}_+$ and $G^{IJ}_-$ 
are self-dual and anti-self-dual, respectively, and nowhere vanishing, 
$G_{\pm}^2 := \hf G_{\pm}^{IJ} G_{\pm IJ} \neq 0$.
The anti-self-dual field strength $G^{IJ}_-$ can equivalently be realized as a 
left $\cO(2)$ multiplet $G_\rL (v_\rL) := G_{ij} v^i v^j$. 
The self-dual field strength $G^{IJ}_+$ can equivalently be realized as a 
right $\cO(2)$ multiplet $G_\rR (v_\rR) := G_{\bar i \bar j} v^{\bar i} v^{\bar j}$. 
The vector multiplet with field strength $G^{IJ}_+$ can be described 
in terms of a gauge prepotential $\cV_\rL(v_\rL)$,  which is a left weight-0
tropical multiplet with gauge freedom \eqref{left-gauge}. 
The right $\cO(2)$ multiplet $G_\rR (v_\rR)$ is constructed in terms
of $\cV_\rL$ according to \eqref{D.10} and proves to be 
a gauge invariant field strength. Similar properties hold 
for the vector multiplet  field strength $G^{IJ}_-$ except all 
 `left' objects have to be replaced by `right' ones and vice versa.

As demonstrated in Appendix \ref{AppendixD}, 
the left action can be recast in the $BF$ form \eqref{D.28}, 
where ${\bm G}_{\rm R}^{(2)}(v_{\rm R}) 
= v_{\bar i} v_{\bar j} {\bm G}^{\bar{i} \,\bar{j}} $ is a composite right
$\cO(2)$ multiplet defined by \eqref{D.29}.
Eq. \eqref{D.28} defines the {\it right linear multiplet action}, $S_{\rm RLM}$.  
Obvious modifications lead to the {\it left linear multiplet action}, $S_{\rm LLM}$. 
One of our goals in this section is to reduce the actions
$S_{\rm RLM}$ and $S_{\rm LLM}$  to components.

\subsection{Left linear multiplet action}

The left linear multiplet action is given by
\be S_{\rm LLM} = \frac{1}{2 \pi} \oint (v_{\rm L} , \rd v_{\rm L}) \int \rd^3 x \rd^8 \theta \ C_{\rm L}^{(-4)} \cV_{\rm L} \bm G_{\rm L}^{(2)} \ ,
\ee
where 
 ${\bm G}_{\rm L}^{(2)}(v_{\rm L}) 
= v_{ i} v_{ j} {\bm G}^{{i} {j}} $ is a composite left
$\cO(2)$ multiplet, and $\cV_\rL(v_\rL) $ is the tropical prepotential 
 of the vector multiplet with field strength $G_+^{IJ}$. 
The  composite left $\cO(2)$ multiplet, ${\bm G}_{\rm L}^{(2)}(v_{\rm L}) $,
can be equivalently realized as the anti-self-dual SO(4) bivector 
${\bm G}_-^{IJ}$.

It turns out that the  action $S_{\rm LLM}$ 
may be reformulated in terms of 
the closed three-form
\be
J = V \wedge \bm F - \S \ ,
\ee
where $\bm F$ is the super two-form  associated with $\bm G_-^{IJ}$ 
and $V = E^A V_A$ is the gauge one-form 
associated with the field strength $G_{+}^{IJ}$. 
The three-form $\S = \frac{1}{3!} E^C \wedge E^B \wedge E^A \S_ {ABC}$ is 
\bsubeq
\begin{align}
\S_{}{}_\a^I{}_\b^J{}_\g^K &= 0 \ , \\
\S_{}{}_a{}_\b^J{}_\g^K &= - 2 \ri (\g_a)_{\b\g} 
G_{+}{}^{P(J} \bm G_{-}{}^{K)}{}_P
 \ , \\
\S_{}{}_{ab}{}_\g^K &= - \frac{1}{3}\eps_{abc} (\g^c)_\g{}^\d (\nabla_{\d I} G_+^{IJ} \bm G_{- J}{}^K + \nabla_{\d I} \bm G_-^{IJ} G_{+ J}{}^K) \ , \\
\S_{}{}_{abc} &= \ri \eps_{abc} \big(\frac{1}{24} \nabla^{\g}_J \nabla_{\g}^I \bm G_{- IK} G_+^{KJ} 
+ \frac{1}{24} \nabla^{\g}_J \nabla_{\g}^I G_{+ IK} \bm G_-^{KJ} + \frac{1}{9} \nabla^{\g}_I \bm G_-^{IJ} \nabla_\g^K G_{+ JK}\big) \ .
\end{align}
\esubeq
The corresponding component action is
\begin{align}
S_{\rm LLM} =& \hf \int \rd^3 x \,e \Big( \eps^{abc} v_{a} \bm f_{bc} + \frac{1}{4} \hat{\bm h}_+{}^{IJ} g_{+ IJ} + \frac{1}{4} \hat{h}_-{}^{IJ} \bm g_{-IJ} - \frac{\ri}{2} \l^{\a I} \bm \l_{\a I} \non\\
&\quad- \frac{1}{2} (\g^a)_{\g\d} \psi_a{}^\g_I (\l^{\d J} \bm g_{-J}{}^I + \bm \l^{\d J} g_{+ J}{}^I ) \non\\
&\quad + \frac{\ri}{2} \eps^{abc} (\g_a)_{\g\d} \psi_b{}^\g_K \psi_c{}^\d_L \, g_+{}^{KP} \bm g_-{}^L{}_P \Big) \ ,
\label{N=4LMA-left}
\end{align}
where the component fields are defined as in section \ref{ComponentActions}:
\bsubeq
\begin{align}
g_+^{IJ} &= G_+^{IJ}| \ , \quad \l{}_\a^I = \frac{2}{3} \nabla_{\a J} G_+^{IJ}| \ , \quad \hat{h}_{-}{}^{IJ} = \frac{\ri}{3} \nabla^{\g [I} \nabla_{\g K} G_+^{J] K}| - 2 w g_{+}{}^{IJ} \ , \\
\bm g_-^{IJ} &= \bm G_-^{IJ}| \ , \quad \bm \l{}_\a^I = \frac{2}{3} \nabla_{\a J} \bm G_-^{IJ}| \ , \quad \hat{\bm h}_{+}{}^{IJ} = \frac{\ri}{3} \nabla^{\g [I} \nabla_{\g K} \bm G_-^{J] K}| + 2 w \bm g_{-}{}^{IJ} \ , \\
v_{a} &= e_a{}^m V_{m}| = V_{a}| + \hf \psi_a{}^\a_I V_\a^I | \ , \\
\bm f_{ab} &= \bm F_{ab}| - \hf (\psi_{[a}{}^K \g_{b]} \bm \l_{K}) + \frac{\ri}{2} \psi_a{}^{\g K} \psi_b{}_\g^L \, \bm g_{- KL} \non\\
&= - \frac{\ri}{24} \eps_{abc} (\g^c)^{\a\b} \nabla_\a^K \nabla_\b^L \bm G_{- KL}| - \hf (\psi_{[a}{}^K \g_{b]} \bm \l_{K}) + \frac{\ri}{2} \psi_a{}^{\g K} \psi_b{}_\g^L \, \bm g_{- KL} \ .
\end{align}
\esubeq
The component fields are defined so that $\hat{h}_{-}^{IJ}$ is anti-self-dual and $\hat{\bm h}_{+}^{IJ}$ is self-dual, see eq. \eqref{compDuality}.

\subsection{Right linear multiplet action}

The right linear multiplet action 
is given by
\be S_{\rm RLM} = \frac{1}{2 \pi} \oint (v_{\rm R} , \rd v_{\rm R}) \int \rd^3 x \rd^8 \theta \ C_{\rm R}^{(-4)} \cV_{\rm R} \bm G_{\rm R}^{(2)} \ ,
\ee
where  ${\bm G}_{\rm R}^{(2)}(v_{\rm R}) 
= v_{\bar i} v_{\bar j} {\bm G}^{\bar{i} \,\bar{j}} $ is a composite right
$\cO(2)$ multiplet,
and $\cV_\rR(v_\rR) $ is the tropical prepotential 
 of the vector multiplet with field strength $G_-^{IJ}$. 
The  composite right $\cO(2)$ multiplet, ${\bm G}_{\rm R}^{(2)}(v_{\rm R}) $,
can be equivalently realized as the self-dual SO(4) bivector 
$\bm G_+^{IJ}$. 

The action $S_{\rm RLM}$ 
may be reformulated in terms of the closed three-form
\be
J = V \wedge \bm F - \S \ ,
\ee
where $\bm F$ is the two-form field strength associated with 
$\bm G_+^{IJ}$ and $V = E^A V_{A}$ is the gauge one-form 
associated with the field strength $G_{-}^{IJ}$. 
The three-form $\S = \frac{1}{3!} E^C \wedge E^B \wedge E^A \S_{ ABC}$ 
is given by
\bsubeq
\begin{align}
\S_{}{}_\a^I{}_\b^J{}_\g^K &= 0 \ , \\
\S_{}{}_a{}_\b^J{}_\g^K &= - 2 \ri (\g_a)_{\b\g} 
G_{-}{}^{P(J} \bm G_{+}{}^{K)}{}_P
 \ , \\
\S_{}{}_{ab}{}_\g^K &= - \frac{1}{3}\eps_{abc} (\g^c)_\g{}^\d (\nabla_{\d I} G_-^{IJ} \bm G_{+ J}{}^K + \nabla_{\d I} \bm G_+^{IJ} G_{- J}{}^K) \ , \\
\S_{}{}_{abc} &= \ri \eps_{abc} \big(\frac{1}{24} \nabla^{\g}_J \nabla_{\g}^I \bm G_{+ IK} G_-^{KJ} 
+ \frac{1}{24} \nabla^{\g}_J \nabla_{\g}^I G_{- IK} \bm G_+^{KJ} + \frac{1}{9} \nabla^{\g}_I \bm G_+^{IJ} \nabla_\g^K G_{- JK}\big) \ .
\end{align}
\esubeq
The corresponding component action is
\begin{align}
S_{\rm RLM} =& \hf \int \rd^3 x \,e \Big( \eps^{abc} v_{a} \bm f_{bc} + \frac{1}{4} \hat{h}_+{}^{IJ} \bm g_{+ IJ} + \frac{1}{4} \hat{\bm h}_-{}^{IJ} g_{-IJ} - \frac{\ri}{2} \bm \l^{\a I} \l_{\a I} \non\\
&\quad- \frac{1}{2} (\g^a)_{\g\d} \psi_a{}^\g_I (\bm \l^{\d J} g_{-J}{}^I + \l^{\d J} \bm g_{+ J}{}^I ) \non\\
&\quad + \frac{\ri}{2} \eps^{abc} (\g_a)_{\g\d} \psi_b{}^\g_K \psi_c{}^\d_L \, \bm g_+{}^{KP} g_-{}^L{}_P \Big) \ ,
\label{N=4LMA-right}
\end{align}
where the component fields are defined as in section \ref{ComponentActions}:
\bsubeq
\begin{align}
g_-^{IJ} &= G_-^{IJ}| \ , \quad \l{}_\a^I = \frac{2}{3} \nabla_{\a J} G_-^{IJ}| \ , \quad \hat{h}_{+}{}^{IJ} = \frac{\ri}{3} \nabla^{\g [I} \nabla_{\g K} G_-^{J] K}| - 2 w g_{-}{}^{IJ} \ , \\
\bm g_+^{IJ} &= \bm G_+^{IJ}| \ , \quad \bm \l{}_\a^I = \frac{2}{3} \nabla_{\a J} \bm G_+^{IJ}| \ , \quad \hat{\bm h}_{-}{}^{IJ} = \frac{\ri}{3} \nabla^{\g [I} \nabla_{\g K} \bm G_+^{J] K}| - 2 w \bm g_{+}{}^{IJ} \ , \\
v_{a} &= e_a{}^m V_{m}| = V_{a}| + \hf \psi_a{}^\a_I V_\a^I | \ , \\
\bm f_{ab} &= \bm F_{ab}| - \hf (\psi_{[a}{}^K \g_{b]} \bm \l_{K}) + \frac{\ri}{2} \psi_a{}^{\g K} \psi_b{}_\g^L \, \bm g_{+ KL} \non\\
&= - \frac{\ri}{24} \eps_{abc} (\g^c)^{\a\b} \nabla_\a^K \nabla_\b^L \bm G_{+ KL}| - \hf (\psi_{[a}{}^K \g_{b]} \bm \l_{K}) + \frac{\ri}{2} \psi_a{}^{\g K} \psi_b{}_\g^L \, \bm g_{+ KL} \ .
\end{align}
\esubeq
The component fields are defined so that $\hat{h}_{+}^{IJ}$ is self-dual and $\hat{\bm h}_{-}^{IJ}$ is anti-self-dual, see eq. \eqref{compDuality}.

\subsection{Composite $\cO (2)$ multiplets} 

Similar to the $\cN=3$ construction described in section \ref{subsection7.2}, 
we now present special examples of composite left and right  $\cO(2)$ multiplets. 
To construct ${\bm G}_{\rm R}^{(2)}$
we consider a massless vector multiplet Lagrangian of the form \cite{KLT-M11}
\bea
\cL^{(2)}_\rL \propto
G^{(2)}_\rL \ln \frac{G^{(2)}_\rL }{\ri \Upsilon^{(1)}_\rL \breve\Upsilon^{(1)}_\rL } 
\label{8.11}
\eea
and make use of the representation \eqref{D.29}.\footnote{The arctic 
weight-1 hypermultiplet $ \Upsilon^{(1)}_\rL$ and its smile conjugate  $\breve\Upsilon^{(1)}_\rL$
in \eqref{8.11} are purely gauge degrees of freedom.}
The contour integral in \eqref{D.29} may be evaluated using the technique 
developed in \cite{KLT-M11}. 
A similar  analysis may be used to construct ${\bm G}_{\rm L}^{(2)}$. 
Alternatively, $G_+^{IJ}\leftrightarrow {\bm G}_{\rm R}^{(2)} (v_\rR)  $  and 
$G_-^{IJ} \leftrightarrow {\bm G}_{\rm L}^{(2)} (v_\rL)  $
may also be found by looking for primary superfields 
which obey the Bianchi identity
\bea \label{CompVMBI}
 \nabla_{\g}^{I} {\bm G}_\pm^{ J K}&=&
 \nabla_{\g}^{[I} {\bm G}_\pm^{ J K]}
- \frac{2}{3} \d^{I [J}  \nabla_{\g L} {\bm G}_\pm^{ K] L}~.
\eea

The resulting composite $\cO(2)$  multiplets
are given by
\begin{subequations} \label{8.13}
\bea 
\bm G_{\pm}^{IJ} = X_\mp^{IJ} \pm \hf \eps^{IJKL} X_{\mp KL} \ , \quad
 \hf \eps_{IJKL} \bm G_{\pm}^{KL}  = \pm\bm G_{\pm IJ} 
 \ , 
\eea
where we have defined
\begin{align} X_{\pm}^{IJ} &:= \frac{\ri}{6 G_{\pm}} \nabla^{\g [I} \nabla_{\g K} G_{\pm}^{J] K}
+ \frac{2 \ri}{9 G_{\pm}^3} \nabla^{\a P} G_{\pm KP} \nabla_{\a Q} G_\pm^{Q [I} G_\pm^{J] K} \ .
\end{align}
\esubeq
To show that $\bm G_{\pm}^{IJ}$ is primary and satisfies the Bianchi identity, 
the following identities prove useful
\bsubeq
\begin{align} G_{\pm}^{IK} G_{\pm JK} &= \hf \d^I_J G_{\pm}^2 \ , \label{inverseGpm} \\
\eps^{IJKL} G_{\pm LP} &= \mp 3 \d^{[I}_P G_{\pm}^{JK]} \ .
\end{align}
\esubeq

In the isospinor notation, 
the composite $\cO(2)$  multiplets constructed 
read
\bsubeq
\begin{align}
\bm G{}^{\bar{i}\bar{j}} &= \frac{\ri}{6 G_-} \nabla^{\a i (\bar{i} } \nabla_\a^{j \bar{j})} G_{ij} 
- \frac{2 \ri}{9 G_-^3} \nabla^{\a i (\bar{i}} G_{ij} \nabla_\a^{k \bar{j})} G_{kl} G{}^{jl} \non\\
&= \frac{\ri}{4} \nabla^{\a i (\bar{i}} \nabla_\a^{j \bar{j})} \Big( \frac{G_{ij}}{G_-}\Big) \ ,\\
\bm G{}^{ij} &= \frac{\ri}{6 G_+} \nabla^{\a (i \bar{i}} \nabla_\a^{j) \bar{j}} G_{\bar{i} \bar{j}}
- \frac{2 \ri}{9 G_+^3} \nabla^{\a (i \bar{i}} G_{\bar{i} \bar{j}} \nabla_\a^{j) \bar{k}} G_{ \bar{k} \bar{l}} G{}^{\bar{l} \bar{j}} \non\\
&= \frac{\ri}{4} \nabla^{\a (i \bar{i}} \nabla_\a^{j) \bar{j}} \Big( \frac{G_{\bar{i}\bar{j}}}{G_+}\Big) \ .
\end{align}
\esubeq
These expressions may be compared with the 4D $\cN = 2$ results in \cite{BK11}.

For completeness we give 
the expressions for the composite $\cO(2)$ multiplets in $\rm SO(4)$ superspace:
\bsubeq
\begin{align}
\bm G{}^{\bar{i}\bar{j}} 
&= \frac{\ri}{4} (\cD^{\a i (\bar{i}} \cD_\a^{j \bar{j})} 
+ 8 \ri \cS^{ij\bar{i}\bar{j}} ) \Big( \frac{G_{ij}}{G_-}\Big) \ ,\\
\bm G{}^{ij} 
&= \frac{\ri}{4} (\cD^{\a (i \bar{i}} \cD_\a^{j) \bar{j}} 
+ 8 \ri \cS^{ij\bar{i}\bar{j}}) \Big( \frac{G_{\bar{i}\bar{j}}}{G_+}\Big) \ .
\end{align}
\esubeq
Here $\cS^{ij\bar{i}\bar{j}} = \cS^{( ij )(\bar{i}\bar{j})} $ is  one of the two irreducible components of the torsion superfield
$ S^{i \bar{i} , j \bar{j}}  := S^{IJ} (\t_I)^{i\bar i}  (\t_I)^{j\bar j}$, defined by
\be 
S^{i \bar{i},  j \bar{j}} = \cS^{ij\bar{i}\bar{j}} + \eps^{ij} \eps^{\bar{i} \bar{j}} \cS \ .
\ee

\subsection{Supercurrent}

The remainder of this section is devoted to a study of specific supergravity theories. 
To start with, we would like to discuss the structure of the 
$\cN=4$ supercurrent
(see also \cite{BKNT-M1,KNT-M}). 
Our consideration below is similar  to the $\cN=3$ analysis of section 
 \ref{subsection7.3}.
 
 Consider a dynamical system describing $\cN=4$ conformal supergravity 
 coupled to matter supermultiplets.
 In general, the supergravity-matter action  
 has the form
\be S = \frac{1}{\widetilde{\mu}} S_{\rm CSG} + S_{\rm matter}\ ,
 \label{SUGRA-matteractionN=4}
\ee
where $S_{\rm CSG}$ denotes the $\cN=4$ conformal supergravity action 
\cite{BKNT-M2} and $S_{\rm matter}$ the matter action. 
The equation of motion for conformal supergravity reads
\be 
\frac{1}{\mu} W + T = 0 \ ,
\ee
where $W$ is the $\cN=4$ Cotton superfield and 
$T$ is the matter supercurrent. 
It follows from this equation that the supercurrent must have the same properties as
$W$. Specifically, $T$ must be a primary superfield of dimension 1, 
\bea
S_\a^I T =0~, \qquad {\mathbb D} T = T~,
\eea
and obey the conservation equation 
\bea
\nabla^{\a (I} \nabla_\a^{J)} T = \frac{1}{4} \d^{IJ} \nabla^{\a}_K \nabla_\a^{K} T ~.
\label{N=4scce}
\eea
Of course, the latter holds provided the matter equations of motion  are satisfied. 

Matter-coupled Poincar\'e or anti-de Sitter supergravities can also be described 
by an action of the type \eqref{SUGRA-matteractionN=4}
with $1/\widetilde{\m} = 0$. The matter supermultiplets
have to include two conformal compensators.
As before, these are chosen to be two 
Abelian vector multiplets such that their field strengths $G^{IJ}_+$ and $G^{IJ}_-$ 
are self-dual and anti-self-dual, respectively, and nowhere vanishing, 
$G_{\pm}^2 := \hf G_{\pm}^{IJ} G_{\pm IJ} \neq 0$.
The supergravity equation of motion is 
\bea
T=0~.
\eea

As an example, let us consider $\cN=4$ AdS supergravity. 
it can be described by two Lagrangians, left and right ones, 
which were chosen in \cite{KLT-M11} as 
\begin{subequations} \label{8.22}
\bea
\cL^{(2)}_{\rm SG, \, L } &=& 
\frac{1}{\k} \Big\{ G^{(2)}_\rL \ln \frac{G^{(2)}_\rL }{\ri \Upsilon^{(1)}_\rL \breve\Upsilon^{(1)}_\rL } 
+{\x_\rL }\, \cV_\rL G^{(2)}_\rL \Big\}~, \\
\cL^{(2)}_{\rm SG, \, R} &=& 
\frac{1}{\k} \Big\{G^{(2)}_\rR \ln \frac{G^{(2)}_\rR }{\ri \Upsilon^{(1)}_\rR \breve\Upsilon^{(1)}_\rR } 
+{\x_\rR} \, \cV_\rR G^{(2)}_\rR \Big\}~.
\eea
\end{subequations}
where $\k$ is the gravitational coupling constant and the parameters
$\x_\rL $ and $\x_\rR$ determine a cosmological constant.
We recall that $G^{(2)}_\rL (v_\rL)$ and  $G^{(2)}_\rR (v_\rR)$
are the gauge invariant field strengths for
$\cV_\rR (v_\rR)$ and $\cV_\rL (v_\rL)$ respectively. 
The cosmological term
is described by the left and right  $BF$ terms in  \eqref{8.22}.
It is known \cite{KLT-M11}
that the action does not change if
the $BF$ coupling constants are modified as 
\bea 
\x_\rL \to \x_\rL + a~, \qquad \x_\rR \to \x_\rR - a~,
\eea
for any real constant $a$. For the action to be mirror invariant, 
we have to choose \cite{KLT-M11} 
\bea
 \x_\rL = \x_\rR \equiv \x/2~.
 \eea
 This choice will be assumed in what follows. 

With the left and right Lagrangians given by \eqref{8.22}, 
the supercurrent is
\be \k T = G_{+} - G_{-} \ .
\ee
It may be shown that the equations of motion
for $\cV_\rL$ and $\cV_\rR$ are equivalent to 
\be \bm G_{\pm}^{IJ} + \xi G_{\pm}^{IJ} = 0 \ , \label{N=4EoM}
\ee
where the composite superfields $\bm G_{\pm}^{IJ} $
are defined according to \eqref{8.13}.
Using these equations of motion, 
one can show that the supercurrent satisfies the 
conservation equation \eqref{N=4scce}.


\subsection{(2,2) anti-de Sitter supergravity}

It turns out that 
the model \eqref{8.22} describes the (2,2) AdS supergravity.
We will show this by degauging 
the equations of motion for the compensators, 
eq. \eqref{N=4EoM}, 
and the equation of motion for the Weyl multiplet, 
\be T = 0 \ . \label{AdSEOM}
\ee

Using the results of \cite{BKNT-M1} we degauge $\bm G_{\pm}^{IJ}$ to $\rm SO(4)$ superspace
\begin{align} \bm G_{\pm}^{IJ} &= \frac{\ri}{3 G_{\mp}} \cD^{\g [I} \cD_{\g K} G_{\mp}^{J] K} 
+ \frac{4}{G_{\mp}} S_K{}^{[I} G_{\mp}^{J] K} 
\mp \frac{2}{G_{\mp}} W G_{\mp}^{IJ} \non\\
&\qquad+ \frac{2 \ri}{9 G_{\mp}^3} \cD^{\a P} G_{\mp KP} \cD_{\a Q} G_\mp^{Q [I} G_\mp^{J] K} \non\\
&\qquad \pm \frac{\ri}{9 G_\mp^3} \eps^{IJRS} \cD^{\a P} G_{\mp KP} \cD_\a^Q G_{\mp QR} G_{\mp SK} \ .
\end{align}
We then use the super-Weyl transformations to impose the gauge condition
\be G_{+} = 1 \ .
\label{8.28}
\ee
Taking a spinor derivative of $G_{+}$ gives
\be G_{+ JK} \cD_\a^{[I} G_+^{JK]} - \frac{2}{3} G_+^{IK} \cD_\a^J G_{+KJ} = 0 \ .
\ee
Then using the (anti-)self-duality condition
\be \cD_\a^{[I} G_{\pm}^{JK]} = \mp \frac{1}{3} \eps^{IJKL} \cD_\a^P G_{\pm LP} \ ,
\label{8.30}
\ee
we find
\be \cD_\a^J G_{+ IJ} = 0 \quad \implies \quad \cD_\a^{[I} G_{+}^{JK]} = 0 \ .
\ee
The above tells us that $G_{+}^{IJ}$ is covariantly constant
\be \cD_\a^I G_{+}^{JK} = 0 \ .
\ee

Since the supercurrent vanishes (eq. \eqref{AdSEOM}) we have
\be G_+ = G_- = 1 \ .
\label{8.33}
\ee
Similarly we deduce that $G_{-}^{IJ}$ is covariantly constant
\be \cD_\a^I G_{-}^{JK} = 0 \ .
\ee

The covariant constancy of $G_{\pm}{}^{IJ}$ has immediate consequences on the superspace geometry. In particular, we have
\begin{align}
0 &= \{ \cD_\a^I , \cD_\b^J \} G_\pm^{KL} \non\\
&= 2 \ri \d^{IJ} \cD_{\a\b} G_\pm^{KL} - 4 \ri \eps_{\a\b} S^{K[I} G_\pm^{J] L} + 4 \ri \eps_{\a\b} S^{L[I} G_\pm^{J] K} + 8 \ri \eps_{\a\b} S^{P [I} \d^{J] [K} G_\pm{}_P{}^{L]} \non\\
&\quad - 4 \ri C_{\a\b}{}^{K(I} G_\pm^{J) L} + 4 \ri C_{\a\b}{}^{L(I} G_\pm^{J) K} + 8 \ri C_{\a\b}{}^{P(I} \d^{J) [K} G_\pm{}_P{}^{L]} \non\\
&\quad \mp 4 \ri \eps_{\a\b} W \d^{L [I} G_{\pm}^{J] K} \pm 4 \ri \eps_{\a\b} W \d^{K[I} G_{\pm}^{J] L} \ ,
\end{align}
which gives
\begin{align} S^{K [I} G_\pm^{J] L} - S^{L[I} G_\pm^{J] K} &= 2 S^{P[I} \d^{J][K} G_\pm{}_P{}^{L]} \pm W \d^{K [I} G_{\pm}^{J] L} \mp W \d^{L [I} G_{\pm}^{J] K} \ , \non\\
C_{\a\b}{}^{K(I} G_\pm^{J) L} - C_{\a\b}{}^{L(I} G_\pm^{J) K} &= 2 C_{\a\b}{}^{P(I} \d^{J) [K} G_\pm{}_P{}^{L]} \ .
\end{align}
The above leads to the constraints
\begin{subequations}
\bea 
S^{K (I} G_{\pm}{}^{J)}{}_K &=& 0 \ , \qquad S^K{}_K = \pm 2 W = 0 \ , \label{7.22}
\\
C_{\a\b}{}^{IJ} &=& 0 \ .
\eea
\end{subequations}
Eq. \eqref{inverseGpm} and eq. \eqref{7.22} tells us that $S^{IJ}$ takes the form
\be S^{IJ} = 2 S^{KL} G_{\pm}{}^I{}_K G_{\pm}{}^J{}_L \ , \quad S_K{}^K = 0 \ .
\ee

The composite $\cO(2)$ multiplets now reduce to
\be \bm G_{\pm}^{IJ} = 4 S_K{}^{[I} G_{\mp}^{J] K} \ .
\ee
Combining the above result with the equation of motion \eqref{N=4EoM} gives
\be S_K{}^I G_{\pm}^{JK} = - \frac{\xi}{4} G_{\mp}{}^{IJ} \ .
\ee
Then making use of eq. \eqref{inverseGpm} fixes the form of $S^{IJ}$ as follows
\be S^{IJ} = - \frac{\xi}{2} G_+^{K(I} G_{-}{}^{J)}{}_K \ .
\ee
Therefore $S^{IJ}$ must be covariantly constant
\be \cD_\a^I S^{JK} = 0 \ .
\ee
The above geometry corresponds to $(2 , 2)$ $\rm AdS$ superspace \cite{AdSpq}. 
To see this, we rewrite $S^{IJ}$ in the isospinor notation 
\be S^{i \bar{i},  j \bar{j}} = \hf \x
G^{ij} G^{\bar{i}\bar{j}} \ ,
\ee
where
\be G_-^{i\bar{i} , j \bar{j}} = - \eps^{\bar{i} \bar{j}} G^{ij} \  , 
\quad G_+^{i\bar{i},  j \bar{j}} = - \eps^{i j} G^{\bar{i} \bar{j}} \ .
\ee
As a result,  the algebra of 
covariant derivative  coincides 
with that for $(2,2)$ $\rm AdS$ superspace \cite{AdSpq}.\footnote{The super-Weyl gauge condition 
used in \cite{AdSpq}  was $G_+=G_-=2$, which differs from ours, eq. \eqref{8.33}.
However, this difference is  irrelevant since $G_+=G_- $ may be normalized
whichever way we like.} 

When $\xi = 0$ the covariant derivative algebra corresponds to 
that of $\cN=4$ Minkowski superspace.


\subsection{Topologically massive supergravity}

Topologically massive $\cN=4$ supergravity can be described by the action 
\bea 
S_{\rm TMSG} = \frac{1}{\widetilde \m} S_{\rm CSG} 
+ S_{\rm SG, \, L} + S_{\rm SG, \, R} ~,
\eea
where the left $S_{\rm SG, \, L} $ and right $S_{\rm SG, \, R} $  actions
correspond to the supergravity Lagrangians \eqref{8.22}.
Now, 
the supercurrent is non-zero, since  the equation of motion
for the Weyl multiplet is 
\be T + \frac{1}{\mu} W = 0 \quad \implies \quad G_+ = G_- + \frac{1}{\hat{\mu}} W \ , \label{8.44}
\ee
where $\hat{\mu} = {\mu}/{\k}$. We choose again the super-Weyl gauge condition 
\eqref{8.28},
$G_+ = 1$.
Then using the (anti-)self-duality condition \eqref{8.30} 
we find that $G_+^{IJ}$ is covariantly constant
\be \cD_\a^I G_+^{JK} = 0 \ .
\ee
Following similar reasoning as in the last subsection, we derive the constraints
\be C_{\a\b}{}^{IJ} = 0 \ , \quad S^{IJ} = 2 S^{KL} G_{+}{}^I{}_K G_{+}{}^J{}_L \ , \quad S_K{}^K = 2 W \ .
\ee

The anti-self-dual composite $\cO(2)$ multiplet, $\bm G_-^{IJ}$, becomes
\begin{align} \bm G_{-}^{IJ} &= 4 S_K{}^{[I} G_{+}^{J] K} + 2 W G_{+}^{IJ} \ .
\end{align}
Using the equation of motion 
\be \bm G_-^{IJ} + \xi G_-^{IJ} = 0
\ee
we find the form of $S^{IJ}$ to be
\be S^{IJ} = - \frac{\x}{2} G_{+}^{K(I} G_-^{J)}{}_K + \hf W \d^{IJ} \ .
\ee

Taking a spinor derivative of $G_{-}$ and using eq. \eqref{8.44} and the anti-self-duality condition \eqref{8.30}
gives
\bsubeq \label{7.38}
\begin{align} \cD_\a^J G_{-IJ} &= - \frac{3}{\hat\m G_-} G_{- IJ} \cD_\a^J W \ , \\
\cD_\a^{[I}  G_-^{JK]} &= - \frac{3}{\hat\m G_-} G_-^{[IJ} \cD_\a^{K]} W \ ,
\end{align}
\esubeq
which, due to the Bianchi identity \eqref{CompVMBI},  lead to 
\be \cD_\a^I G_{-}^{JK} = - \frac{3}{\hat\m G_-} G_-^{[IJ} \cD_\a^{K]} W + \frac{2}{\hat\m G_-} \d^{I [J} G_-^{K] L} \cD_{\a L} W \ .
\ee

Upon degauging to $\rm SO(4)$ superspace and using eq. \eqref{7.38}, we find that the composite vector multiplet $\bm G_+^{IJ}$ may be 
expressed as
\begin{align} \bm G_{+}^{IJ} 
&= \frac{\ri}{2 \hat\m^2 G_-^3} G_-^{IJ} \cD^\a_K W \cD_\a^K W  \non\\
&\quad + \frac{\ri}{4 \hat\m G_-^2} G_-^{IJ} \cD^\a_K \cD_\a^K W - \frac{\x W}{\hat{\mu}} G_{+}{}^{IJ} \non\\
&\quad - \x G_- G_{+}^{IJ} - \frac{4 W}{G_-} G_-^{IJ} \ .
\end{align}
Then the equation of motion
\be \bm G_{+}^{IJ} + \x G_+^{IJ} = 0
\ee
leads to
\begin{align} 0 &= \frac{\ri}{2 \hat\m^2 G_-^3} G_-^{IJ} \cD^\a_K W \cD_\a^K W + \frac{\ri}{4 \hat\m G_-^2} G_-^{IJ} \cD^\a_K \cD_\a^K W - \frac{4 W}{G_-} G_-^{IJ} \ .
\end{align}
It follows that the equation of motion on $W$ is
\be \cD^\a_K \cD_\a^K W + 16 \ri \hat\m G_- W + \frac{3}{\hat\m G_-} \cD^\a_K W \cD_\a^K W = 0 \ , \label{8.56}
\ee
with
\be G_- = 1 - \frac{1}{\hat\m} W \ .
\ee
This equation must be solved in conjunction with the Bianchi identity 
\eqref{2.39}. 

For $\x =0$, a solution of the equations of motion for topologically massive supergravity 
is obtained by setting $W =0$ in the above relations. 
This solution describes a flat superspace. 
Linearizing the equation \eqref{8.56} and the Bianchi identity \eqref{2.39}
 around Minkowski superspace, it may be shown that 
$W$ obeys the Klein-Gordon equation
\be 
(\Box -m^2) W
= 0 \ , \qquad m=2\hat \m = 2 \frac{\m}{\k}
\ ,
\ee
with $\Box := \partial^a \partial_a$.


\section{Concluding comments} \label{conclusion}

In this paper we have worked out the linear multiplet action principles 
in $\cN=3$ and $\cN=4$ conformal supergravities. At the component level, 
the $\cN=3$ action is  given by eq. \eqref{7.8}, while the $\cN=4$ action is a sum
of the left and right sectors, given by eqs. \eqref{N=4LMA-left} and \eqref{N=4LMA-right}
respectively. 
Using these locally supersymmetric actions, it is not difficult to construct 
the component off-shell actions for the (2,1) and (2,2) AdS supergravities
and their topologically massive extensions. 
For instance, the composite 
$\cO (2)$ multiplet, which has to be used in the action \eqref{7.8}
in order to describe the  (2,1) AdS supergravity,
proves to be ${\bm G}^I +\hf \x G^I$, where ${\bm G}^I$ is given by eq. \eqref{7.11}. The derivation of 
the component actions will be given elsewhere.

In superspace, the  off-shell formulations for 
(2,1) and (2,2) AdS supergravities were given in \cite{KLT-M11}. 
The specific feature of (2,1) AdS supergravity is that its conformal compensator 
is a vector multiplet that can be described in terms of the tropical prepotential 
$\cV (v)$.\footnote{If $\x=0$, the vector multiplet can be dualized into 
a weight-1/2 polar hypermultiplet \cite{KLT-M11}.}
The specific feature of (2,2) AdS supergravity is that its conformal compensators
are two vector multiplets that can be described using the left  and right  tropical 
prepotentials, $\cV_\rL (v_\rL) $ and $\cV_\rR (v_\rR) $.\footnote{One of the vector 
multiplets can be dualized into a weight-1/2 polar hypermultiplet \cite{KLT-M11}.}
As concerns the (3,0), (3,1) and (4,0)  AdS supergravity theories, 
the structure of the corresponding conformal compensators is not yet known, 
which is an interesting open problem. 

Our procedure of constructing composite $\cO(2)$ multiplets can be used to generate
higher derivative couplings for vector multiplets, similar to the known results 
in 4D $ \cN=2$ supersymmetry \cite{BK11,BKT}.\footnote{For other constructions 
of higher derivative 4D $\cN=2$ supersymmetric invariants, 
see \cite{deWit:2010za,KM13} and references therein.}
To illustrate the idea, let us fix $\cN=3$ and consider 
the composite $\cO(2)$ multiplet defined by  \eqref{7.11}.
Choosing in \eqref{C.9} a composite prepotential of the form
\bea
\cV (v) \quad \longrightarrow \quad   \left[  \frac{ \bm G^{(2)} (v)}{G^{(2)} (v)}\right]^n~, 
\qquad n = 1,2, \dots
\eea
leads to a family of composite real $\cO (2) $ multiplets
\bea
{\bm H}_n{}^{(2)}(v) := {\bm H}_n{}^{ij} v_iv_j
=\D^{(4)}
\oint_{\hat \g} \frac{({\hat v},\rd {\hat v})}{ 2\pi (v, {\hat v} )^2} 
\left[  \frac{ \bm G^{(2)} (\hat v)}{G^{(2)} (\hat v)}\right]^n
~, \qquad
\nabla^{(2)}_\a {\bm H}_{n}{}^{(2)} =0~.
\eea
Here the contour integral can be computed using the technique of \cite{BK11}. 
Now, we have two types of composite $\cO(2)$ multiplets, 
${\bm G}^{(2)}(v) $ and ${\bm H}_n{}^{(2)}(v) $, which differ by the number of spinor derivatives involved. 
Both of them can be used to generate new composite $\cO(2)$ multiplets
\bea
\D^{(4)}
\oint_{\hat \g} \frac{({\hat v},\rd {\hat v})}{ 2\pi (v, {\hat v} )^2} 
\left[  \frac{ \bm G^{(2)} (\hat v)}{G^{(2)} (\hat v)}\right]^p
\left[  \frac{ {\bm H}_n{}^{(2)} (\hat v)}{G^{(2)} (\hat v)}\right]^q~,
\eea
with $p$ and $q$ non-negative integers. 
The above composite $\cO(2)$ multiplets are expected to appear in low-energy 
effective actions for quantum $\cN=3$ supersymmetric gauge theories. 
In general, such an affective action is given by \eqref{InvarAc}
with a Lagrangian of the form
\bea
\cL^{(2)} =  G^{(2)} \cL \Big( G^{(2)} , \bm G^{(2)} ,\bm H_n{}^{(2)} , \dots \Big)~,
\label{9.4}
\eea
where $\cL $ is a homogeneous function of degree zero. 

In the $\cN=4$ case, we need two vector multiplets, $G_\rL^{(2)}(v_\rL)$ and 
$G_\rR^{(2)}(v_\rR)$, in order to generate higher derivative composite 
$\cO(2)$ multiplets. 

In the rigid supersymmetric case, Zupnik has derived, 
building on the earlier work by Howe and Leeming \cite{HL},
harmonic superspace formulations for the $\cN=5$ vector multiplet 
and corresponding Chern-Simons actions \cite{Z2007,Z2008}. 
In this setting, the off-shell  vector multiplet involves an infinite number of 
bosonic and fermionic degrees of freedom, which makes possible 
the construction of Chern-Simons actions. 

It is known that the harmonic superspace approach is the most elaborated 
scheme to do supergraph calculations in off-shell theories with six and eight supercharges. It would be interesting to see how 
quantum corrections of the type \eqref{9.4} are generated 
within the background field formulation for quantum 3D $\cN=3$ 
super Yang-Mills theories \cite{Buchbinder:2009dc}. 
\\

\noindent
{\bf Acknowledgements:}\\
We are grateful to Daniel Butter for comments on the manuscript. 
This work was supported in part by the Australian Research Council,
projects DP1096372 and DP140103925. 
The work of JN was also supported in part by the Australian 
Research Council's Discovery Early Career 
Award (DECRA), project  DE120101498.


\appendix


\section{Geometry of $\cN$-extended conformal superspace} \label{geometry}

Here we collect the essential details of the $\cN$-extended 
superspace geometry of \cite{BKNT-M1}. 
We refer the reader to \cite{KLT-M11,BKNT-M1} for our conventions for 3D spinors.

We begin with a curved three-dimensional $\cN$-extended superspace
 $\cM^{3|2 \cN}$ parametrized by
local bosonic $(x^m)$ and fermionic coordinates $(\theta^\m_I)$:
\be z^M = (x^m, \ \q^\mu_I) \ ,
\ee
where $m = 0, 1, 2$, $\mu = 1, 2$ and $I = 1, \cdots , \cN$. The 
structure group is chosen to be ${\rm OSp}(\cN|4, {\mathbb R})$ and 
the covariant derivatives are postulated to have the form
\be
\nabla_A = E_A{}^M \pa_M - \o_A{}^{\underline b} X_{\underline b} 
= E_A{}^M \pa_M - \hf \Omega_A{}^{bc} M_{bc} - \hf \Phi_A{}^{PQ} N_{PQ} - B_A \mathbb D - \mathfrak{F}_A{}^B K_B \ .
\ee
Here $E_A = E_A{}^M \partial_M$ is the inverse vielbein, 
$M_{ab}$ are the Lorentz generators, $N_{IJ}$ are generators of the 
$\rm SO(\cN)$ group, $\mathbb D$ is the dilatation generator and $K_A = (K_a , S_\a^I)$ are the special superconformal 
generators.\footnote{As usual, we refer to $K_a$ as the special conformal generator and $S_\a^I$ as the $S$-supersymmetry 
generator.}

The Lorentz generators obey
\bsubeq \label{SCA}
\begin{align}
[M_{ab} , M_{cd}] &= 2 \eta_{c[a} M_{b] d} - 2 \eta_{d [a} M_{b] c} \ , \\
[M_{ab} , \nabla_c ] &= 2 \eta_{c [a} \nabla_{b]} \ , \quad [M_{\a\b} , \nabla_\g^I] = \eps_{\g(\a} \nabla_{\b)}^I \ .
\end{align}
The $\rm SO(\cN)$ and dilatation generators obey
\begin{align}
[N_{KL} , N^{IJ}] &= 2 \d^I_{[K} N_{L]}{}^J - 2 \d^J_{[K} N_{L]}{}^I \ , \quad [N_{KL} , \nabla_\a^I] = 2 \d^I_{[K} \nabla_{\a L]} \ ,  \\
[\mathbb D , \nabla_a] &= \nabla_a \ , \quad [\mathbb D , \nabla_\a^I] = \hf \nabla_\a^I \ .
\end{align}
The Lorentz and $\rm SO(\cN)$ 
generators
act
on the special conformal generators $K_A$ as
\begin{align}
[M_{ab} , K_c] &= 2 \eta_{c[a} K_{b]} \ , \quad [M_{\a\b} , S_\g^I] = \eps_{\g(\a} S_{\b)}^I \ , \\
[N_{KL} , S_\a^I] &= 2 \d^I_{[K} S_{\a L]} \ ,
\end{align}
while the dilatation generator acts on  $K_A$ as
\begin{align}
[\mathbb D , K_a] = - K_a \ , \quad [\mathbb D, S_\a^I] &= - \hf S_\a^I \ .
\end{align}
Among themselves, the generators $K_A$ obey the algebra
\begin{align}
\{ S_\a^I , S_\b^J \} = 2 \ri \d^{IJ} (\g^c)_{\a\b} K_c \ .
\end{align}
Finally, the algebra of $K_A$ with $\nabla_A$ is given by
\begin{align}
[K_a , \nabla_b] &= 2 \eta_{ab} \mathbb D + 2 M_{ab} \ , \\
[K_a , \nabla_\a^I ] &= - \ri (\g_a)_\a{}^\b S_\b^I \ , \\
[S_\a^I , \nabla_a] &= \ri (\g_a)_\a{}^\b \nabla_{\b}^I \ , \\
\{ S_\a^I , \nabla_\b^J \} &= 2 \eps_{\a\b} \d^{IJ} \mathbb D - 2 \d^{IJ} M_{\a \b} - 2 \eps_{\a \b} N^{IJ} \ .
\end{align}
\esubeq

The covariant derivatives obey the (anti-)commutation relations of the form
\begin{align}
[ \nabla_A , \nabla_B \}
	&= -T_{AB}{}^C \nabla_C
	- \frac{1}{2} \RM_{AB}{}^{cd} M_{cd}
	- \frac{1}{2} \RN_{AB}{}^{PQ} N_{PQ}
	\non \\ & \quad
	- \RD_{AB} \mathbb D
	- \RS_{AB}{}^\g_K S_\g^K
	- \RK_{AB}{}^c K_c~, \label{nablanabla}
\end{align}
where $T_{AB}{}^C$ is the torsion, and $\RM_{AB}{}^{cd}$, $\RN_{AB}{}^{PQ}$, $\RD_{AB}$, $\RS_{AB}{}^\g_K$ and $\RK_{AB}{}^c$ 
are the curvatures corresponding to the Lorentz, $\rm SO(\cN)$, dilatation, $S$-supersymmetry and special conformal boosts respectively.

The full gauge group  of conformal supergravity, $\cG$, 
is generated by 
{\it covariant general coordinate transformations}, 
$\delta_{\rm cgct}$, associated with a parameter $\xi^A$ and 
{\it standard superconformal transformations}, 
$\delta_{\cH}$, associated with a parameter $\L^{\ul a}$. 
The latter include
the dilatation,
Lorentz, 
$\rm SO(\cN)$, 
and special conformal
(bosonic and fermionic) transformations.
The covariant derivatives transform as
\bea 
\d_\cG \nabla_A &=& [\cK , \nabla_A] \ ,
\label{TransCD}
\eea
where $\cK$ denotes the first-order differential operator
\bea
\cK = \xi^C \nabla_C + \hf \L^{ab} M_{ab} + \hf \L^{IJ} N_{IJ} + \L \mathbb D + \L^A K_A ~.
\eea
Covariant (or tensor) superfields transform as
\bea 
\d_{\cG} T &=& \cK T~.
\eea

In order to describe the Weyl multiplet of conformal supergravity, 
some of the components of the torsion and curvatures must be constrained. Following \cite{BKNT-M1}, the 
spinor derivative torsion and curvatures are chosen to resemble super-Yang Mills
\be \{ \nabla_\a^I , \nabla_\b^J \} =  -2 \ri \ve_{\a\b} W^{IJ} \ ,
\ee
where $W^{IJ}$ is some operator that takes values 
in the superconformal algebra, with $P_A$ replaced by $\nabla_A$. In 
\cite{BKNT-M1} it was shown how to constrain $W^{IJ}$ entirely in terms of the super Cotton tensor 
(for each value of $\cN$). Remarkably, for all $\cN$ the torsion tensor takes its constant flat space value, 
while the Lorentz and dilatation
curvatures always vanish:
\begin{subequations}
\begin{align}
T^a &= -\ri \, E^\beta \wedge E^\gamma (\gamma^a)_{\gamma \beta}~, \quad T^\alpha_I = 0 \ ,
\\
\RM^{ab} &= 0~, \qquad \RD = 0  ~. \label{RMDConstraint}
\end{align}
\end{subequations}
We now summarize the resulting covariant derivative algebra for all values of $\cN$.


\subsection{The $\cN = 1$ case}

The $\cN=1$ super Cotton tensor $W_{\a\b\g}$ is a symmetric primary superfield of dimension-$5/2$
\be S_\d W_{\a\b\g} = 0 \ , \quad K_d W_{\a\b\g}=0\ , \quad 
\mathbb D W_{\a\b\g} = \frac{5}{2} W_{\a\b\g} ~.
\ee
The algebra of covariant derivatives is given by
\begin{subequations}
\begin{align} \{ \nabla_\a , \nabla_\b \} &= 2 \ri \nabla_{\a\b} \ , \\
[ \nabla_a , \nabla_\a ] &= \frac{1}{4} (\g_a)_\a{}^\b W_{\b \g\d} K^{\g\d} \ , \\
[\nabla_a , \nabla_b] &= - \frac{\ri}{8} \eps_{abc} (\g^c)^{\a\b} \nabla_\a W_{\b\g\d} K^{\g\d} - \frac{1}{4} \eps_{abc} (\g^c)^{\a\b} W_{\a\b\g} S^\g \ ,
\end{align}
\end{subequations}
The Bianchi identities imply an additional constraint on $W_{\a\b\g}$, the vanishing of its spinor divergence,
\be \nabla^\a W_{\a \b\g} = 0 \ .
\ee


\subsection{The $\cN = 2 $ case}

The $\cN = 2$ super Cotton tensor $W_{\a\b}$ is a symmetric primary superfield of dimension-$2$
\be S_\g^I W_{\a\b} = 0 \ , \quad K_c W_{\a\b} =0\ , \quad
\mathbb D W_{\a\b} = 2 W_{\a\b} ~.
\ee
As in the $\cN=1$ case, its spinor divergence vanishes,
\be \nabla^{\a I} W_{\a\b} = 0 \ .
\ee
The algebra of covariant derivatives is 
\begin{subequations} \label{N=2Algebra}
\begin{align} \{ \nabla_\a^I , \nabla_\b^J \} &= 2 \ri \d^{IJ} \nabla_{\a\b} - \ri \eps^{IJ} \eps_{\a\b} W_{\g\d} K^{\g\d} \ , \\
[ \nabla_a , \nabla_\b^J ] &= \hf (\g_a)_\b{}^\g \eps^{JK} \nabla_{\g K} W^{\a\d} K_{\a\d} + \ri (\g_a)_{\b\g} \eps^{JK} W^{\g\d} S_{\d K} \ , \\
[\nabla_a , \nabla_b] &= - \frac{\ri}{8} \eps_{abc} (\g^c)^{\g\d} \Big( \eps^{KL} (\nabla_{\g K} \nabla_{\d L} W_{\a\b} K^{\a\b} + 4 \ri \nabla_{\g K} W_{\d\b} S^\b_L) - 8 W_{\g\d} \cJ \Big) \ , 
\end{align}
\end{subequations}
where the $\rm U(1)$ generator $\cJ$ obeys
\be N_{KL} = \ri \eps_{KL} \cJ \ , \quad \cJ = - \frac{\ri}{2} \eps^{KL} N_{KL} ~, \qquad
[\cJ , \nabla_\a^I] = - \ri \eps^{IJ} \nabla_{\a J} \ .
\ee


\subsection{The $\cN = 3$ case}

The $\cN = 3$ super Cotton tensor $W_{\a}$ is a primary superfield of dimension-$3/2$,
\be S_\b^I W_{\a} = 0 \ , \quad K_b W_\a =0 \ , \quad
\mathbb D W_{\a} = \frac{3}{2} W_{\a}\ , 
\ee
with vanishing spinor divergence,
\be
\nabla^{\alpha I} W_\alpha = 0~.
\ee
The algebra of covariant derivatives is
\begin{subequations} \label{N=3Algebra}
\begin{align} 
\{ \nabla_\a^I , \nabla_\b^J \} &= 2 \ri \d^{IJ} \nabla_{\a\b} 
- 2 \eps_{\a\b} \eps^{IJL} W^\g S_{\g L} + \ri \eps_{\a\b} (\g^c)^{\g\d} \eps^{IJK} (\nabla_{\g K} W_\d) K_c \ , \label{A.18a} \\
 [\nabla_a , \nabla_\b^J ] &= \ri \eps^{JKL} (\g_a)_{\b\g} W^\g N_{KL} + \ri \eps^{JKL} (\g_a)_{\b\g} (\nabla^\g_K W^\d) S_{\d L} \non\\
 &\qquad + \frac{1}{4} \eps^{JKL} (\g_a)_{\b\g} (\g^c)_{\d\rho} (\nabla^\g_K \nabla^\d_L W^\rho) K_c \ , \\
[\nabla_a , \nabla_b] &= \eps_{abc} (\g^c)_{\a\b} \Big[ - \hf \eps^{IJK} (\nabla^\a_I W^\b) N_{JK} - \frac{1}{4} \eps^{IJK} (\nabla^\a_I \nabla^\b_J W^\g ) S_{\g K} \non\\
&\qquad +\frac{\ri}{24} \eps^{IJK} (\g^d)_{\g\d} (\nabla^\a_I \nabla^\b_J \nabla^\g_K W^\d) K_d \Big] \ .
\end{align}
\end{subequations}

In order to define a large class of matter multiplets coupled to supergravity, 
it is often useful to  switch to an isospinor notation 
using the isomorphism ${\rm SO}(3) \cong {\rm SU}(2)/{\mathbb Z}_2$.
As usual, this is achieved by replacing 
any SO(3) vector  index by a  symmetric pair of 
SU$(2)$ spinor indices, $\nabla^I_\a \to \nabla^{ij}_\a = \nabla^{ji}_\a$.
The details of this correspondence are available in \cite{KLT-M11}.
Here we only give the results essential for our discussion. 
Converting the indices of the SO(3) generator $N_{KL}$ into isospinor 
indices gives
\be N_{ij , kl} = \hf \eps_{jl} \cJ_{ik} + \hf \eps_{ik} \cJ_{jl} \ , \quad 
N^{ij , kl} = - \hf \eps^{jl} \cJ^{ik} - \hf \eps^{ik} \cJ^{jl} \ ,
\ee
where the SU(2) generator $\cJ^{kl}$ 
acts on the spinor covariant derivatives as
\be [\cJ^{kl} , \nabla_\a^{ij}] = \eps^{i (k} \nabla_\a^{l) j} + \eps^{j(k} \nabla_\a^{l)i} \ .
\ee
Eq. \eqref{A.18a} turns into
\begin{align} \{ \nabla_\a^{ij} , \nabla_\b^{kl} \} &= - 2 \ri \eps^{i(k} \eps^{l) j} \nabla_{\a\b} + \eps_{\a\b} \eps^{jl} W^\g S_{\g}{}^{ik} + \eps_{\a\b} \eps^{ik} W^\g S_{\g}{}^{jl} \non\\
&- \frac{\ri}{2} \eps_{\a\b} \eps^{jl} (\g^c)^{\g\d} (\nabla_{\g}{}^{ik} W_\d) K_c - \frac{\ri}{2} \eps_{\a\b} \eps^{ik} (\g^c)^{\g\d} (\nabla_{\g}{}^{jl} W_\d) K_c \ .
\label{A.21}
\end{align}
We also have
\begin{align}
\{ S_\a^{ij} , \nabla_\b^{kl} \}  = - 2 \eps_{\a\b} \eps^{i(k} \eps^{l) j} \mathbb D + 2 \eps^{i(k} \eps^{l) j} M_{\a\b} + \eps_{\a\b} \eps^{jl} \cJ^{ik} + \eps_{\a\b} \eps^{ik} \cJ^{jl} \ .
\label{A.22}
\end{align}


\subsection{The $\cN > 3$ case}

For all values of $\cN>3$, we introduce the super Cotton tensor $W^{IJKL}$,
which is a totally antisymmetric primary superfield of dimension-1,
\be
S_\alpha^P W^{IJKL} = 0~, \quad K_a W^{IJKL} =0 \ , \quad
\mathbb D W^{IJKL} = W^{IJKL} \ .
\ee
The algebra of covariant derivatives is\footnote{The algebra for $\cN \leq 3$ can be deduced 
from that for $\cN > 3$ \cite{BKNT-M1}.}
\begin{subequations} \label{covDN>3}
\begin{align} 
\{ \nabla_\a^I , \nabla_\b^J \} &= 2 \ri \d^{IJ} \nabla_{\a\b} + \ri \eps_{\a\b} W^{IJKL} N_{KL} - \frac{\ri}{\cN - 3} \eps_{\a\b} (\nabla^\g_K W^{IJKL}) S_{\g L} \non\\
&\qquad + \frac{1}{2 (\cN - 2)(\cN - 3)} \eps_{\a\b} (\g^c)^{\g\d} (\nabla_{\g K} \nabla_{\d L} W^{IJKL}) K_c \ , \\
[\nabla_a , \nabla_\b^J ] &= \frac{1}{2 (\cN - 3)} (\g_a)_{\b\g} (\nabla^\g_K W^{JPQK}) N_{PQ} \non\\
&\qquad - \frac{1}{2 (\cN - 2) (\cN - 3)} (\g_a)_{\b\g} (\nabla^\g_L \nabla^\d_P W^{JKLP}) S_{\d K} \non\\
&\qquad - \frac{\ri}{4 (\cN - 1) (\cN - 2)(\cN - 3)} (\g_a)_{\b\g} (\g^c)_{\d\rho} (\nabla^\g_K \nabla^\d_L \nabla^\rho_P W^{JKLP}) K_c \ , \\
[\nabla_a , \nabla_b] &
=    \frac{1}{4 (\cN - 2) (\cN - 3)}   \eps_{abc} (\g^c)_{\a\b} 
\Big( \ri (\nabla^\a_I \nabla^\b_J W^{PQIJ}) N_{PQ} \non\\
&\qquad + \frac{\ri}{ \cN - 1} (\nabla^\a_I \nabla^\b_J \nabla^\g_K W^{LIJK}) S_{\g L} \non\\
&\qquad + \frac{1}{2 \cN (\cN - 1) } (\g^d)_{\g\d} (\nabla^\a_I \nabla^\b_J \nabla^\g_K \nabla^\d_L W^{IJKL}) K_d \Big) \ ,
\end{align}
\end{subequations}
where $W^{IJKL}$ satisfies the Bianchi identity
\be \nabla_{\a}^I W^{JKLP} = \nabla_\a^{[I} W^{JKLP]} - \frac{4}{\cN - 3} \nabla_{\a Q} W^{Q [JKL} \d^{P] I} \ . \label{CSBIN>3}
\ee

For $\cN=4$, the equation \eqref{CSBIN>3} is trivially satisfied,
and instead  a fundamental Bianchi identity occurs at  dimension-2. 
Rewriting the super Cotton tensor as a scalar superfield,
as in  \eqref{5.7}, 
the Bianchi identity is
\bea
\nabla^{\a I}\nabla_{\a}^JW=\frac{1}{4}\d^{IJ}\nabla^{\a}_P\nabla_{\a}^PW~. \label{2.39}
\eea

We now we turn to a discussion of special features of the $\cN = 4$ case.

\subsection{The $\cN=4$ case}
For each $\rm SO(4)$ vector $V_I$ we can associate a second-rank isospinor $V_{i\bar{i}}$
\be V_I \leftrightarrow V_{i \bar{i}} := (\t^I)_{i \bar{i}} V_{i \bar{i}} \ , \quad (V_{i \bar{i}})^* = V^{i \bar{i}} \ .
\ee

The original $\rm SO(4)$ connection turns into a sum of two $\rm SU(2)$ connections
\be \Phi_A = (\Phi_{\rm L})_A + (\Phi_{\rm R})_A \ , \quad (\Phi_{\rm L})_A = \Phi_A{}^{kl} \bm {\rm L}_{kl} \ , \quad (\Phi_{\rm R})_A = \Phi_A{}^{\bar{k} \bar{l}} \bm {\rm R}_{kl} \ .
\ee
Note that
\be N_{KL} \rightarrow N_{k \bar{k},  l \bar{l}} = \eps_{\bar{k} \bar{l}} \bm {\rm L}_{kl} + \eps_{kl} \bm {\rm R}_{\bar{k} \bar{l}} \ .
\ee

The left and right operators act on the covariant derivatives as
\be [ \bm {\rm L}^{kl} ,  \nabla_\a^{i\bar{i}}] = \eps^{i (k} \nabla_\a^{l) \bar{i}} \ , \quad [\bm {\rm R}^{kl} , \nabla_\a^{i\bar{i}} ] = \eps^{\bar{i}( \bar{k}} \nabla_\a^{i \bar{l})} \ .
\ee

In the isospinor notation, 
the Bianchi identity on $W$ becomes
\be \nabla^{\a i \bar{i}} \nabla_\a^{j \bar{j}} W
= \frac{1}{4} \eps^{ij} \eps^{\bar{i} \bar{j}} 
\nabla^\a_{k \bar{k}} \nabla_\a^{k \bar{k}} W \ .
\ee

The algebra of spinor covariant derivatives becomes
\begin{align}
\{ \nabla_\a^{i \bar{i}} , \nabla_\b^{j \bar{j}}\} &= 2 \ri \eps^{ij} \eps^{\bar{i} \bar{j}} \nabla_{\a\b} 
+ 2 \ri \eps_{\a\b} \eps^{\bar{i} \bar{j}} W \bm {\rm L}^{ij} - 2 \ri \eps_{\a\b} \eps^{ij} W \bm {\rm R}^{\bar{i} \bar{j}} \non\\
&\quad- \ri \eps_{\a\b} \eps^{ij} \nabla^\g{}_k{}^{\bar{i}} W S_\g^{k \bar{j}} + \ri \eps_{\a\b} \eps^{\bar{i} \bar{j}} \nabla^\g{}^i{}_{\bar{k}} W S_\g^{j \bar{k}} \non\\
&\quad + \frac{1}{4} \eps_{\a\b} \Big( \eps^{ij} \nabla_\g{}_k{}^{\bar{i}} \nabla_\d^{k \bar{j}} W 
- \eps^{\bar{i} \bar{j}} \nabla_\g{}^j{}_{\bar{k}} \nabla_\d^{i \bar{k}} W \Big) K^{\g\d} \ .
\label{A.33}
\end{align}

Note that
\be \{ S_\a^{i \bar{i}} , \nabla_\b^{j \bar{j}} \} = 2 \eps_{\a\b} \eps^{ij} \eps^{\bar{i}\bar{j}} \mathbb{D} - 2 \eps^{ij} \eps^{\bar{i}\bar{j}} M_{\a\b} 
+ 2 \eps_{\a\b} \eps^{\bar{i} \bar{j}} L^{ij} + 2 \eps_{\a\b} \eps^{ij} R^{\bar{i} \bar{j}} \ .
\ee


\section{Supersymmetry transformations} \label{SUSYTrans}

In this appendix we give the supersymmetry transformations of the component fields for 
vector multiplets with $\cN < 5$. For the supersymmetry transformations of the Weyl multiplet we refer the reader to \cite{BKNT-M2}. In general 
there are additional auxiliary fields coming from the super Cotton tensor $W^{IJKL}$. 
These are defined for $\cN > 3$ as follows \cite{BKNT-M2}
\bsubeq \label{Wcomps}
\begin{align}
w_{IJKL} &:= W_{IJKL}| \ , \\
w_\a{}^{IJK} &:= - \frac{\ri}{2 (\cN - 3)} \nabla_{\a L} W^{IJKL}| \ , \\
y^{I J K L} &:= \frac{\ri}{\cN - 3} \nabla^{\g [I}  \nabla_{\g P} W^{JKL]P}| \ , \\
X_{\a_1 \cdots \a_n}{}^{I_1 \cdots I_{n+4}} &:= I(n) \nabla_{(\a_1}^{[I_1} \cdots \nabla_{\a_n)}^{I_n} W^{I_{n+1} \cdots I_{n+4}]}| \ , \label{eq:defHigherX}
\end{align}
\esubeq
where $I(n)$ is defined by eq. \eqref{Ifunct}. Expressions involving the component fields for lower values of $\cN$ may be derived via the truncation procedure 
given in \cite{BKNT-M2}.

One readily finds the $Q$-supersymmetry and $S$-supersymmetry transformations of $v_m$ to be
\bsubeq
\begin{align}
\d_Q (\xi) v_m &= - \hf \xi^\g_K e_m{}^b (\g_b)_\g{}^\d \l_\d^K - \ri \psi_m{}^\b_J \xi_{\b K} g^{JK} \ , \\
\d_S (\eta) v_m &= 0 \ .
\end{align}
\esubeq
The $S$-supersymmetry transformations of the non-gauge fields are
\bsubeq
\begin{align} \d_S(\eta) g^{IJ} &= 0 \ , \\
\d_S(\eta) \l_\a^I &= 4 \eta_{\a J} g^{JI} \ , \\
\d_S(\eta) h^{IJ} &= -2 \ri \eta^{\a [I} \l_\a^{J]} - 2 \eta^\a_K \chi_\a{}^{IJK} \ , \\
\d_S(\eta) \chi_\a{}^{IJK} &= - 6 \ri \eta_\a^{[I} g^{JK]} \ .
\end{align}
\esubeq
Their $Q$-supersymmetry transformations are
\bsubeq
\begin{align} \d_Q(\xi) g^{IJ} &= - \ri \xi^\g_K \chi_\g{}^{IJK} - \xi^{\g [I} \l_\g^{J]} \ , \\
\d_Q(\xi) \l_\a^I &= \ri \xi_{\a J} h^{JI} - 2 \ri \xi^\b_J \hat{\bm \nabla}_{\b\a} g^{JI} + 2 \ri \xi^{\b I} \hat{F}_{\b\a} \ , \\
\d_Q(\xi) h^{IJ} &= 2 \ri \xi^\a_K \hat{\bm \nabla}_\a{}^\g \chi_\g{}^{IJK} - 2 \xi^{\a [I} \hat{\bm \nabla}_\a{}^\g \l_\g^{J]} - 8 \ri \xi^\a_K w_\a{}^{IJL} g_L{}^K \non\\
&\quad- 8 \ri \xi^\a_K w_\a{}^{PK[I} g_P{}^{J]} + 2 \xi^\a_K w^{IJKL} \l_{\a L} + 4 \ri \xi^\a_K [g^{K[I} , \l_\a^{J]}] \non\\
&\quad+ \ri \xi^\a_K [g^{IJ} , \l_\a^K ] \ , \\
\d_Q(\xi) \chi_\a{}^{IJK} &= \xi^\b_L \chi_{\a\b}{}^{LIJK} - \frac{3}{2} \xi_{\a}^{[I} h^{JK]} - 3 \xi^{\b [I} \hat{\bm \nabla}_{\b\a} g^{JK]} + 6 \xi_{\a L} w^{PL[IJ} g^{K]}{}_P \non\\
&\quad+ 3 \ri \xi_{\a L} [g^{L [I} , g^{JK]}] \ , 
\end{align}
\esubeq
where we have made use of the covariant derivative
\be
\bm \cD_a = e_a{}^m \bm \cD_m = e_a{}^m (\partial_m - \hf \omega_m{}^{bc} M_{bc} - \hf V_m{}^{IJ} N_{IJ} - b_m \mathbb D - \ri v_m)
\ee
and defined\footnote{The component $S$-supersymmetry connection is defined as in \cite{BKNT-M2}, $\phi_a{}_\b^J := e_a{}^m \phi_m{}_\b^J$.}
\bsubeq
\begin{align}
\hat{\bm \nabla}_a g^{IJ} &:= \bm \cD_a g^{IJ} + \frac{\ri}{2} \psi_a{}^\b_K \chi_\b{}^{IJK} + \hf \psi_a{}^{\b[I} \l_\b^{J]} \ , \\
\hat{\bm \nabla}_a \l_\a^I &:= \bm \cD_a \l_\a^I - \frac{\ri}{2} \psi_{a \a J} h^{JI} + \ri \psi_a{}^\b_J \hat{\bm \nabla}_{\a\b} g^{JI} - \ri \psi_a{}^{\b I} \hat{F}_{\a\b} 
- 2 \phi_{a \a J} g^{JI} \ , \\
\hat{\bm \nabla}_a \chi_\a{}^{IJK} &:= \bm \cD_a \chi_\a{}^{IJK} + \hf \psi_a{}^\b_L \chi_{\b\a}{}^{IJKL} + \frac{3}{4} \psi_a{}_\a^{[I} h^{JK]}
+ \frac{3}{2} \psi_a{}^{\b [I} \hat{\bm \nabla}_{\b\a} g^{JK]} \non\\
&\qquad- 3 \psi_a{}_{\a L} w^{PL[IJ} g^{K]}{}_P - \frac{3 \ri}{2} \psi_{\a L} [g^{L[I} , g^{JK]}] + 3 \ri \phi_a{}_\a^{[I} g^{JK]} \ .
\end{align}
\esubeq

In the above we have derived the supersymmetry transformations of the component fields for general $\cN$. However, we are still missing the 
supersymmetry transformations of
\be
\chi_{\a_1 \cdots \a_n}{}^{I_1 \cdots I_{n+2}} \ , \quad n > 1 \ .
\ee
These fields only appear for $\cN > 3$, while for $\cN = 4$ $\chi_{\a\b}{}^{IJKL}$ is composite once one imposes the (anti-)self-dual condition \eqref{N=4duality}, 
see eq. \eqref{compDuality}. Keeping in mind the definition of the component fields, eqs. \eqref{N=3Comps} and \eqref{N=4comps}, and the 
truncation procedure, we see that all the supersymmetry transformations for $\cN < 5$ are specified.


\section{Action principle in $\cN=3$ supergravity}
\label{AppendixC} 

As demonstrated in \cite{KLT-M11}, general off-shell $\cN=3$ 
supergravity-matter systems
are naturally formulated in curved $\cN=3$ projective superspace 
$\cM^{3|6} \times {\mathbb C}P^1$ in terms of covariant projective multiplets. 
These multiplets were defined in \cite{KLT-M11} 
in SO(3) superspace. Here we briefly extend those definitions to $\cN=3$ conformal 
superspace. 

Let $v^i \in {\mathbb C}^2 \setminus  \{0\}$
be homogeneous coordinates for ${\mathbb C}P^1$.
We use these variables to define a subset of spinor covariant derivatives
\bea
\nabla^{(2)}_\a := v_i v_j \nabla^{ij}_\a~.
\eea
It follows from \eqref{A.21} that 
the operators $\nabla_\a^{(2)}$ anticommute with each other, 
\bea 
\{ \nabla_\a^{(2)} , \nabla_\b^{(2)} \} = 0 \ .
\eea
This property allows us to define a family of constrained superfields. 

By definition, a covariant projective multiplet of weight $n$, $Q^{(n)} (z,v)$, 
is a Lorentz-scalar superfield on $\cM^{3|6}$
that is holomorphic 
on an open domain of $ {\mathbb C}^2 \setminus  \{0\}$ with respect to $v^i$, 
 and is characterized by the following properties: 
\begin{enumerate}
\item it obeys the analyticity constraint 
\bea
\nabla_\a^{(2)} Q^{(n)} =0~;
\label{C.3}
\eea
\item it is a homogeneous   function 
of $v^i$ of degree $n$, 
\bea
Q^{(n)}(c\,v)\,=\,c^n\,Q^{(n)}(v)~, \qquad c \in   {\mathbb C} \setminus  \{0\}~;
\eea
\item its SU(2) transformation is
\bea
\d_\L Q^{(n)} &=& \hf \L^{ij} \cJ_{ij} Q^{(n)} \ , 
\qquad 
\L^{ij} \cJ_{ij} Q^{(n)} = - (\L^{(2)} \bm \partial^{(-2)} - n \L^{(0)}) Q^{(n)} \ .~~~
\eea
\end{enumerate} 
Here we have defined
\bea \L^{(2)} := \L^{ij} v_i v_j \ , \quad \L^{(0)} := \frac{v_i u_j}{(v , u)} \L^{ij}~, 
\quad (v , u) := v^i u_i \ 
\eea
and introduced the differential operator
\be \bm \partial^{(-2)} := \frac{1}{(v,u)} u^i \frac{\partial}{\partial v^i} \ .
\ee
These relations involve a fixed isospinor  $u_{i}$ which is subject to
the condition $(v,u)\ne0$, but otherwise completely arbitrary.
For the covariant projective multiplet, one can define the operation of 
{\it smile conjugation} which takes $Q^{(n)} (v)$ to its 
smile-conjugate $\breve{Q}^{(n)} (v)$, which is also a covariant weight-$n$ 
projective multiplet, see \cite{KLT-M11} for the details. Its property is
\bea
\breve{ \breve{Q}}^{(n)}(v) =(-1)^n {Q}^{(n)}(v)~.
\label{smile-iso2}
\eea
Therefore, if  $n$ is even, one can define real projective multiplets, 
 $\breve{Q}^{(2n)} = {Q}^{(2n)}$.

A weight-$n$ isotwistor superfield $U^{(n)} (z,v)$ is defined to share with 
$Q^{(n)} (z,v)$ all its properties except the analyticity constraint \eqref{C.3}.

In this paper, all  covariant projective multiplets are assumed to be 
primary, 
\bea
S^{ij}_\a Q^{(n)} =0~, \qquad K_a Q^{(n)} =0~,
\eea
and hence $\{ S^{ij}_\a , \nabla^{(2)} \} Q^{(n)}=0$.  
Then it follows from \eqref{A.22} that the dimension of $Q^{(n)} $ is 
equal to $n / 2$, 
\bea
{\mathbb D} Q^{(n)} =\frac{n}{2} Q^{(n)} ~.
\eea

An important example of covariant projective multiplets is 
a real $\cO(2n) $ multiplet, with $n=1,2,\dots$
It  is described by a real weight-$2n$ projective superfield $H^{(2n)} (v)$ 
of the form:
\bea
H^{(2n)} (v) &=& H^{i_1 \dots i_{2n}} v_{i_1} \dots v_{i_{2n}} 
=\breve{H}^{(2n)} (v) ~.
\eea
The analyticity constraint (\ref{C.3}) is equivalent to 
\bea
\nabla_\a^{(ij} H^{k_1 \dots k_{2n} )} =0~,
\eea
while the reality condition $\breve{H}^{(2n)}  = {H}^{(2n)} $ is equivalent to 
\bea
\overline{ H^{i_1 \dots i_{2n}} } &=& H_{i_1 \dots i_{2n}}
=\ve_{i_1 j_1} \cdots \ve_{i_{2n} j_{2n} } H^{j_1 \dots j_{2n}} ~.
\eea
The field strength of an Abelian vector multiplet, $G^{(2)}$, 
is a real $\cO(2) $ multiplet.

To describe the dynamics of a supergravity-matter system, one has to specify 
a Lagrangian, $\cL^{(2)}(v)$, which is postulated to be 
a real weight-two projective multiplet. 
Associated with $\cL^{(2)}$ is the supersymmetric action \cite{KLT-M11}
\bea
S&=&
\frac{1}{2\pi\ri} \oint_\g (v, \rd v)
\int \rd^3 x \,{\rm d}^6\q\,E\, C^{(-4)} \cL^{(2)}~, 
\qquad E^{-1}= {\rm Ber}(E_A{}^M)~.
\label{InvarAc}
\eea
Here the {\it model-independent}  
isotwistor superfield $C^{(-4)} (v)$ of weight $-4$  
is required to be conformally primary and of dimension $-1$, 
\bea
S_\a^{ij}  C^{(-4)} =0~, \qquad K_a  C^{(-4)} =0~, 
\qquad {\mathbb D}  C^{(-4)} = -  C^{(-4)}~,
\eea
and obey the condition
\bea
\D^{(4)}C^{(-4)}&=&1~,
\label{AcComp-b}
\eea 
where 
\bea
\D^{(4)} := \frac{\ri}{4} \nabla^{(2)\a} \nabla^{(2)}_\a~.
\label{C.8}
\eea
As shown in \cite{KLT-M11}, the action \eqref{InvarAc}
does not change under an arbitrary infinitesimal variation of  $C^{(-4)} $, 
and thus  \eqref{InvarAc} is actually  independent of $ C^{(-4)} $.  

The second-order operator \eqref{C.8} allows us to engineer covariant 
projective multiplets. The point is that the superfield 
$\D^{(4)} U^{(n-4)} (v)$
is a covariant weight-$n$ projective multiplet, for any primary isotwistor superfield
$U^{(n-4)}$ of dimension $n/2 -1$.

We now derive a new representation for the action \eqref{InvarAc}  
that is valid under the assumption that there is an Abelian vector multiplet 
such that its gauge invariant field strength $G^{ij}$ is nowhere vanishing, 
$G:=\sqrt{G^{ij}G_{ij}}\neq 0$. 
Let $\cV(v)$ be the tropical prepotential for this multiplet. 
 By definition, $\cV(v)$  is a real weight-zero projective multiplet.  
The  superfield $G^{ij} $ is a real $\cO(2)$ multiplet which is 
related to $\cV (v)$ as follows:
\bea
G^{(2)}(v) := G_{ij} v^i v^j 
= \D^{(4)} \oint_{\hat \g} \frac{ ({\hat v},\rd {\hat v})}{  2\p(v, {\hat v})^2}\,\cV( {\hat v})
~.
\label{C.9}
\eea
The right-hand side on \eqref{C.9}
  is invariant under the gauge transformations \eqref{7.3}.

In the action \eqref{InvarAc}, we first  replace 
$\cL^{(2)} \to  G^{(2)} [\cL^{(2)}/G^{(2)}]$ and make use of the representation 
\eqref{C.9} for the first multiplier. As a next step, we can integrate by parts
in order to let $\D^{(4)} $ hit $C^{(-4)}$ and then use \eqref{AcComp-b}. 
Finally, we can change the order of contour integrations to result in
\bea
S&=&
\frac{1}{2\pi\ri} 
 \oint_{\hat \g}  ({\hat v},\rd {\hat v})\int \rd^3 x \,{\rm d}^6\q\,E\,\cV( {\hat v})
\oint_\g \frac{ (v, \rd v)}{  2\p(v, {\hat v})^2}\,
\frac{ \cL^{(2)}(v)}{G^{(2)}(v)} ~.
\eea
In this functional, we first re-label $v \leftrightarrow \hat v$, then insert the unity 
$1 = \D^{(4)} C^{(-4)} (v) $ and finally integrate $\D^{(4)}$ by parts. 
Since $\cV(v)$ obeys the constraint \eqref{C.3}, 
the projection operator $\D^{(4)} $ commutes with $\cV$, and we end up with the 
representation 
\begin{subequations} \label{C.15}
\bea
S&=&
\frac{1}{2\pi\ri} \oint_\g (v, \rd v)
\int \rd^3 x \,{\rm d}^6\q\,E\, 
C^{(-4)}
 \cV \,\bm G^{(2)}~, 
\eea
where
\bea
{\bm G}^{(2)}(v) := {\bm G}_{ij} v^iv^j
=\D^{(4)}
\oint_{\hat \g} \frac{({\hat v},\rd {\hat v})}{ 2\pi (v, {\hat v} )^2}  \frac{\cL^{(2)} ( {\hat v}) }
{G^{(2)}  ( {\hat v})} ~, \qquad
\nabla^{(2)}_\a {\bm G}^{(2)} =0
\label{C15.b}
\eea
\end{subequations}
is a composite real $\cO (2) $ multiplet. Eq. \eqref{C.15}
is our new representation for the action \eqref{InvarAc}. 
It is the main result of this section.

We conclude with an example that provides evidence of the universality of the 
projective superspace action  \eqref{InvarAc}. 
Let us consider the conventional $\cN=3$ locally supersymmetric action 
\bea
S = \ri \int \rd^3 x \,{\rm d}^6\q\,E\, \cL ~, \qquad {\mathbb D} \cL=0~,
\eea
where the Lagrangian $\cL$ is a dimensionless primary scalar superfield. 
It turns out that this action can be recast in the form \eqref{InvarAc} if we define
\bea
\cL^{(2)} =  2 \D^{(4)} \frac{G \cL}{ G^{(2)} }~.
\eea
This may be proved using the the contour integration techniques of \cite{BK11}.

\section{Action principle in $\cN=4$ supergravity}\label{AppendixD} 

Within the approach \cite{KLT-M11}, off-shell $\cN=4$ supergravity-matter systems
are formulated in curved $\cN=4$ projective superspace 
$\cM^{3|8} \times {\mathbb C}P^1_\rL \times {\mathbb C}P^1_\rR$ in terms of covariant projective multiplets. These multiplets were defined in \cite{KLT-M11} 
in SO(4) superspace. Here we briefly extend those definitions to $\cN=4$ conformal 
superspace. Our presentation is similar to the $\cN=3$ story of the previous section. 

Let $v_\rL:=v^i \in {\mathbb C}^2 \setminus  \{0\}$ and 
$v_\rR:=v^{\bar i} \in {\mathbb C}^2 \setminus  \{0\}$ 
be homogeneous coordinates for ${\mathbb C}P^1_\rL $ and
$ {\mathbb C}P^1_\rR$ respectively.
We use these variables to define two different subsets,  
$\nabla_\a^{(1)\bar i}$ and $\nabla_\a^{(\bar 1)i}$,
in the set of spinor covariant  derivatives $\nabla^{i\bar i}_\a$,  
\bea
\nabla_\a^{(1)\bar i}:=v_i\nabla_\a^{i\bar i}~,~~~~~~
\nabla_\a^{(\bar 1)i}:=v_{\bar i}\nabla_\a^{i\bar i}~.
\eea
It follows from \eqref{A.33} that the operators $\nabla_\a^{(1)\bar i}$
obey the anti-commutation relations:
\bea
\big\{ \nabla_\a^{(1)\bar i} , \nabla_\b^{(1)\bar j} \big\} &=& 2 \ri \eps_{\a\b} \eps^{\bar{i} \bar{j}} W \bm {\rm L}^{(2)} 
+ \ri \eps_{\a\b} \eps^{\bar{i} \bar{j}} \nabla^{\g (1)}{}_{\bar{k}} W S_\g^{(1)\bar{k}} \non\\
&&- \frac{1}{4} \eps_{\a\b} \eps^{\bar{i} \bar{j}} \nabla_\g{}^{(1)}{}_{\bar{k}} \nabla_\d^{(1) \bar{k}} W K^{\g\d} \ .
\eea

There are two types of covariant projective multiplets, 
the left and right ones. A left projective multiplet of weight $n$, $Q^{(n)}_\rL (v_\rL)$,
is defined to obey the constraint 
\bea
\nabla_\a^{(1)\bar i} Q^{(n)}_\rL =0
\eea
and is required to be a holomorphic and homogeneous   function 
of $v_\rL$ of degree $n$, 
\be
Q_\rL^{(n)}(c\,v_\rL)\,=\,c^n\,Q_\rL^{(n)}(v_\rL)\ ,
\qquad c \in   {\mathbb C} \setminus  \{0\} \ ,
\ee 
on some open domain of $ {\mathbb C}^2 \setminus  \{0\}$.
The left projective multiplet is inert with respect to $\rm SU(2)_{\rm R}$ 
and transforms under $\rm SU(2)_{\rm L}$ as
\bsubeq
\bea
\d_\L Q_{\rm L}^{(n)} &=& \L^{ij} \bm {\rm L}_{ij} Q_{\rm L}^{(n)} \ , \\
\bm {\rm L}_{ij} Q_{\rm L}^{(n)} &=& - (\L^{(2)}_{\rm L} \bm \partial_{\rm L}^{(-2)} - n \L_{\rm L}^{(0)}) Q_{\rm L}^{(n)} \ ,
\eea
\esubeq
where we have defined
\be \L_{\rm L}^{(2)} := \L^{ij} v_i v_j \ , \quad \L_{\rm L}^{(0)} = \frac{v_i u_j}{(v_{\rm L} , u_{\rm L})} \L^{ij}
\ee
and introduced the differential operator
\be \bm \partial_{\rm L}^{(-2)} = \frac{1}{(v_{\rm L} , u_{\rm L})} u^i \frac{\partial}{\partial v^i} \ , \quad (v_{\rm L} , u_{\rm L}) = v^i u_i \ .
\ee
The right projective multiplets are defined similarly. 

In $\cN=4$ conformal superspace,  we can also introduce 
{\it hybrid projective multiplets} 
and {\it isotwistor projective multiplets}. The corresponding definitions are completely analogous to those given in \cite{KLT-M11}. 

All left and right projective multiplets are assumed to be primary, in particular
\be S_\a^{i \bar{i}} Q_{\rm L}^{(n)} = 0 \ , \quad K_a Q_{\rm L}^{(n)} = 0 \ .
\ee
Hence we have the condition
\be \{ S_\a^{i \bar{i}} , \nabla_\b^{(1) \bar{j}} \} Q_{\rm L}^{(n)} = 0 \ ,
\ee
which fixes the dimension of $Q_{\rm L}^{(n)}$
\be \mathbb{D} Q_{\rm L}^{(n)} = \frac{n}{2} Q_{\rm L}^{(n)} \ .
\ee

In general, the $\cN=4$ supersymmetric action may be represented as a sum of two terms, the 
left $S_\rL$ and right $S_\rR$ ones,\footnote{There exist different action principles, in particular the one with a hybrid Lagrangian \cite{KLT-M11}. However, 
they may be always reduced to the form \eqref{D.12}.}
\bea
S= S_{\rm L} + S_{\rm R}~.
\label{D.12}
\eea 
The left action has the form 
\bea
S_{\rL}  &=& \frac{1}{2\pi} \oint_{\g_\rL} (v_\rL, \rd v_\rL)
\int \rd^3 x \,{\rm d}^8\q\,E\, C_\rL^{({-4})} \cL_\rL^{(2)}~, 
\qquad E^{-1}= {\rm Ber}(E_A{}^M)~,
\label{Action-left} 
\eea
where the Lagrangian $\cL_\rL^{(2)}(v_\rL)$ is a real left projective multiplet 
of weight 2.
The action involves a {\it model-independent} primary
 isotwistor superfield  $C_\rL^{(-4)} (v_\rL) $ of  dimension $-2$, 
 ${\mathbb D} C_\rL^{(-4)} =-2 C_\rL^{(-4)}$. 
It is defined to be real with respect to the smile-conjugation
and obey the differential equation
\bea
\D_\rL^{(4)}C_\rL^{(-4)}=1~.
\label{N=4AcComp-L}
\eea
Here $\D_\rL^{(4)}$ denotes the following fourth-order operator\footnote{The operator
$\D_\rL^{(4)}$ is a covariant projection operator. Given  a covariant left projective multiplet $Q_\rL^{(n)} (v_\rL) $ of weight $n$, 
it may be represented in the form 
$Q_\rL^{(n)}=\D_\rL^{(4)} T_\rL^{(n-4)}$, 
for some left isotwistor superfield $T_\rL^{(n-4)} (v_\rL )$, see  \cite{KLT-M11}
for the details.}
\bea
\D_\rL^{(4)}&=&\frac{1}{96}\Big(
\nabla^{ (2) \bar i \bar j}  \nabla^{(2)}_{\bar i\bar j}
-\nabla^{(2)\a\b}  \nabla^{(2)}_{\a\b}
\Big) = \frac{1}{48}\nabla^{ (2) \bar i \bar j}  \nabla^{(2)}_{\bar i \bar j}
~,
\eea
with
\bea
\nabla^{(2)}_{\bar i\bar j}:=\nabla^{(1)\g}_{(\bar i}\nabla^{(1)}_{\g \bar j)}~,\qquad
\nabla^{(2)}_{\a\b}:=\nabla^{(1)\bar k}_{(\a}\nabla^{(1)}_{\b) \bar k}
~.
\eea
The action \eqref{Action-left} is independent of $ C_\rL^{(-4)} $ in the sense that 
it does not change under an arbitrary infinitesimal variation of  $C_\rL^{(-4)} $.

An Abelian vector multiplet with self-dual field strength $G_+^{IJ}$
can be described by
a left tropical prepotential 
$\cV_{\rm L}(v_{\rm L}) $ defined modulo gauge transformations 
\bea
\d  V_\rL = \l_\rL + \breve{\l}_\rL~,
\label{left-gauge}
\eea
where the gauge parameter $\l_\rL$ is an arbitrary left arctic multiplet of weight zero.
The corresponding gauge invariant field strength, $G^{\bar i\bar j}$,
 is a right $\cO(2) $ multiplet related to $\cV_\rL$ as follows:
\bea 
G_{\rm R}^{(2)}(v_{\rm R}) = v_{\bar i} v_{\bar j} G^{\bar{i} \,\bar{j}} 
= \frac{\ri}{4}  v_{\bar i} v_{\bar j}
\oint \frac{(v_{\rm L} , \rd v_{\rm L})}{2 \pi} 
\frac{u_iu_j}{(v_{\rm L}, u_{\rm L})^2}
 \nabla^{\a i \bar i} \nabla_\a{}^{j\bar j}
\cV_{\rm L}(v_{\rm L}) \ .
\label{D.10}
\eea
Here $u_\rL = u^i$ is a constant isospinor such that $(v_{\rm L}, u_{\rm L}) \neq 0$
along the closed integration contour. One may show that the right-hand side of 
\eqref{D.10} is independent of $u_\rL$.

The left $\cO(2)$ multiplet
$G^{ij}$ is 
associated with a  right tropical prepotential $\cV_{\rm R} (v_{\rm R})$
according to the rule:
\bea
G_{\rm L}^{(2)}(v_{\rm L}) = v_{ i} v_{ j} G^{ij} 
=  \frac{\ri}{4}  v_{ i} v_{ j}
\oint \frac{(v_{\rm {\rm R}} , \rd v_{\rm {\rm R}})}{2 \pi} 
\frac{u_{\bar i}u_{\bar j}}{(v_{\rm R}, u_{\rm R})^2}
 \nabla^{\a i \bar i} \nabla_\a{}^{j\bar j}
\cV_{\rm R}(v_{\rm R}) \ .
\label{D.11}
\eea

The prepotential $\cV_{\rm R}$ can always be represented as 
\bea
\cV_{\rm R}(v_{\rm R}) = \D_\rR^{(4)} T_\rR^{(-4)} (v_\rR)~, 
\eea
for some isotwistor superfield $T_\rR^{(-4)} (v_\rR)$.\footnote{See \cite{KLT-M11}
for the definition of $\cN=4$ isotwistor superfields.} Here $\D_{\rm R}^{(4)}$ is defined similar to $\D_{\rm L}^{(4)}$,
\be \D_\rR^{(4)} = \frac{1}{48}\nabla^{ (2) i j}  \nabla^{(2)}_{i j} \ , \quad \nabla^{(2)}_{ij} := \nabla^{\g(\bar 1)}_{(i} \nabla_{\g j)}^{(\bar 1)} \ .
\ee
There  exists an isotwistor superfield
$T^{(-2,-4)} (\hat{v}_\rL , v_\rR )$ such that\footnote{For instance, 
we can choose  $T^{(-2,-4)} (\hat{v}_\rL , v_\rR ) 
= -2 T_\rR^{(-4)} (v_\rR)  \frac{ G_\rL}{  G^{(2)}_\rL (\hat{v}_\rL) }$.}
\bea
T_\rR^{(-4)} (v_\rR) = 
\oint_{\hat{\g}_\rL}  \frac{(\hat v_{\rm L} , \rd \hat v_{\rm L})}{2 \pi} 
T^{(-2,-4)} (\hat{v}_\rL , v_\rR )~.
\eea
Then, the field strength \eqref{D.11} can be rewritten in the form \cite{KLT-M11}
\bea
G_{\rm L}^{(2)}(v_{\rm L}) 
=  \D^{(4)}_\rL
\oint_{\g_\rR} \frac{(v_{\rm {\rm R}} , \rd v_{\rm {\rm R}})}{2 \pi } 
\oint_{\hat{\g}_\rL}  \frac{(\hat v_{\rm L} , \rd \hat v_{\rm L})}
{2 \pi   (v_{\rm L}, \hat{v}_{\rm L})^2  } \D^{(\hat 2, 2)} 
T^{(-2,-4)} (\hat{v}_\rL , v_\rR )~,
\label{D.14}
\eea
where we have defined
\bea
 \D^{(\hat 2, 2)} = \frac{\ri}{4}  \hat{v}_i \hat{v}_j v_{\bar i} v_{\bar j} 
 \nabla^{\a i \bar i} \nabla^{j \bar j}_\a~.
 \label{D.15}
 \eea

We now obtain an alternative representation for the left action \eqref{Action-left}. 
The idea is to insert the unity 
$1= G_{\rm L}^{(2)}(v_{\rm L}) /G_{\rm L}^{(2)}(v_{\rm L}) $
into the integrand \eqref{Action-left}, make use of 
the expression \eqref{D.14} for the field strength in the numerator and 
then integrate by parts in order to let $ \D^{(4)}_\rL$ hit 
$ C_\rL^{({-4})}$.  This gives
\bea
S_\rL &=& \oint_{\g_\rR} \frac{(v_{\rm {\rm R}} , \rd v_{\rm {\rm R}})}{2 \pi } 
\int \rd^3 x \,{\rm d}^8\q\,E\, 
 \oint_{\g_\rL}   \frac{ (v_\rL, \rd v_\rL) }{2\pi}
\frac{ \cL_\rL^{(2)} (v_\rL ) }{ G^{(2)}_\rL (v_\rL )} \non \\
&& \times \oint_{\hat{\g}_\rL}  \frac{(\hat v_{\rm L} , \rd \hat v_{\rm L})}
{2 \pi   (v_{\rm L}, \hat{v}_{\rm L})^2  } \D^{(\hat 2, 2)} 
T^{(-2,-4)} (\hat{v}_\rL , v_\rR )~,
\eea
where we have changed the order of contour integrals. 
The next step is to insert the unity $\D_\rR^{(4)}C_\rR^{(-4)}=1$ into the integrand
and then integrate by parts. This gives
\bea
S_\rL &=& \oint_{\g_\rR} \frac{(v_{\rm {\rm R}} , \rd v_{\rm {\rm R}})}{2 \pi } 
\int \rd^3 x \,{\rm d}^8\q\,E\, C_\rR^{(-4)} 
\non \\
&& \times
\D^{(4)}_\rR
 \oint_{\g_\rL}   \frac{ (v_\rL, \rd v_\rL) }{2\pi}
\frac{ \cL_\rL^{(2)} (v_\rL ) }{ G^{(2)}_\rL (v_\rL )} 
 \oint_{\hat{\g}_\rL}  \frac{(\hat v_{\rm L} , \rd \hat v_{\rm L})}
{2 \pi   (v_{\rm L}, \hat{v}_{\rm L})^2  } \D^{(\hat 2, 2)} 
T^{(-2,-4)} (\hat{v}_\rL , v_\rR )~.
\eea
Taking the explicit form of $\D^{(4)}_\rR$ into account, this expression can be 
replaced with the following:
\bea
S_\rL &=& - \oint_{\g_\rR} \frac{(v_{\rm {\rm R}} , \rd v_{\rm {\rm R}})}{2 \pi } 
\int \rd^3 x \,{\rm d}^8\q\,E\, C_\rR^{(-4)} \oint_{\g_\rL}   \frac{ (v_\rL, \rd v_\rL) }{2\pi}
\non \\
&& \times
\D^{(-2,2)} \left\{ 
\frac{ \cL_\rL^{(2)} (v_\rL ) }{ G^{(2)}_\rL (v_\rL )} 
 \oint_{\hat{\g}_\rL}  \frac{(\hat v_{\rm L} , \rd \hat v_{\rm L})}
{2 \pi   (v_{\rm L}, \hat{v}_{\rm L})^2  }  \D^{(2,2)} \D^{(\hat 2, 2)} 
T^{(-2,-4)} (\hat{v}_\rL , v_\rR ) \right\}~.
\label{D.18}
\eea
Here the operator $\D^{(-2,2)} $ is defined by 
\bea
\D^{(-2,2)} := \frac{u_{ i}u_{ j}}{(v_{\rm L}, u_{\rm L})^2}
v_{\bar i} v_{\bar j}  \nabla^{\a i \bar i} \nabla_\a{}^{j\bar j}~, 
\eea
for an isospinor $u_i$ such that  $(v_{\rm L}, u_{\rm L}) \neq 0$.
The operator $\D^{(2, 2)} $ in \eqref{D.18} is obtained from 
$\D^{(\hat 2, 2)} $, eq. \eqref{D.15}, by replacing $\hat{v}_i \to v_i$.
Now, one may notice that  
$\D^{(2,2)} \D^{(\hat 2, 2)} $ in \eqref{D.18} is equivalent to 
$ (v_{\rm L}, \hat{v}_{\rm L})^2\D^{(4)}_\rR  $, and therefore the action turns into
\bea
S_\rL &=&  \oint_{\g_\rR} \frac{(v_{\rm {\rm R}} , \rd v_{\rm {\rm R}})}{2 \pi } 
\int \rd^3 x {\rm d}^8\q\,E\, C_\rR^{(-4)} \oint_{\g_\rL}   \frac{ (v_\rL, \rd v_\rL) }{2\pi}
\D^{(-2,2)} \left\{ 
\frac{ \cL_\rL^{(2)} (v_\rL ) }{ G^{(2)}_\rL (v_\rL )} \cV_\rR  (v_\rR )\right\}~.~~~~~
\eea
Since $\cV_\rR $ is a right projective multiplet, it commutes with 
the operator $\D^{(-2,2)} $, and we end  up with the following representation 
for $S_\rL$:
\bea
S_\rL &=&  \oint_{\g_\rR} \frac{(v_{\rm {\rm R}} , \rd v_{\rm {\rm R}})}{2 \pi } 
\int \rd^3 x {\rm d}^8\q\,E\, C_\rR^{(-4)} 
\cV_\rR \,{\bm G}^{(2)}_\rR~, 
\label{D.28}
\eea
where
\bea 
{\bm G}_{\rm R}^{(2)}(v_{\rm R}) 
= v_{\bar i} v_{\bar j} {\bm G}^{\bar{i} \,\bar{j}} 
= \frac{\ri}{4}  v_{\bar i} v_{\bar j}
\oint \frac{(v_{\rm L} , \rd v_{\rm L})}{2 \pi} 
\frac{u_iu_j}{(v_{\rm L}, u_{\rm L})^2}
 \nabla^{\a i \bar i} \nabla_\a{}^{j\bar j}
\left\{  \frac{ \cL_\rL^{(2)} (v_\rL ) }{ G^{(2)}_\rL (v_\rL )}\right\}
\label{D.29}
 \eea
 is a composite right $\cO(2)$ multiplet.

Eq. \eqref{D.28} is our new representation for the left action \eqref{D.12}. 
The important point is that the integration in \eqref{D.12} and \eqref{D.28} 
is carried out over different subspaces of the curved projective superspace. 
The original left action \eqref{D.12} is given as an integral over
$\cM^{3|8} \times {\mathbb C}P^1_\rL $. In the final action \eqref{D.28}, 
the integration is carried out over $\cM^{3|8}  \times {\mathbb C}P^1_\rR$.

Since  \eqref{D.28} involves the composite right $\cO(2)$ multiplet 
${\bm G}_{\rm R}^{(2)}$, it will be referred to as the right linear multiplet action.


\begin{footnotesize}

\end{footnotesize}


\begin{thebibliography}{66}

\bibitem{Sohnius1}
  M.~F.~Sohnius,
``Supersymmetry and central charges,''
Nucl.\ Phys.\  B {\bf 138}, 109 (1978).

\bibitem{Fayet}
P.~Fayet, ``Fermi-Bose hypersymmetry,''
Nucl.\ Phys.\ B {\bf 113}, 135 (1976).

\bibitem{GSW}
  R.~Grimm, M.~Sohnius and J.~Wess,
  ``Extended supersymmetry and gauge theories,''
  Nucl.\ Phys.\  B {\bf 133}, 275 (1978).

\bibitem{BS}
P.~Breitenlohner and M.~F.~Sohnius,
``Superfields, auxiliary fields, and tensor calculus for N=2 extended
supergravity,''
Nucl.\ Phys.\  B {\bf 165}, 483 (1980).

\bibitem{deWvHVP3}
  B.~de Wit, J.~W.~van Holten and A.~Van Proeyen,
  ``Central charges and conformal supergravity,''
  Phys.\ Lett.\ B {\bf 95}, 51 (1980).

\bibitem{deWvHVP}
 B.~de Wit, J.~W.~van Holten and A.~Van Proeyen,
 ``Transformation rules of N=2 supergravity multiplets,''
Nucl.\ Phys.\  B {\bf 167}, 186 (1980).

\bibitem{BdeRdeW}
  E.~Bergshoeff, M.~de Roo and B.~de Wit,
  ``Extended conformal supergravity,''
  Nucl.\ Phys.\  B {\bf 182}, 173 (1981).

\bibitem{deWvHVP2}
B.~de Wit, J.~W.~van Holten and A.~Van Proeyen,
  ``Structure of N=2 supergravity,''
  Nucl.\ Phys.\  B {\bf 184}, 77 (1981)h
  [Erratum-ibid.\  B {\bf 222}, 516 (1983)].
 
\bibitem{FVP} 
  D.~Z.~Freedman and A.~Van Proeyen,
 {\it Supergravity},
  Cambridge University Press,  Cambridge, 2012.

\bibitem{Zucker} 
  M.~Zucker,
  ``Gauged N=2 off-shell supergravity in five-dimensions,''
  JHEP {\bf 0008}, 016 (2000)
  [hep-th/9909144];
   ``Off-shell supergravity in five
   dimensions and supersymmetric brane world scenarios,''
  Fortsch.\ Phys.\  {\bf 51}, 899 (2003).

\bibitem{Ohashi}
T.~Kugo and K.~Ohashi,
 ``Supergravity tensor calculus in 5D from 6D,''
  Prog.\ Theor.\ Phys.\  {\bf 104}, 835 (2000)
  [hep-ph/0006231];
``Off-shell d = 5 supergravity coupled to matter-Yang-Mills system,''
Prog.\ Theor.\ Phys.\  {\bf 105}, 323 (2001)   {[hep-ph/0010288]};
  T.~Fujita and K.~Ohashi, ``Superconformal tensor calculus in five dimensions,''
  Prog.\ Theor.\ Phys.\  {\bf 106}, 221 (2001)
 {[hep-th/0104130]}.

\bibitem{Bergshoeff}
 E.~Bergshoeff, S.~Cucu, M.~Derix, T.~de Wit, R.~Halbersma and A.~Van Proeyen,
  ``Weyl multiplets of N = 2 conformal supergravity in five dimensions,''
  JHEP {\bf 0106}, 051 (2001)
  [hep-th/0104113];
  E.~Bergshoeff, S.~Cucu, T.~de Wit, J.~Gheerardyn, R.~Halbersma, 
  S.~Vandoren and A.~Van Proeyen,
  ``Superconformal N = 2, D = 5 matter with and without actions,''
  JHEP {\bf 0210}, 045 (2002)
  {[hep-th/0205230]};
E.~Bergshoeff, S.~Cucu, T.~de Wit, J.~Gheerardyn, S.~Vandoren and A.~Van Proeyen,
  ``N = 2 supergravity in five dimensions revisited,''
  Class.\ Quant.\ Grav.\  {\bf 21}, 3015 (2004)
 {[hep-th/0403045]}. 
 
\bibitem{BSV} 
  E.~Bergshoeff, E.~Sezgin and A.~Van Proeyen,
  ``Superconformal tensor calculus and matter couplings in six dimensions,''
  Nucl.\ Phys.\ B {\bf 264}, 653 (1986)
  [Erratum-ibid.\ B {\bf 598}, 667 (2001)].
  
\bibitem{KT-M08}
S.~M.~Kuzenko and G.~Tartaglino-Mazzucchelli,
  ``Five-dimensional superfield supergravity,''
    Phys.\ Lett.\ B {\bf 661}, 42 (2008)
 [arXiv:0710.3440];
  ``5D supergravity and projective superspace,''
  JHEP {\bf 0802}, 004 (2008) [arXiv:0712.3102];
  ``Super-Weyl invariance in 5D supergravity,''
  JHEP {\bf 0804}, 032 (2008)
  [arXiv:0802.3953].

\bibitem{SSW2}
  M.~F.~Sohnius, K.~S.~Stelle and P.~C.~West,
 ``Representations of extended supersymmetry,''
in {\it Superspace and Supergravity}, S. W. Hawking and M. Ro\v{c}ek (Eds.), 
Cambridge University Press, Cambridge, 1981, p. 283.  

\bibitem{LR2008}
U.~Lindstr\"om and M.~Ro\v{c}ek,
``Properties of hyperk\"ahler manifolds and their twistor spaces,''
Commun.\ Math.\ Phys.\  {\bf 293}, 257 (2010)
[arXiv:0807.1366 [hep-th]].

\bibitem{Kuzenko_lectures} 
  S.~M.~Kuzenko,
  ``Lectures on nonlinear sigma models in projective superspace,''
  J.\ Phys.\ A {\bf 43}, 443001 (2010)
  [arXiv:1004.0880 [hep-th]].

\bibitem{Wess}
J.~Wess, ``Supersymmetry and internal symmetry,''
Acta Phys.\ Austriaca {\bf 41} (1975) 409.

\bibitem{KLRT-M}
S.~M.~Kuzenko, U.~Lindstr\"om, M.~Ro\v cek and G.~Tartaglino-Mazzucchelli,
``4D N=2 supergravity and projective superspace,'' 
JHEP {\bf 0809}, 051 (2008) [arXiv:0805.4683];
%
S.~M.~Kuzenko,
``On N = 2 supergravity and projective superspace: Dual formulations,''
Nucl.\ Phys.\  B {\bf 810}, 135 (2009)
[arXiv:0807.3381 [hep-th]];
%
 S.~M.~Kuzenko, U.~Lindstr\"om, M.~Ro\v{c}ek and G.~Tartaglino-Mazzucchelli,
``On conformal supergravity and projective superspace,''
JHEP {\bf 0908}, 023 (2009)
  [arXiv:0905.0063 [hep-th]].

\bibitem{KT-M09} 
  S.~M.~Kuzenko and G.~Tartaglino-Mazzucchelli,
  ``Different representations for the action principle in 4D N = 2 supergravity,''
  JHEP {\bf 0904}, 007 (2009)
  [arXiv:0812.3464 [hep-th]].

\bibitem{KT}
S.~M.~Kuzenko and S.~Theisen,
 ``Correlation functions of conserved currents in N = 2 superconformal
theory,''  Class.\ Quant.\ Grav.\  {\bf 17}, 665 (2000)  [hep-th/9907107]. 

\bibitem{DIKST}
N.~Dragon, E.~Ivanov, S.~Kuzenko, E.~Sokatchev and U.~Theis,
 ``N=2 rigid supersymmetry with gauged central charge,''
Nucl.\ Phys.\  B {\bf 538}, 411 (1999)
  [arXiv:hep-th/9805152].

\bibitem{LT-M} 
  W.~D.~Linch III and G.~Tartaglino-Mazzucchelli,
``Six-dimensional supergravity and projective superfields,''
  JHEP {\bf 1208}, 075 (2012)
  [arXiv:1204.4195 [hep-th]].

\bibitem{GIKOS}
A.~S.~Galperin, E.~A.~Ivanov, S.~N.~Kalitsyn, V.~Ogievetsky, E.~Sokatchev, 
``Unconstrained N=2 matter, Yang-Mills and supergravity theories in harmonic
superspace,''
Class.\ Quant.\ Grav.\  {\bf 1}, 469 (1984).

\bibitem{GIOS}
A.~S.~Galperin, E.~A.~Ivanov, V.~I.~Ogievetsky and E.~S.~Sokatchev,
{\it Harmonic Superspace}, Cambridge University Press, 
Cambridge, 2001.

\bibitem{KLR}
A. Karlhede, U. Lindstr\"om and M. Ro\v cek,
``Self-interacting tensor multiplets in N=2 superspace,''
Phys.\ Lett.\ B {\bf 147}, 297 (1984).

\bibitem{LR-proj}
  U.~Lindstr\"om and M.~Ro\v{c}ek,
  ``New hyperk\"ahler  metrics  and new supermultiplets,''
  Commun.\ Math.\ Phys.\  {\bf 115}, 21 (1988);
%
  ``N=2 super Yang-Mills theory in projective superspace,''
  Commun.\ Math.\ Phys.\  {\bf 128}, 191 (1990).

\bibitem{BKN} 
D.~Butter, S.~M.~Kuzenko and J.~Novak,
``The linear multiplet and ectoplasm,'' JHEP {\bf 1209}, 131 (2012)
  [arXiv:1205.6981 [hep-th]].

\bibitem{Butter-N=2}
  D.~Butter,
  ``N=2 conformal superspace in four dimensions,''
  JHEP {\bf 1110}, 030 (2011).
  [arXiv:1103.5914 [hep-th]]. 
  
\bibitem{Castellani} 
  L.~Castellani, R.~D'Auria and P.~Fre,
{\it Supergravity and superstrings: A Geometric perspective. Vol. 2: Supergravity},
World Scientific,  Singapore, 1991, pp. 680--684. 
  
\bibitem{Hasler}
M.~F.~Hasler,
  ``The three form multiplet in N=2 superspace,''
  Eur.\ Phys.\ J.\ C {\bf 1}, 729 (1998)
  [hep-th/9606076].
  
\bibitem{Ectoplasm} 
S.~J.~Gates, Jr., ``Ectoplasm has no topology: The prelude,''
in {\it Supersymmetries and Quantum Symmetries},
 J. Wess and E. A. Ivanov (Eds.), Springer, Berlin, 1999, p. 46, arXiv:hep-th/9709104.
 
\bibitem{GGKS}
S.~J.~Gates, Jr., M.~T.~Grisaru, M.~E.~Knutt-Wehlau and W.~Siegel,
``Component actions from curved superspace: Normal coordinates and
ectoplasm,'' Phys.\ Lett.\  B {\bf 421}, 203 (1998)
[hep-th/9711151].
  
\bibitem{KT-M12} 
  S.~M.~Kuzenko and G.~Tartaglino-Mazzucchelli,
  ``Conformal supergravities as Chern-Simons theories revisited,''
  JHEP {\bf 1303}, 113 (2013)
  [arXiv:1212.6852 [hep-th]].

\bibitem{BKNT-M2}
D.~Butter, S.~M.~Kuzenko, J.~Novak and G.~Tartaglino-Mazzucchelli,
``Conformal supergravity in three dimensions: Off-shell actions,''
JHEP {\bf 1310}, 073 (2013)
  [arXiv:1306.1205 [hep-th]].

\bibitem{KNT-M} 
  S.~M.~Kuzenko, J.~Novak and G.~Tartaglino-Mazzucchelli,
  ``N=6 superconformal gravity in three dimensions from superspace,''
  arXiv:1308.5552 [hep-th].

\bibitem{NT} 
  M.~Nishimura and Y.~Tanii,
  ``N=6 conformal supergravity in three dimensions,''
  JHEP {\bf 1310}, 123 (2013)
  [arXiv:1308.3960 [hep-th]].
  
  \bibitem{vN}
P.~van Nieuwenhuizen,
``D = 3 conformal supergravity and Chern-Simons terms,''
Phys.\ Rev.\  D {\bf 32}, 872 (1985).
  
\bibitem{RvN} 
  M.~Ro\v{c}ek and P.~van Nieuwenhuizen,
  ``N $\geq$ 2 supersymmetric Chern-Simons terms as d = 3 extended conformal supergravity,''
  Class.\ Quant.\ Grav.\  {\bf 3}, 43 (1986).
  
  \bibitem{BKNT-M1} 
  D.~Butter, S.~M.~Kuzenko, J.~Novak and G.~Tartaglino-Mazzucchelli,
  ``Conformal supergravity in three dimensions: New off-shell formulation,''
 JHEP {\bf 1309}, 072 (2013)
  [arXiv:1305.3132 [hep-th]].
  
\bibitem{HIPT}
P.~S.~Howe, J.~M.~Izquierdo, G.~Papadopoulos and P.~K.~Townsend,
``New supergravities with central charges and Killing spinors in 2+1 dimensions,''
Nucl.\ Phys.\  B {\bf 467}, 183 (1996)
  [arXiv:hep-th/9505032].
  
\bibitem{KLT-M11} 
  S.~M.~Kuzenko, U.~Lindstr\"om and G.~Tartaglino-Mazzucchelli,
  ``Off-shell supergravity-matter couplings in three dimensions,''
  JHEP {\bf 1103}, 120 (2011)
  [arXiv:1101.4013 [hep-th]].

\bibitem{Siegel}
  W.~Siegel,  ``Unextended superfields in extended supersymmetry,''
  Nucl.\ Phys.\  B {\bf 156}, 135 (1979).
  
\bibitem{GGRS}
 S.~J.~Gates, Jr., M.~T.~Grisaru, M.~Ro\v{c}ek and W.~Siegel,
{\it Superspace, or One Thousand and One Lessons in Supersymmetry},
Front.\ Phys.\  {\bf 58}, 1 (1983) [arXiv:hep-th/0108200].

\bibitem{ZP}
  B.~M.~Zupnik and D.~G.~Pak,
  ``Superfield formulation of the simplest three-dimensional gauge theories and
  conformal supergravities,''  Theor.\ Math.\ Phys.\  {\bf 77} (1988) 1070
  [Teor.\ Mat.\ Fiz.\  {\bf 77} (1988) 97].
  
\bibitem{Ivanov91} 
 E.~A.~Ivanov,
  ``Chern-Simons matter systems with manifest N=2 supersymmetry,''
  Phys.\ Lett.\ B {\bf 268}, 203 (1991).
  
  \bibitem{ZH}
B.~M.~Zupnik and D.~V.~Hetselius,
``Three-dimensional extended supersymmetry in harmonic superspace,''
Sov.\ J.\ Nucl.\ Phys.\  {\bf 47}, 730 (1988)
  [Yad.\ Fiz.\  {\bf 47}, 1147 (1988)].

\bibitem{BrooksG}
  R.~Brooks and S.~J.~Gates Jr.,
  ``Extended supersymmetry and super-BF gauge theories,''
  Nucl.\ Phys.\  B {\bf 432}, 205 (1994)
  [arXiv:hep-th/9407147].

\bibitem{KS}
  A.~Kapustin and M.~J.~Strassler,
  ``On mirror symmetry in three dimensional Abelian gauge theories,''
  JHEP {\bf 9904}, 021 (1999)
  [arXiv:hep-th/9902033].
  
 \bibitem{Zupnik99}
 B.~Zupnik,
 ``Harmonic superpotentials and symmetries in gauge theories 
with eight  supercharges,''
  Nucl.\ Phys.\ B {\bf 554},  365 (1999)
  [Erratum-ibid.\ B {\bf 644},  405  (2002)]
{[hep-th/9902038]}.

\bibitem{Zupnik2009} 
  B.~M.~Zupnik,
  ``Three-dimensional N=4 superconformal superfield theories,''
  Theor.\ Math.\ Phys.\  {\bf 162}, 74 (2010)
  [arXiv:0905.1179 [hep-th]].
 
\bibitem{Zupnik2010} 
  B.~M.~Zupnik,
  ``Three-dimensional $\mathcal {N}=4 $ supersymmetry in harmonic $\mathcal {N}=3$ superspace,''
  Theor.\ Math.\ Phys.\  {\bf 165}, 1315 (2010)
  [Teor.\ Mat.\ Fiz.\  {\bf 165}, 97 (2010)]
  [arXiv:1005.4750 [hep-th]].
  
\bibitem{dWNT}
  B.~de Wit, H.~Nicolai and A.~K.~Tollsten,
  ``Locally supersymmetric D = 3 nonlinear sigma models,''
  Nucl.\ Phys.\  B {\bf 392}, 3 (1993)
  [arXiv:hep-th/9208074].

\bibitem{dWHS}
  B.~de Wit, I.~Herger and H.~Samtleben,
 ``Gauged locally supersymmetric D = 3 nonlinear sigma models,''
  Nucl.\ Phys.\  B {\bf 671}, 175 (2003)
  [arXiv:hep-th/0307006].

\bibitem{dWNS} 
  B.~de Wit, H.~Nicolai and H.~Samtleben,
 ``Gauged supergravities in three-dimensions: A Panoramic overview,''
  hep-th/0403014.
  
  \bibitem{BCSS} 
  E.~Bergshoeff, S.~Cecotti, H.~Samtleben and E.~Sezgin,
  ``Superconformal sigma models in three dimensions,''
  Nucl.\ Phys.\ B {\bf 838}, 266 (2010)
  [arXiv:1002.4411 [hep-th]].
  
  \bibitem{NR02} 
  H.~Nishino and S.~Rajpoot,
  ``Supersymmetric E(8(+8))/ SO(16) sigma model coupled to N=1 supergravity in three-dimensions,''
  Phys.\ Lett.\ B {\bf 535}, 337 (2002)
  [hep-th/0203102].
  
\bibitem{HitchinKLR}
N.~J.~Hitchin, A.~Karlhede, U.~Lindstr\"om and M.~Ro\v cek,
``Hyperk\"ahler metrics and supersymmetry,''
Commun.\ Math.\ Phys.\  {\bf 108}, 535 (1987).  

\bibitem{Sohnius}
  M.~F.~Sohnius,
``Bianchi identities for supersymmetric gauge theories,''
  Nucl.\ Phys.\  B {\bf 136}, 461 (1978).

\bibitem{GGHN} 
  U.~Gran, J.~Greitz, P.~Howe and B.~E.~W.~Nilsson,
  ``Topologically gauged superconformal Chern-Simons matter theories,''
  JHEP {\bf 1212}, 046 (2012)
  [arXiv:1204.2521 [hep-th]].
  
\bibitem{BPT} 
  L.~Bonora, P.~Pasti and M.~Tonin,
  ``Chiral anomalies in higher dimensional supersymmetric theories,''
  Nucl.\ Phys.\ B {\bf 286}, 150 (1987).

\bibitem{BHLSW} 
  G.~Bossard, P.~S.~Howe, U.~Lindstr\"om, K.~S.~Stelle and L.~Wulff,
  ``Integral invariants in maximally supersymmetric Yang-Mills theories,''
  JHEP {\bf 1105}, 021 (2011)
  [arXiv:1012.3142 [hep-th]].
    
\bibitem{BHS13} 
  G.~Bossard, P.~S.~Howe and K.~S.~Stelle,
  ``Invariants and divergences in half-maximal supergravity theories,''
  arXiv:1304.7753 [hep-th].
  
\bibitem{Novak1} 
  J.~Novak,
  ``Superform formulation for vector-tensor multiplets in conformal supergravity,''
  JHEP {\bf 1209}, 060 (2012)
  [arXiv:1205.6881 [hep-th]].
  
\bibitem{NG93} 
  H.~Nishino and S.~J.~Gates Jr.,
  ``Chern-Simons theories with supersymmetries in three-dimensions,''
  Int.\ J.\ Mod.\ Phys.\ A {\bf 8}, 3371 (1993).
  
\bibitem{WZ2}
J.~Wess and B.~Zumino,
``The component formalism follows from 
the superspace formulation of supergravity,''
Phys.\ Lett.\ B {\bf 79},  394 (1978).

\bibitem{WB}
J.~Wess and J.~Bagger,
{\it Supersymmetry and Supergravity},
Princeton University Press, Princeton, 1992.

\bibitem{KT-M11}
S.~M.~Kuzenko and G.~Tartaglino-Mazzucchelli,
``Three-dimensional N=2 (AdS) supergravity and associated supercurrents,''
JHEP {\bf 1112}, 052 (2011)
[arXiv:1109.0496 [hep-th]].

\bibitem{Kuzenko12} 
S.~M.~Kuzenko,
``Prepotentials for N=2 conformal supergravity in three dimensions,''
JHEP {\bf 1212}, 021 (2012)  [arXiv:1209.3894 [hep-th]].  

\bibitem{KLRST-M} 
S.~M.~Kuzenko, U.~Lindstr\"om, M.~Ro\v{c}ek, I.~Sachs and G.~Tartaglino-Mazzucchelli,
``Three-dimensional N=2 supergravity theories: From superspace to components,''
  arXiv:1312.4267 [hep-th].
  
\bibitem{BN} 
  D.~Butter and J.~Novak,
 ``Component reduction in N=2 supergravity: the vector, tensor, and vector-tensor multiplets,''
JHEP {\bf 1205}, 115 (2012)
  [arXiv:1201.5431 [hep-th]].
  
\bibitem{deWPV}
B.~de Wit, R.~Philippe and A.~Van Proeyen,
``The improved tensor multiplet in N = 2 supergravity,''
Nucl.\ Phys.\ B {\bf 219}, 143 (1983).

\bibitem{LR83}
U.~Lindstr\"om and M.~Ro\v{c}ek,
``Scalar tensor duality and N = 1, 2 nonlinear sigma models,''
Nucl.\ Phys.\  B {\bf 222}, 285 (1983).
  
\bibitem{deWitRocek} 
  B.~de Wit and M.~Ro\v{c}ek,
  ``Improved tensor multiplets,''
  Phys.\ Lett.\ B {\bf 109}, 439 (1982).
  
\bibitem{BK11} 
  D.~Butter and S.~M.~Kuzenko,
  ``New higher-derivative couplings in 4D N = 2 supergravity,''
  JHEP {\bf 1103}, 047 (2011)
  [arXiv:1012.5153 [hep-th]].
  
\bibitem{AT}
A.~Ach\'ucarro and P.~K.~Townsend,
``A Chern-Simons action for three-dimensional anti-de Sitter supergravity theories,''
Phys.\ Lett.\  B {\bf 180}, 89 (1986).

\bibitem{AdSpq} 
  S.~M.~Kuzenko, U.~Lindstr\"om and G.~Tartaglino-Mazzucchelli,
  ``Three-dimensional (p,q) AdS superspaces and matter couplings,''
  JHEP {\bf 1208}, 024 (2012)
  [arXiv:1205.4622 [hep-th]].
  
\bibitem{DK} 
S.~Deser and J.~H.~Kay,
``Topologically massive supergravity,''
Phys.\ Lett.\ B {\bf 120}, 97 (1983).

\bibitem{Deser}
  S.~Deser,
  ``Cosmological topological supergravity,''
 in {\it Quantum Theory Of Gravity}, S. M. Christensen (Ed.), 
 Adam Hilger, Bristol, 1984, pp. 374-381. 
  
\bibitem{BKT}
I.~L.~Buchbinder, S.~M.~Kuzenko and A.~A.~Tseytlin,
``On low-energy effective actions in N = 2,4 superconformal theories in  four
dimensions,'' Phys.\ Rev.\  D {\bf 62}, 045001 (2000) [arXiv:hep-th/9911221].

\bibitem{deWit:2010za}
  B.~de Wit, S.~Katmadas and M.~van Zalk,
``New supersymmetric higher-derivative couplings: Full N=2
    superspace does not count!,'' 
  JHEP {\bf 1101} (2011) 007
  [arXiv:1010.2150 [hep-th]].

\bibitem{KM13} 
  S.~Katmadas and R.~Minasian,
  ``N=2 higher-derivative couplings from strings,''
  arXiv:1311.4797 [hep-th].


\bibitem{HL} 
  P.~S.~Howe and M.~I.~Leeming,
  ``Harmonic superspaces in low dimensions,''
  Class.\ Quant.\ Grav.\  {\bf 11}, 2843 (1994)
  [hep-th/9408062].

\bibitem{Z2007} 
  B.~M.~Zupnik,
  ``Chern-Simons D=3, N=6 superfield theory,''
  Phys.\ Lett.\ B {\bf 660}, 254 (2008)
  [arXiv:0711.4680 [hep-th]].

\bibitem{Z2008} 
  B.~M.~Zupnik,
  ``Chern-Simons theory in SO(5)/U(2) harmonic superspace,''
  Theor.\ Math.\ Phys.\  {\bf 157}, 1550 (2008)
  [arXiv:0802.0801 [hep-th]].


\bibitem{Buchbinder:2009dc} 
  I.~L.~Buchbinder, E.~A.~Ivanov, O.~Lechtenfeld, N.~G.~Pletnev, I.~B.~Samsonov and B.~M.~Zupnik,
  ``Quantum N=3, d=3 Chern-Simons matter theories in harmonic superspace,''
  JHEP {\bf 0910}, 075 (2009)
  [arXiv:0909.2970 [hep-th]].

\end{thebibliography}
\end{document}